\documentclass[letterpaper]{article} 

\usepackage{aaai23}  

\usepackage{times}  
\usepackage{helvet}  
\usepackage{courier}  
\usepackage[hyphens]{url}  
\usepackage{graphicx} 
\usepackage{natbib}  
\usepackage{caption} 
\usepackage{algorithm}
\usepackage{algorithmic}
\usepackage{comment}
\usepackage{amsthm}
\usepackage{pgfplots}
\pgfplotsset{compat=newest}
\usetikzlibrary{intersections}
\usepgfplotslibrary{fillbetween}
\usepgfplotslibrary{groupplots}
\usepgfplotslibrary{dateplot}
\usepackage{bm}
\usepackage{bbm}
\usepackage{tikz}
\usepackage{soul}
\usepackage{tabularx}
\usepackage{booktabs}
\usepackage{mathtools}
\usepackage{multirow}
\usepackage{subcaption}
\usepackage{thm-restate}
\usepackage{nicefrac}

\usepackage{soul}

\usepackage[hidelinks]{hyperref}

\newtheorem{theorem}{Theorem}
\newtheorem*{theorem*}{Theorem}
\newtheorem{lemma}{Lemma}
\newtheorem{corollary}{Corollary}

\newtheorem{proposition}{Proposition}

\usepackage{newfloat}
\usepackage{listings}
\DeclareCaptionStyle{ruled}{labelfont=normalfont,labelsep=colon,strut=off} 
\floatstyle{ruled}
\newfloat{listing}{tb}{lst}{}
\floatname{listing}{Listing}
\title{Online Platforms and the Fair Exposure Problem under Homophily}

\author{
    Jakob Schoeffer,\textsuperscript{\rm 1}\equalcontrib\ 
    Alexander Ritchie,\textsuperscript{\rm 2}\equalcontrib\ 
    Keziah Naggita,\textsuperscript{\rm 3}\equalcontrib\ 
    Faidra Monachou,\textsuperscript{\rm 4}\equalcontrib\\
    Jessie Finocchiaro,\textsuperscript{\rm 4,\rm 5}\equalcontrib\ 
    Marc Juarez\textsuperscript{\rm 6}
}
\affiliations {
    \textsuperscript{\rm 1}Karlsruhe Institute of Technology (KIT) \\
    \textsuperscript{\rm 2}University of Michigan \\
    \textsuperscript{\rm 3}Toyota Technological Institute at Chicago \\
    \textsuperscript{\rm 4}Harvard University \\
    \textsuperscript{\rm 5}Center for Research on Computation and Society (CRCS) \\
    \textsuperscript{\rm 6}University of Edinburgh\\
    jakob.schoeffer@kit.edu, aritch@umich.edu, knaggita@ttic.edu,
    monachou@stanford.edu, jessie@seas.harvard.edu, marc.juarez@ed.ac.uk
}

\begin{document}

\maketitle

\begin{abstract}
    In the wake of increasing political extremism, online platforms have been criticized for contributing to polarization.
    One line of criticism has focused on echo chambers and the recommended content served to users by these platforms.
    In this work, we introduce the \textit{fair exposure problem}: given limited intervention power of the platform, the goal is to enforce balance in the spread of content (e.g., news articles) among two groups of users through constraints similar to those imposed by the \emph{Fairness Doctrine} in the United States in the past.
    Groups are characterized by different affiliations (e.g., political views) and have different preferences for content.
    We develop a stylized framework that models intra- and intergroup content propagation under homophily, and we formulate the platform's decision as an optimization problem that aims at maximizing user engagement, potentially under fairness constraints.
    Our main notion of fairness requires that each group see a mixture of their preferred and non-preferred content, encouraging information diversity.
    Promoting such information diversity is often viewed as desirable and a potential means for breaking out of harmful echo chambers.
    We study the solutions to both the fairness-agnostic and fairness-aware problems. We prove that a fairness-agnostic approach inevitably leads to group-homogeneous targeting  by the platform.
    This is only partially mitigated by imposing fairness constraints: we show that there exist optimal fairness-aware solutions which target one group with different types of content and the other group with only one type that is not necessarily the group's most preferred.
    Finally, using simulations with real-world data, we study the system dynamics and quantify the price of fairness.
\end{abstract}

\section{Introduction}\label{sec:introduction}

In the wake of increasing political extremism~\citep{doj2021riots}, online platforms (e.g., social media networks) have been extensively criticized for exacerbating political polarization in the United States~\citep{boxell2017internet,Bail9216,hawdon2020social,Yarchi2020PoliticalPO} and elsewhere.
This phenomenon is often attributed to platform designs that aim to generate revenue by maximizing user engagement with promoted or shared content (e.g., news articles, opinions, ads).
Motivated by the need to promote pluralism online, this paper focuses on understanding the spread of information under a limited platform intervention scheme, where the platform exposes (a subset of) users of the same affiliation to content of contrasting views.
We introduce this problem and its study as the \textit{fair exposure problem}.

From a historical perspective, parallels can be drawn between the fair exposure problem and the \textit{Fairness Doctrine} \citep{NYT_FairnessDoctrine_2021, WP_FairnessDoctrine_2021}, a past media policy which required that news media cover issues of public importance by presenting diverse, opposing perspectives in an attempt to ensure media diversity.
Over the decades, the effectiveness and ethical use of this policy was questioned \citep{WP_FairnessDoctrine_2021}: for example, the doctrine enabled activists to help combat racist broadcasting, but it also helped promote the Anti-Equal Rights Amendment campaign \citep{WP_FairnessDoctrine_2021}.
As  history has shown, interventions aimed at balancing the exposure of the public to opposing views might have ambiguous results.
Thus, the goal of this paper is to shed light on the trade-offs that the adoption of such policies may introduce for online platforms.

Towards this goal, we develop a stylized model to understand the impact of platform interventions on the propagation of different articles over time to understand the effects of \emph{positional polarization} (cf.~\citet{Yarchi2020PoliticalPO}).
Our model considers two groups of users with different affiliations and different preferences for articles.
Among two opposing articles, we assume that each group tends to like more the article that aligns with the group's views.
Moreover, due to homophily in social networks, users in a given group see mostly articles shared by other users in the same group.
In this framework, the platform wishes to maximize user exposure (measured through the aggregate number of clicks and likes), potentially subject to fairness constraints.
We only consider interventions where the platform chooses the articles that an \emph{initial set} of users in each group sees.
Our main fairness notion aims at approximately equalizing the relative exposure to a mixture of preferred and non-preferred articles across groups, by imposing certain lower and upper bounds.
We analyze the platform's optimization problem and compare the solutions for its unconstrained (fairness-agnostic) version to the solutions for its constrained (fairness-aware) version. 
We prove that the fairness-agnostic solution always targets each group with one article. 
When the platform must abide by the fairness constraints, we show that at least one group will be targeted with a mixture of articles. 
However, depending on the model parameters, it may be optimal that the other group is targeted with only one article type; interestingly, the selected article may not be the group's preferred article. Thus, one group incurs the ``cost of fairness,'' whereas the other one the ``cost of maximizing engagement.'' When the content refers to high-stakes procedures (e.g., referendums, elections), such an outcome can be problematic.

We supplement our theoretical results with empirical results to gain additional insights by estimating our model parameters from real-world datasets collected from Twitter and Facebook~\citep{balancedExposure,ReplicationExposure}.
Moreover, we measure the \emph{price of fairness}, i.e., the difference in the platform's utility between the fairness-aware and the fairness-agnostic settings. 
Using parameters estimated from~\citet{ReplicationExposure}, we observe an optimal fairness-aware solution that heavily favors one group.

\section{Related Work}\label{sec:background_and_related_work}

The spread of information in social networks is well-studied; the structure of these social networks tends to be homophilous~\citep{mcpherson2001birds,lazarsfeld1948mass}.
\citet{Yarchi2020PoliticalPO} formalize three notions of polarization that emerge from social networks: \emph{interactional, positional, and affective} polarization.
We study a model most suited to study positional polarization through information exposure.
Balancing information exposure has also been studied through several different technical methods; however, to our knowledge, the impact of platform interventions to ensure balanced exposure via fairness constraints has not been studied before.
\citet{celis2019controlling} study a similar problem of controlling polarization in bandit settings, though our model differs by assuming that intervention is only possible at the first time step; their constrained problem is similar to our approximately fair average exposure constraint in \eqref{eq:average_exposure}.
Our model is sequential like the social learning models of~\citet{banerjee1992simple,bikhchandani1992theory}, which also study information spread, but without balancing content exposure.
\citet{papanastasiou2020fake,candogan2020optimal} study stylized models for fact-checking news articles in social networks when the platform can intervene to inspect the content or incentivize fact-checking by users through information design; 
\citet{cisternas2020fake} take a market design approach. 
\citet{allon2019information} further show that polarization arises due to  uncertainty in content accuracy. 

\citet{starbird2018ecosystem} demonstrate the emergence of echo chambers by a mixed methods analysis of perceptions of the White Helmets, particularly enabled by content sharing platforms such as Twitter, and \citet{jeon2021chamberbreaker} gamify the balance of seeking influence and reputability simultaneously on Twitter.
Our setting is also similar to influence maximization literature~\citep{kempe2003maximizing} in the sense that platform interventions are limited.
However, our model is sequential and aims for balance in article exposure, while the influence maximization literature  seeks to maximize information diffusion~\citep{fish2019gaps,stoica2020seeding, ali2019fairness}.
Finally, balancing information propagation is well-studied in literature on recommender systems ~\citep{content_recommendation,adaptive_recommendation,NIPS2016_b0904096} and the emergence of echo chambers~\citep{barbera2015tweeting,mukerjee2020ideological,dubois2018echo,hosseinmardi2020evaluating}.
\citet{ReplicationExposure} and~\citet{balancedExposure} empirically study the extent of disparity in intragroup exposure of ideas and do not aim to balance it.
In general, although previous works~\citep{ReplicationExposure,balancedExposure} investigate the empirics of information flow in similar models, they do not study the mechanisms that lead to (imbalanced) exposure; our model addresses this.

Many of the standard metrics of group fairness are not applicable in our setting as we work with heterogeneous preferences of outcomes: members of one group prefer seeing content that aligns with their group identity.
Graph-based models of opportunity flow have considered similar, yet inherently different, fairness constraints and problems.
For example,~\citet{liu2021rawlsnet} consider fair equality of opportunity in settings where flow of opportunity proceeds along an acyclic graph and everyone is striving for the same desired outcome.
Similarly,~\citet{arunachaleswaran2021pipeline} approximately optimize social welfare in settings where opportunity flows along an acyclic graph.
Recently, \citet{chen2022fair} apply fairness constraints to other online platform operations, specifically assortment planning.
However, neither of the approaches in this paper are directly applicable to our setting.
The definitions of \emph{fair exposure} presented in \S~\ref{sec:model} are stylized for this particular setting.

\section{Model}\label{sec:model}

\subsection{General Setup}
We consider a platform with a finite mass $M$ of users with  affiliation group $g\in\{A,B\}$.
Let $\pi_g\in(0,1)$ denote the fraction of users from group $g$ (at any time).
We assume that $\pi_A = 1-\pi_B = \pi$.
Time is discrete with $t=1,2, \ldots, T$, $T\leq M$.
All notation is summarized in Table~\ref{tab:overview_of_notation}.

\begin{table*}[ht]
  \centering
  \begin{tabular}{ll}
    \toprule
    Symbol & Definition \\
    \midrule
    $M$ & Finite mass of users  \\
    $g\in\{A,B\}$ & Affiliation group \\
    $\pi_g\in (0,1)$ & Fraction of users in group $g$  \\
    $t\in\{1,\dots,T\}$ & Time step (discrete) with horizon $T\leq M$ \\
    $s\in \{a,b\}$ & Article sources affiliated with groups $A,B$ \\
    $\theta_{g,s} \in [0,1]$ & Fraction of group $g$ users who are shown article $s$ by the platform at $t=1$\\
    $p_{g,s} \sim F_{g,s}$ & Probability for users of group $g$ to like an article of source $s$ \\
    $F_{g,s}$ & Distribution with support $[0,1]$ \\
    $c_{g,s}>0$ & Cost users in $g$ occur when clicking on an article $s$ \\
    $v_{g,s}>0$ & Valuation of users in $g$ when liking an article $s$ \\
    $q_g \in (\nicefrac{1}{2},1)$ & Probability of intragroup propagation \\
    $l_{g,s}(t,\theta)$ & Mass of users in $g$ born at time $t$ who have clicked and liked an article $s$ \\
    $e_{g,s}(t)$ & Exposure of users in $g$ to article $s$ at time $t$ \\
    $\underline{\delta} <1 < \overline{\delta}$ & Fairness lower and upper bound parameters\\ 
    \bottomrule
  \end{tabular}  
  \caption{Overview of notation.}
  \label{tab:overview_of_notation}
\end{table*}

Before time $t=1$, the platform receives two articles representing different views $a$, $b$ that are aligned with groups $A$, $B$, respectively (e.g., sponsored posts on Facebook or Twitter).
For simplicity, we refer to the two articles as $a$ and $b$, where $a$ (resp. $b$) is the \emph{in-group}/\emph{preferred} (resp. \emph{out-group}/\emph{non-preferred}) article type of group $A$, and similarly for group $B$.
At time $t=1$, the platform decides how many users in group $g$ to show an article $s$ to. 
Let $\theta_{g,s}$  denote the fraction of users in group $g$ who are shown article $s$ by the platform at time $t=1$.

Each user observes the source $s\in\{a,b\}$ of the article they are shown.
Users of group $g$ have a probability $p_{g,s} \sim F_{g,s}$ of ``liking'' an article of source $s$, where $F_{g,s}$ is a known distribution with support $[0,1]$. 
Each user in group $g$ 
knows their own realized probabilities $p_{g,s}$ for $s \in \{a,b\}$.
Users from group $A$ have a higher preference for articles of source $a$; the same holds for users of group $B$ and articles of source $b$.
To model this (stochastically) biased behavior of users in each group $g$, we assume that $p_{A,a} \succ_{FSD} p_{A, b}$ and $p_{B,b} \succ_{FSD} p_{B, a}$.\footnote{Recall that a random variable $X$ with CDF $F_X$ \textit{first-order stochastically dominates} $Y$ with CDF $F_Y$, that is $X \succ_{FSD} Y$, if $F_X(z) < F_Y(z)$ for all $z$.}

At every time period $t >0$, a unit mass of users arrives.
At time $t$, each user in group $g$ sees one article $s$ and decides whether to click with probability dependent on $p_{g,s} \sim F_{g,s}$.
If the user clicks on the article, they incur a constant cost $c_{g,s}>0$ for reading the article.
If they like it, they get valuation $v_{g,s}>0$ (minus the cost $c_{g,s}$), so their final payoff is $v_{g,s}-c_{g,s}>0$. 
If they do not like it, their final payoff is $-c_{g,s}<0$.

At the next period $t+1$, an equal mass of users arrives.
This modeling choice reflects the general format of content sharing on social platforms, in which at different time steps, there are different batches of people on the online platform. 
We assume synchronicity in individual arrivals rather than time-step measures so we can reduce to a discretized time analysis without loss of generality.
Specifically, each user from group $g$ gives their position to a user\footnote{Our model and analysis can be directly extended to the case where a user in period $t$ is replaced by $n_{t+1}>1$ users in period $t+1$. The current assumption is made for clarity of exposition.} from the same group $g$ with probability $q_g\in (\nicefrac{1}{2},1)$, where the lower bound comes from homophily assumptions; we refer to this event as \emph{intragroup propagation}.
With probability $1-q_g$, this user is replaced by a user in $g' \neq g$ (\emph{intergroup propagation}).
In order to ensure consistency with the fraction $\pi_g$ of each group $g$ over time, we require the parameters $q_A$, $q_B$, and $\pi$  satisfy $ q_A \pi_A + (1-q_B) \pi_B = \pi_A.$\footnote{This is necessary for theoretical results, but it does not hold for the parameters used in \S~\ref{sec:simulations}, and does not affect results there.}
If a user $i$ arriving at time $t$ liked the article, then the new user $i'$, replacing  user $i$ at time $t+1$, sees the same article as $i$.
If user $i$ did not like an article, then user $i'$ is not shown any article at time $t+1$.

For $t\geq 1$, let $l_{g,s}(t,\theta)$ denote the mass of users born at time $t$ who belong to group $g$ and have clicked and liked an article $s$.
The objective of the platform is to maximize user exposure  over time, i.e.,
\footnotesize
\begin{equation}
     \max_{\theta_{A,a}, \theta_{B,a} \in [0,1]} \, \sum_{t = 1}^T \sum_{g\in \{A,B\}}\sum_{s\in \{a,b\}} l_{g,s}(t,\theta), \label{eq:platform-optimization-problem-general}
\end{equation}
\normalsize
potentially subject to  fair exposure constraints.
We measure \emph{user exposure} in the number of users who click and like an article.
The strengths of this metric are two-fold: first, because the platform has to plan for $T$ time steps, ensuring an article is liked means it will continue to propagate in the next time step.
Second, we assume that liking an article is a proxy for more \emph{meaningful} engagement than simply clicking on it.

While our model makes many simplifying assumptions, this strengthens our negative results (e.g., Lemma~\ref{lemma:possibility_of_fair_exposure}) as they do not hold even in an oversimplified model.
Moreover, while our model is not graph-based, it is an abstraction of the Erd\H{o}s-Rényi random graph in expectation with different attachment parameters for each group. 
As many social networks generally closely resemble preferential attachment models rather than Erd\H{o}s-Rényi graphs~\citep{clauset_larremore}, we compare our model’s performance to graph-based simulations in \S~\ref{sec:one-to-one}, and observe similar results.

\subsection{Notions of Fair Exposure}
Broadly speaking, we define \emph{fair exposure} as a situation where users of different affiliation are similarly exposed to non-preferred content.
Promoting such information diversity---as opposed to \emph{selective exposure}~\citep{freedman1965selective}---is often viewed as desirable and a potential means for breaking out of harmful echo chambers that are detrimental to ``the quality, safety, and diversity of discourse online,'' as~\citet{gillani2018me} put it.
\citet{garrett2011resisting}, among others, likewise suggest that ``software designers ought to create tools that
encourage and facilitate consumption of diverse news streams, making users, and society, better off.''
Diversity of perspectives might also help users to see things from novel perspectives or become aware that they might be already stuck in an echo chamber.
We operationalize fair exposure through two types of constraints: first, we ask that exposure rates for both types of content be \emph{equal at each point in time} (``constant fair exposure'').
Acknowledging that this is a rather restrictive constraint, we also examine fair \emph{average} exposure, where we further allow a certain deviation from equality (``approximately fair average exposure'').

\paragraph{Constant Fair Exposure}
The rate of exposure of users  to their preferred article $s$ is constant at level $e\in [0,1]$ at each time step and equal across groups, i.e.,
\footnotesize
\begin{equation}
    \label{eq:constant_exposure}
    \frac{l_{A,s}(t,\theta)}{\pi_A} = \frac{l_{B,s'}(t,\theta)}{\pi_B}=e\quad 
    \forall t \leq T,\ \forall s,s'\in\{a,b\}, s\neq s'.
\end{equation}
\normalsize

\paragraph{Approximately Fair Average Exposure}
The total exposure of users to their preferred article $s$ (resp. non-preferred article $s'$) is approximately equal across groups, i.e.,  for given parameters $\underline{\delta} <1 < \overline{\delta}$,
\footnotesize
\begin{equation}
\label{eq:average_exposure}
    \underline{\delta} \leq \frac{\sum_{t=1}^T l_{A,a}(t,\theta)}{\sum_{t=1}^T l_{B,b}(t,\theta)}  \leq \overline{\delta}\quad \text{and} \quad \underline{\delta}\leq \frac{\sum_{t=1}^T l_{A,b}(t,\theta)}{\sum_{t=1}^T l_{B,a}(t,\theta)}  \leq \overline{\delta}.
\end{equation}
\normalsize

\section{Theoretical Analysis}\label{sec:analysis}

\subsection{Preliminaries}
We begin with preliminaries.
We define the users' decision problem, analytically describe the system dynamics, and finally transform them to a tractable non-recursive form.

\paragraph{Users' Decision Problem}
A user in group $g$ with realized probability $p_{g,s}$ of liking an article shown to them clicks on the article if and only if their expected utility is non-negative, that is
\footnotesize
\begin{equation}
    v_{g,s}p_{g,s} \geq c_{g,s}.
\end{equation}
\normalsize
Therefore, the fraction of users in $g$ who click on  article  $s$ shown to them is $1-F_{g,s}(\frac{c_{g,s}}{v_{g,s}})$.
Since $p_{A,a} \succ_{FSD} p_{A, b}$ and $p_{B,b} \succ_{FSD} p_{B, a}$, users tend to click more on their in-group articles.

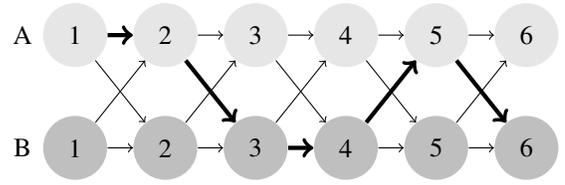
\begin{figure}[t]
\centering
\begin{tikzpicture}[scale=1,auto=center]
    \node (a0) at (0.5,0) {A}; 
    \node (b0) at (0.5, -1.5) {B};
    
    \node[style={minimum size=0.85cm, circle,fill=gray!20}] (a1) at (1.2,0) {1}; 
    \node[style={minimum size=0.85cm,circle,fill=gray!50}] (b1) at (1.2, -1.5) {1}; 
    
    \node[style={minimum size=0.85cm,circle,fill=gray!20}] (a2) at (2.4,0) {2}; 
    \node[style={minimum size=0.85cm,circle,fill=gray!50}] (b2) at (2.4, -1.5) {2}; 
    
    \node[style={minimum size=0.85cm,circle,fill=gray!20}] (a3) at (3.6,0) {3}; 
    \node[style={minimum size=0.85cm,circle,fill=gray!50}] (b3) at (3.6, -1.5) {3}; 
      
    \node[style={minimum size=0.85cm,circle,fill=gray!20}] (a4) at (4.8,0) {4}; 
    \node[style={minimum size=0.85cm,circle,fill=gray!50}] (b4) at (4.8, -1.5) {4}; 
    
    \node[style={minimum size=0.85cm,circle,fill=gray!20}] (a5) at (6,0) {5}; 
    \node[style={minimum size=0.85cm,circle,fill=gray!50}] (b5) at (6, -1.5) {5}; 
    
    \node[style={minimum size=0.85cm,circle,fill=gray!20}] (a6) at (7.2,0) {6}; 
    \node[style={minimum size=0.85cm,circle,fill=gray!50}] (b6) at (7.2, -1.5) {6};
    
    \draw[->,ultra thick] (a1) -- (a2);
    \draw[->] (a2) -- (a3);
    \draw[->] (a3) -- (a4);
    \draw[->] (a4) -- (a5);
    \draw[->] (a5) -- (a6);
    \draw[->] (b1) -- (b2);
    \draw[->] (b2) -- (b3);
    \draw[->,ultra thick] (b3) -- (b4);
    \draw[->] (b4) -- (b5);
    \draw[->] (b5) -- (b6);
    \draw[->] (b1) -- (a2);
    \draw[->] (b2) -- (a3);
    \draw[->] (b3) -- (a4);
    \draw[->,ultra thick] (b4) -- (a5);
    \draw[->] (b5) -- (a6);
    \draw[->] (a1) -- (b2);
    \draw[->,ultra thick] (a2) -- (b3);
    \draw[->] (a3) -- (b4);
    \draw[->] (a4) -- (b5);
    \draw[->,ultra thick] (a5) -- (b6);
\end{tikzpicture}  
\caption{\textit{Article sharing over time with $T=6$:} The horizontal and diagonal edges represent intragroup and intergroup propagation, respectively. With thicker edges, we give an example of how an article $s$ initially read by a user in $A$ is propagated through the network.}
\label{fig:propagation_example}
\end{figure}

\paragraph{Understanding System Dynamics}
As a warm-up, we show how the different masses of users evolve in the first time period.
We then generalize to any $t>1$.

\emph{Time ${t=1}$.}
Fix the fractions $\theta_{A,a}$ and $\theta_{B,a}$ of users in groups $A$ and $B$, respectively, who are shown article $a$ at time $t=1$; recall that $\theta_{A,a}$ and $\theta_{B,a}$ are the platform's decision.
Let $L$ denote the Bernoulli random variable that a user likes the article after clicking on it.
Then, the mass of users in $g$ who clicked on the article from source $a$ and liked it during period $t=1$ is
\footnotesize
\begin{align*}
        l_{g,s}(1, \theta)
        =&\pi_g \theta_{g,s} \int_{c_{g,s}/v_{g,s}}^1 \Pr[L=1 \mid p_{g,s}] d F_{g,s} (p_{g,s}) \\
        =& \pi_g \theta_{g,s} \int_{c_{g,s}/v_{g,s}}^1 p d F_{g,s}(p).
\end{align*}
\normalsize
Symmetrically, the mass of users in $g$ who clicked on  article $s$ but did \emph{not} like it equals $\pi_g \theta_{g,s} \int_{c_{g,s}/v_{g,s}}^1 (1-p) d F_{g,s}(p)$. 
The rest of  users  in group $g$ who were shown article $s$ did not click on it; their mass equals $\pi_g \theta_{g,s} F_{g,s}\left(\frac{c_{g,s}}{v_{g,s}}\right)$.

\emph{Time $t>1$.} 
For general $t>1$, recall that a user in group $g$ who was shown article $s$ is replaced by a user also in $g$ in the next time period with probability $q_g$ (and by a user in the opposite group $g' \neq g$ with probability $1-q_g$).
For brevity, we refer to the new user as the \textit{replacing user}. 
Figure~\ref{fig:propagation_example} illustrates how an article ``travels'' throughout the network via intra- and intergroup propagation.

Generalizing the system dynamics for $t>1$, we obtain the following recursive formula:
\footnotesize
\begin{equation}
\label{eq:mass_sh}
\begin{split}
    l_{g, s}(t+1,\theta) 
    &= \psi_{g,s}(q_g l_{g,s}(t,\theta) + (1-q_{g'}) l_{g', s}(t,\theta)),
\end{split}
\end{equation}
\normalsize
where we used\footnote{We assume $\psi_{g,s} > 0$, i.e., we do not consider the trivial case of $\psi_{g,s} = 0$.}
\[
    \psi_{g,s} \coloneqq \int_{c_{g,s}/v_{g,s}}^1 p d F_{g,s}(p).
\]
For a visual demonstration, see Figure~\ref{fig.visualization_propagation} in \S~\ref{sec:visualization_network}.
The next lemma follows from~\eqref{eq:mass_sh}.
All proofs can be found in \S~\ref{sec:proofs}.

\begin{lemma}
\label{lemma:incr_linear}
     The mass function can be written as $l_{g,s}(t,\theta) = \theta_{g,s}w_{g,s}(t) + \theta_{g',s}u_{g,s}(t)$, where $u_{g,s}(1) = 0$, $u_{g,s}(t) > 0$ for $t\geq 2$, and $w_{g,s}(t) > 0$ for $t \geq 1$.
\end{lemma}
Lemma \ref{lemma:incr_linear} says that $l_{g,s}$ is a strictly increasing linear function of $\theta_{g,s}$ and $\theta_{g',s}$ except at time $t=1$, when $l_{g,s}$ is not a function of $\theta_{g',s}$.
We note that $w_{g,s}(t)$ corresponds to the mass of intragroup propagation and $u_{g,s}(t)$ to that of intergroup propagation.

Unfortunately, the recursive expression for the mass function given in \eqref{eq:mass_sh} is intractable.
Thus, in Theorem \ref{theorem:closed_form_Z_transform_of_coeffs} we derive an equivalent non-recursive expression using the one-sided $\mathcal{Z}$-transform.

\begin{theorem}\label{theorem:closed_form_Z_transform_of_coeffs}
For all $t\geq 1$, regardless of group $g$ and article $s$, we have
\footnotesize
\begin{align}
    w_{g,s}(t) &= 
    A^w_{1,g,s}a_{1,s}^{t-1} + A^w_{2,g,s}a_{2,s}^{t-1}, \quad  t\geq 1, \\
    u_{g,s}(t) &= 
    A^u_{g,s}(a_{1,s}^{t-1} - a_{2,s}^{t-1}), \quad  t\geq 1,
\end{align}
\normalsize
with $w_{g,s}(t)$ and $u_{g,s}(t)$ as introduced in Lemma~\ref{lemma:incr_linear}, and
\footnotesize
\begin{align*}
    a_{1,s} \coloneqq &\frac{1}{2} \Big( \psi_{g,s}q_g + \psi_{g',s}q_{g'} + \big[ (\psi_{g,s}q_g + \psi_{g',s}q_{g'})^2 \\
    &- 4 \psi_{g,s}\psi_{g',s}(q_g + q_{g'} -1) \big]^{\frac{1}{2}} \Big) \\
    a_{2,s} \coloneqq &\psi_{g,s}q_g + \psi_{g',s}q_{g'} - a_{1,s} \\
    A^w_{1,g,s} \coloneqq &\frac{ \pi_g \psi_{g,s} + \psi_{g',s}\left( \pi_{g'} \psi_{g',s}(1-q_{g'}) - \pi_g \psi_{g,s} q_{g'}\right)a_{1,s}^{-1}}{1-a_{2,s}a_{1,s}^{-1}} \\
    A^w_{2,g,s} \coloneqq &\frac{ \pi_g \psi_{g,s} + \psi_{g',s}\left( \pi_{g'} \psi_{g',s}(1-q_{g'}) - \pi_g \psi_{g,s} q_{g'}\right)a_{2,s}^{-1}}{1-a_{2,s}^{-1}a_{1,s}} \\
    A^u_{g,s} \coloneqq &\frac{ \psi_{g',s}\left( \pi_{g'} \psi_{g',s}(1-q_{g'}) - \pi_g \psi_{g,s} q_{g'}\right)a_{1,s}^{-1}}{1-a_{2,s}a_{1,s}^{-1}}.
\end{align*}
\normalsize
\end{theorem}

We note that all quantities in Theorem \ref{theorem:closed_form_Z_transform_of_coeffs} are real numbers, which is shown in Lemma \ref{lemma:as_real} in \S~\ref{appendix:additional_theoretical_results}. 
An intuitive interpretation of these quantities is as follows: the terms $a_{1,s}$ and $a_{2,s}$ are the roots of a quadratic in $\mathcal{Z}$-space that roughly corresponds to a kinematic equation describing the homophilic sharing process.
$A^u_{g,s}$ corresponds roughly to the difference between contributions to the mass that would have been realized if intergroup propagation had not occurred and those that would have been realized if intragroup propagation had not occurred.
The quantities $A^w_{1,g,s}$ and $A^w_{2,g,s}$ correspond roughly to $A^u_{g,s}$ and $-A^u_{g,s}$, respectively, plus an additional term relating to the mass generated by propagation \emph{within} group $g$ and propagation \emph{from} group $g'$ to $g$.

\subsection{Platform's Optimization Problem}\label{sec:platform_optimization_problem}
Building upon our previous results, in this section we proceed to formulate the  platform's problem, i.e., the maximization of user exposure, as a linear program subject to \emph{approximately fair average exposure} constraints.
%
More specifically, at time $t=1$ the platform needs to decide the fraction of users in each group to show articles $a$ and $b$.
Recall that we denote the proportion of users in $g$ that are shown article $s$ by $\theta_{g,s}$.
The platform wants to maximize the total number of users across all groups that click on and like the two articles, but also faces
a fair exposure constraint (see \eqref{eq:average_exposure}).
Thus, the platform's optimization problem  becomes:
\footnotesize
\begin{align*}
    \tag{P}
    \max_{\theta_{A,a}, \theta_{B,a}\in [0,1]} \ &\sum_{t = 1}^T \sum_{g\in \{A,B\}}\sum_{s\in \{a,b\}} l_{g,s}(t,\theta) \label{eq:platform-objective}\\
    \tag{C1}
    \text{s.t.} \ & 
    \underline{\delta} \leq \frac{\sum_{t=1}^T l_{A,a}(t,\theta)}{\sum_{t=1}^T l_{B,b}(t,\theta)}  \leq \overline{\delta}
    \label{eq:platform-constraints-1}  \\
    \tag{C2}
    &\underline{\delta}\leq \frac{\sum_{t=1}^T l_{A,b}(t,\theta)}{\sum_{t=1}^T l_{B,a}(t,\theta)}  \leq \overline{\delta}.
    \label{eq:platform-constraints-2} \\
\end{align*}
\normalsize
Intuitively, to avoid the extreme, but feasible, case where each group $g$ is only shown their preferred article, definition \eqref{eq:average_exposure} introduces constraints \eqref{eq:platform-constraints-1} and \eqref{eq:platform-constraints-2}.
These constraints require that each group is exposed to their preferred article and their non-preferred article in a balanced way, i.e., the exposure ratio is similar for both articles within a group (within bounds $\underline{\delta} <1 < \overline{\delta}$).

From Lemma \ref{lemma:incr_linear}, we know that, given $t \in \{1,\dots,T\}$, $l_{g,s}(t,\theta)$ is a linear and strictly increasing function in $\theta_{g,s}$, $\theta_{g',s}$.
Thus, the objective function of \eqref{eq:platform-objective} is  linear in two dimensions; similarly, the exposure constraints can also be transformed to linear inequalities.
Consequently, \eqref{eq:platform-objective} is a linear program. 

\paragraph{Fairness-Agnostic Optimization Problem}
As a natural benchmark, we first consider the optimization problem \eqref{eq:platform-objective} \textit{without} exposure constraints \eqref{eq:platform-constraints-1} and \eqref{eq:platform-constraints-2}, while retaining the constraint $\theta_{g,s} \in [0,1]$ for all $g,s$.
We refer to this as the \emph{fairness-agnostic} problem.
We show that the exclusion of fairness constraints \emph{always} results in all members of the same group being shown the same article by the platform  at time $t=1$.
Specifically, the solution to the fairness-agnostic exposure problem is given in the following proposition.

\begin{proposition}
\label{prop:unconstrained_optimization}
The solution to the fairness-agnostic optimization problem is
\footnotesize
\begin{align*}
    \theta_{A,a}^* &=  \mathbf{1} \Big\{ \sum_{t=1}^T \left(w_{A,a}(t) - w_{A,b}(t) + u_{B,a}(t) - u_{B,b}(t)\right) > 0 \Big\}, \\
    \theta_{B,a}^* &=  \mathbf{1} \Big\{\sum_{t=1}^T \left( w_{B,a}(t) - w_{B,b}(t) + u_{A,a}(t) - u_{A,b}(t)\right) > 0 \Big\}.
\end{align*}
\normalsize
\end{proposition}
From a theoretical perspective, this result follows from the linearity of~\eqref{eq:platform-objective}.
From a practical perspective, Proposition~\ref{prop:unconstrained_optimization} suggests that targeting a group with their preferred article is not necessarily optimal for maximizing user engagement.
Albeit counter-intuitive, it might be optimal for the platform to ignore group preferences and target the whole user network with a single article.
Two additional implications of Proposition~\ref{prop:unconstrained_optimization} are given in Corollary~\ref{cor:feasible_never_optimal} and Lemma~\ref{lemma:condition_for_higher_intergroup_exposure} below. 

\begin{corollary}\label{cor:feasible_never_optimal}
The feasible solution $\theta_{A,b}=1$, $\theta_{B,a}=1$ is never optimal for \eqref{eq:platform-objective}.
\end{corollary}

\begin{lemma}\label{lemma:condition_for_higher_intergroup_exposure}
Assume $\theta_{A,a} = 1$, $\theta_{B,b}=1$. If
\footnotesize
\begin{equation}
\label{eq:conditions_A_more_exposed_to_b}
    \frac{q_A \pi_A  }{(1-q_B) \pi_B} <1,\quad  \frac{\psi_{A,a} \psi_{B,a}}{\pi_B \psi_{A,b} \psi_{B,b}}<1,
\end{equation}
\normalsize
then group $A$ is exposed more to article $b$ than $a$ over time, i.e., $e_{A,b}(T) = \frac{l_{A,b}(T,\theta)}{\pi_A} > e_{A,a}(T) = \frac{l_{A,a}(T,\theta)}{\pi_A}$ for any $T>2$. 
\end{lemma}

Under homophily, one might expect a group to be preferentially exposed to in-group articles.
However, as Lemma~\ref{lemma:condition_for_higher_intergroup_exposure} shows, this may not be the case if group sizes are radically different or if one group displays much lower levels of homophily than the other.
Lemma \ref{lemma:condition_for_higher_intergroup_exposure} can shed light on several counter-intuitive possibilities for article exposure over time. 
More specifically, it suggests that, due to the network structure and the dynamics of propagation, targeting each group with their preferred article might \textit{not always} bring the intended targeting  and thus potentially lead to suboptimal outcomes for the platform.
Even if the platform targets each group only with their preferred (in-group) article, one group may---after several rounds---be exposed to their non-preferred (out-group) article. 
For example, given a significantly larger group $B$, weak homophily for both groups ($q_A \simeq q_B \simeq \nicefrac{1}{2}$) and similar preferences for compatible articles ($\psi_{A,a} \simeq \psi_{B,b}$), group $A$ is exposed more to article $b$ even if $\theta_{A,a}=1$ and $\theta_{B,b}=1$.
A similar property holds when there is an extreme preference for article $b$ in group $B$ compared to moderate preference in group $A$, i.e., $\psi_{B,b} \gg \psi_{A,a}$.

In contrast to Lemma~\ref{lemma:condition_for_higher_intergroup_exposure}, Corollary~\ref{cor:feasible_never_optimal} offers a quite intuitive insight, showing that the opposite strategy (i.e., targeting both groups with their out-group article) is \textit{never} optimal.
Indeed, depending on the model parameters, either the network will eventually favor the article with the largest sharing rate in total or the users in each group will start clicking more on their in-group article.
In both cases, the platform's initial targeting $\theta_{A,b}=1$, $\theta_{B,a}=1$ would only manage to delay any of these events thus leading to a suboptimal number of clicks and likes at the initial stages of propagation.

\paragraph{Fairness-Aware Optimization Problem: Constant Fair Exposure}
In this section, we explore the feasibility of a natural but stricter fairness notion, i.e., constant fair exposure, as defined in~\eqref{eq:constant_exposure}.
As detailed in Lemma \ref{lemma:possibility_of_fair_exposure} below, we show that it is generally not possible to achieve  equal and constant exposure at every time step unless certain restrictive conditions hold.

\begin{lemma}\label{lemma:possibility_of_fair_exposure}
Let $e \in (0,1)$ be the platform's targeted fair exposure level.
Achieving constant fair exposure is possible if and only if for both $s\in\{a,b\}$,
\footnotesize
\begin{align}
    \label{eq:condition_equal_constant_exposure}
    \begin{split}
        \psi_{A,s}&\left(q_A  + \frac{1-\pi_A}{\pi_A} (1-q_B) \right) \\
    = \psi_{B,s'}&\left( \frac{1-\pi_A}{\pi_A} q_B  + (1-q_A)\right)
    \end{split}
\end{align}
\normalsize
and the platform sets $\theta_{A,a} = 1-\theta_{B,a}=e $ at time $t=1$.
\end{lemma}

The conditions of Lemma \ref{lemma:possibility_of_fair_exposure} guarantee that the mass of users clicking on a given article will be the same across all groups and time steps.
However, this will almost certainly never occur in practice due to differing preferences in content across groups.
Therefore, we ask if average exposure over time can be equalized across groups, i.e., if $\frac{1}{T}\sum_{t=1}^T\frac{l_{A,s}(t,\theta)}{\pi_A} = \frac{1}{T}\sum_{t=1}^T\frac{l_{B,s'}(t,\theta)}{\pi_B}=e$ is possible.
Lemma \ref{lemma:avg_exp_1group} shows that it is very difficult to achieve any desired average exposure rate:

\begin{lemma} \label{lemma:avg_exp_1group}
For any $\pi_g \in (0,1)$, average exposure levels for group $g$ to article $s$ are achievable only in the range $0 \leq e \leq \frac{1}{T \pi_g} \sum_{t=1}^T \left(w_{g,s}(t) + u_{g,s}(t) \right)$.
\end{lemma}

\paragraph{Fairness-Aware Optimization Problem: Approximately Fair Average Exposure}
Given the restrictive nature of constant fair exposure, we  turn to a relaxed notion.
Specifically, we explore the feasibility of the optimization problem \eqref{eq:platform-objective} \emph{with} fairness constraints \eqref{eq:platform-constraints-1} and \eqref{eq:platform-constraints-2}, and analytically describe the solution by deriving expressions for the extreme points of the constraint polytope.
Let
\footnotesize
\begin{align*}
    \overline{m}_{g,s} &\coloneqq \sum_{t=1}^T u_{g,s}(t) + \overline{\delta}\sum_{t=1}^T w_{g',s'}(t), \\
    \underline{m}_{g,s} &\coloneqq \sum_{t=1}^T u_{g,s}(t) + \underline{\delta}\sum_{t=1}^T w_{g',s'}(t), \\
    \overline{n}_{g,s} &\coloneqq \sum_{t=1}^T w_{g,s}(t) + \overline{\delta}\sum_{t=1}^T u_{g',s'}(t), \\
    \underline{n}_{g,s} &\coloneqq \sum_{t=1}^T w_{g,s}(t) + \underline{\delta}\sum_{t=1}^T u_{g',s'}(t), \\
    m_{g,s} &\coloneqq \sum_{t=1}^T u_{g,s}(t) + \sum_{t=1}^T w_{g,s}(t).
\end{align*}
\normalsize
From constraints \eqref{eq:platform-constraints-1} and \eqref{eq:platform-constraints-2} and using Theorem \ref{theorem:closed_form_Z_transform_of_coeffs}, we can infer the feasible bounds on $\theta_{B,a}$ (dependent on $\theta_{A,a}$), in addition to $\theta_{A,a}, \theta_{B,a} \in [0,1]$.
We state these bounds as well as the axes intersects of the hyperplanes that induce the half-spaces containing the feasible region in \S~\ref{sec:addendum_constrained_problem}.
Evaluating the relative positions of these hyperplanes, we can then infer when the fairness-aware optimization problem is infeasible:

\begin{theorem}\label{theorem:feasibility}
The fairness-aware optimization problem is infeasible if and only if one of the following holds:
\footnotesize
    \begin{align*}
    \frac{\underline{\delta}m_{B,b}}{\underline{m}_{A,a}} > \frac{m_{A,b}}{\underline{m}_{A,b}}\quad &\text{and}\quad \frac{\underline{\delta}m_{B,b}}{\underline{n}_{A,a}} > \frac{m_{A,b}}{\underline{n}_{A,b}}; \\
    \frac{m_{A,b}}{\overline{m}_{A,b}} > \frac{\overline{\delta}m_{B,b}}{\overline{m}_{A,a}}\quad &\text{and}\quad \frac{m_{A,b}}{\overline{n}_{A,b}} > \frac{\overline{\delta} m_{B,b}}{\overline{n}_{A,a}}; \\
    \underline{\delta} \sum_{t=1}^T w_{B,b}(t) > \sum_{t=1}^T w_{A,a}(t)\quad &\text{and}\quad \frac{\underline{\delta} m_{B,b}}{\underline{m}_{A,a}} > \frac{\underline{\delta} m_{B,b}}{\underline{\delta} m_{B,b} - \underline{n}_{A,a}} ; \\
    \sum_{t=1}^T u_{A,b}(t) > \overline{\delta} \sum_{t=1}^T u_{B,a}(t)\quad &\text{and}\quad \frac{m_{A,b}}{\overline{m}_{A,b}} > \frac{m_{A,b}}{m_{A,b} - \overline{n}_{A,b}}.
\end{align*}
\end{theorem}
It follows that we can always make the problem feasible by setting $\underline{\delta}$ and $\overline{\delta}$ accordingly.
As noted in \S~\ref{sec:addendum_constrained_problem}, letting $\underline{\delta} \to 0$ and $\overline{\delta} \to \infty$, the fairness-agnostic problem is recovered.
By the intermediate value theorem, there exist infinitely many values of $\underline{\delta},\overline{\delta}$ that define a non-empty feasible region strictly contained in the unit box.
Otherwise, if the problem is feasible, the fundamental theorem of linear programming states that an optimal solution will occur at a corner point of the feasible region, or on a line segment between two corner points.
Theorem \ref{thm:constrained_soln} (deferred to \S~\ref{appendix:additional_theoretical_results}) states the collection of possible solutions $\theta_{g,s}^i$ to the \emph{fairness-aware} optimization problem.
In particular, note that all of these solutions may not be feasible for a particular problem instance.
Which of these solutions is feasible and optimal will depend on the true problem parameters.
In particular, define $c_{g,s} \coloneqq \sum_{t=1}^T (w_{g,s}(t) - w_{g,s'}(t) + u_{g',s}(t) - u_{g',s'}(t))$ and write the objective as $\theta_{A,a}c_{A,a} + \theta_{B,a}c_{B,a}$.
As in Proposition~\ref{prop:unconstrained_optimization}, the particular solution then depends on the signs and relative magnitudes of $c_{A,a}$ and $c_{B,a}$.
For example, if $c_{A,a} \gg c_{B,a} > 0$, then the largest feasible value of $\theta_{A,a}^i$ and the corresponding $\theta_{B,a}^i$ will be the optimal solution; see Figure~\ref{fig:graphical_solution} in \S~\ref{sec:fairness_constraints} for an illustration.

The main difference to Proposition~\ref{prop:unconstrained_optimization} is that, due to the imposed fairness constraints, some of the optimal unconstrained solutions might be out of the feasible region.
At a higher level, the more restrictive the bounds $\underline{\delta}, \overline{\delta}$ get, the further we move from the optimal binary solution of the fairness-agnostic problem.
Thus, some solutions correspond to a mixture of articles shown to each group, and no group is targeted with one article type.
However, others may correspond to cases where exactly one group is targeted with only one article, while the other sees both articles at unequal rates. 
Observe that it is still possible that a group is only shown their out-group article.

Our results offer novel insights for platform design. 
Even though satisfying \eqref{eq:platform-constraints-1} and \eqref{eq:platform-constraints-2} imposes a significant restriction on the platform and ostensibly seems to ensure a balanced exposure up to some extent, extreme solutions may still arise.
Introducing fairness constraints does not automatically imply that the final outcome is \textit{truly} fair---or even balanced.
Furthermore, in any solution $i$ where $\theta_{g,s}^i \in \{0,1\}$ while $0<\theta_{g',s}^i<1$, only one group incurs the ``price of fairness'' whereas the other group, which is targeted with only one article, serves the platform's major goal of maximizing clicks.
(Note that one can verify that half of the solutions in Theorem~\ref{thm:constrained_soln} have this property.)
Thus, when the content is related to sensitive or high-stakes procedures (e.g., a referendum), ensuring fair exposure is not just a technical challenge; if the interventions are not carefully designed (e.g., choosing $\underline{\delta}$, $\overline{\delta}$ thoughtfully), they can lead to unintended outcomes, potentially with legal consequences. 

\section{Simulations}\label{sec:simulations}

We use our model to empirically study the effects of different model parameters from real-world click data.\footnote{All code is available at \url{https://github.com/jfinocchiaro/fair-exposure}.}
Tables~\ref{tab:exp-param-vals} and~\ref{tab:default-params-data-indep} in \S~\ref{app:experimental} describe the parameters used, such as number of runs, proportional representation of each group, among others.
We use maximum likelihood estimation to fit parameter values from~\citet{ReplicationExposure} in this section, and study three datasets from~\citet{balancedExposure} deferred to \S~\ref{subapp:additional-experiments}.
For $p$, we fit a beta distribution and present the parameters $\alpha$ and $\beta$ in the appendix.
See \S~\ref{app:experimental} for additional experiments evaluating the effects of population-based parameters.

%

\paragraph{Effect of Fairness Bounds $\underline{\delta}$ and $\overline{\delta}$ on $\theta$}
We start by studying the effect of different model parameters on the platform's optimization and outcomes.
In particular, we focus on the change of $\underline{\delta}$ and $\overline{\delta}$, and its impacts on exposure and click rates, both en masse and across groups.
%
%
\begin{figure}[ht!]
\begin{minipage}{0.495\linewidth}
    \centering
    \includegraphics[width=\linewidth]{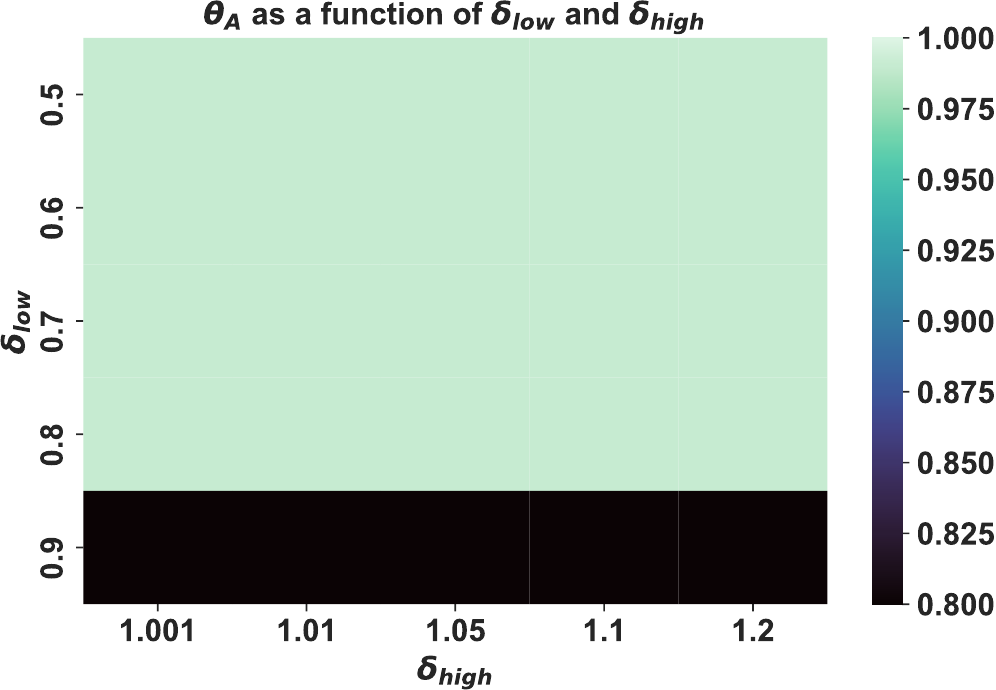}
\end{minipage}
\hfill
\begin{minipage}{0.495\linewidth}
    \centering
    \includegraphics[width=\linewidth]{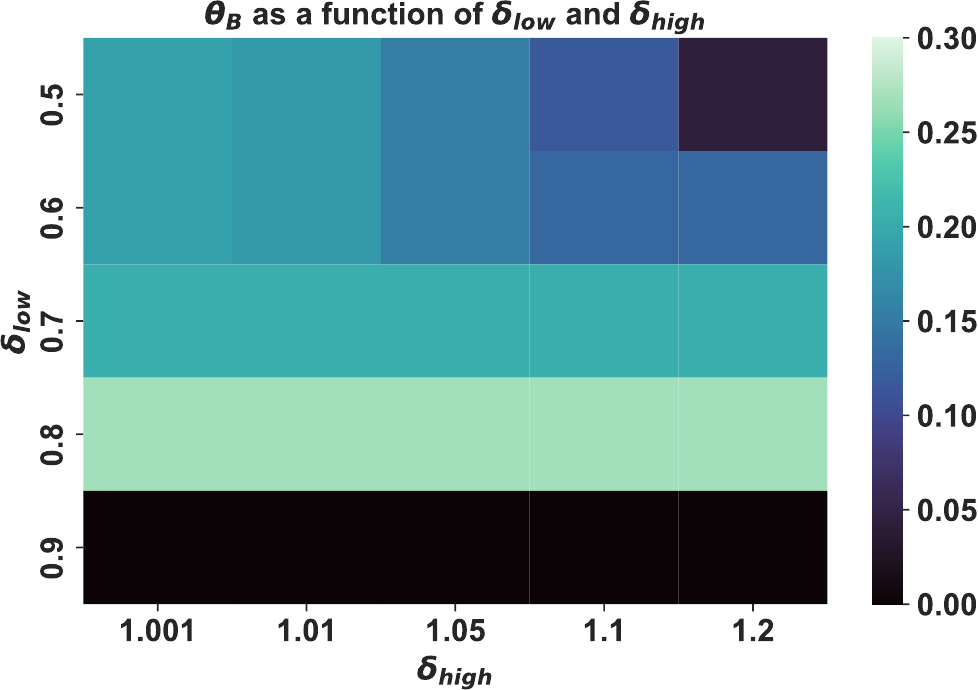}
\end{minipage}
\caption{Calculating $\theta_{A,a}$ (left) and $\theta_{B,a}$ (right) as a function of $\underline{\delta}$ and $\overline{\delta}$ with parameters estimated from~\citet{ReplicationExposure}. Black cells at the bottom indicate no feasible solution to the fairness-constrained problem.}
\label{fig:facebook-theta-heatmaps}
\end{figure}
Figure~\ref{fig:facebook-theta-heatmaps} illustrates that the optimal solution  is to almost always show article $a$ to members of group $A$, and fair exposure is then enforced by restricting how group $B$ is shown articles.
In this case, setting $\overline{\delta}$ closer to $1$ (making the constraint more restrictive) generally increases the proportion of members of group $B$ who are shown article $a$.
It is helpful to understand when the fairness-aware problem is $(i)$ feasible and $(ii)$ restrictive; if $\underline{\delta}$ and $\overline{\delta}$ are too close to $1$, the feasible region may be empty (as in the bottom row of Figure~\ref{fig:facebook-theta-heatmaps}), but if they are too far from $1$, they may not constrain the fairness-agnostic problem around the agnostic optimum.
For intuition on how these parameters may affect the feasible region, see Figure~\ref{fig:graphical_solution} in \S~\ref{sec:addendum_constrained_problem}.

\paragraph{Exposure Disparity}
We are also interested in understanding how imposing balanced exposure constraints  might affect disparity in expected exposure and clicks.
Figure~\ref{fig:facebook-boxplots-dispartity} in \S~\ref{app:experimental} highlights the disparity in exposure for different optimization policies $\theta$.
%
%
There, we observe that a uniformly randomized policy (\textit{random}) and proportional policy (\textit{proportional}; $\theta_{g,s} = \pi_{g}$) yield a large disparity in article exposure between article $a$ and article $b$, while this disparity is lower in the fairness-agnostic (\textit{unconstrained}) and fairness-aware (\textit{fair}) settings, though there is no significant difference between the two.
When evaluating differences in how often the articles get liked, this gap closes across all four policies.
%

\begin{figure*}
     \centering
     \begin{subfigure}[b]{0.33\textwidth}
         \centering
         \includegraphics[width=\textwidth]{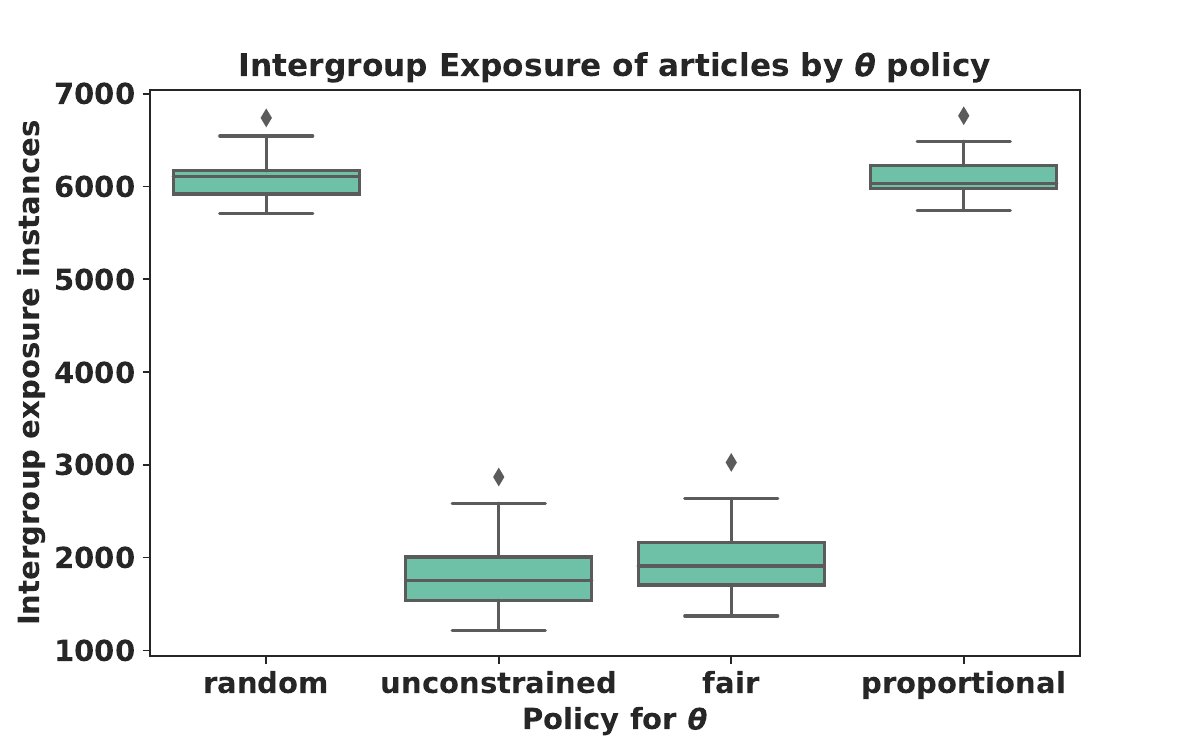}
         \caption{}
         \label{fig:boxplot_facebook_exposure}
     \end{subfigure}
     \hfill
     \begin{subfigure}[b]{0.33\textwidth}
         \centering
         \includegraphics[width=\textwidth]{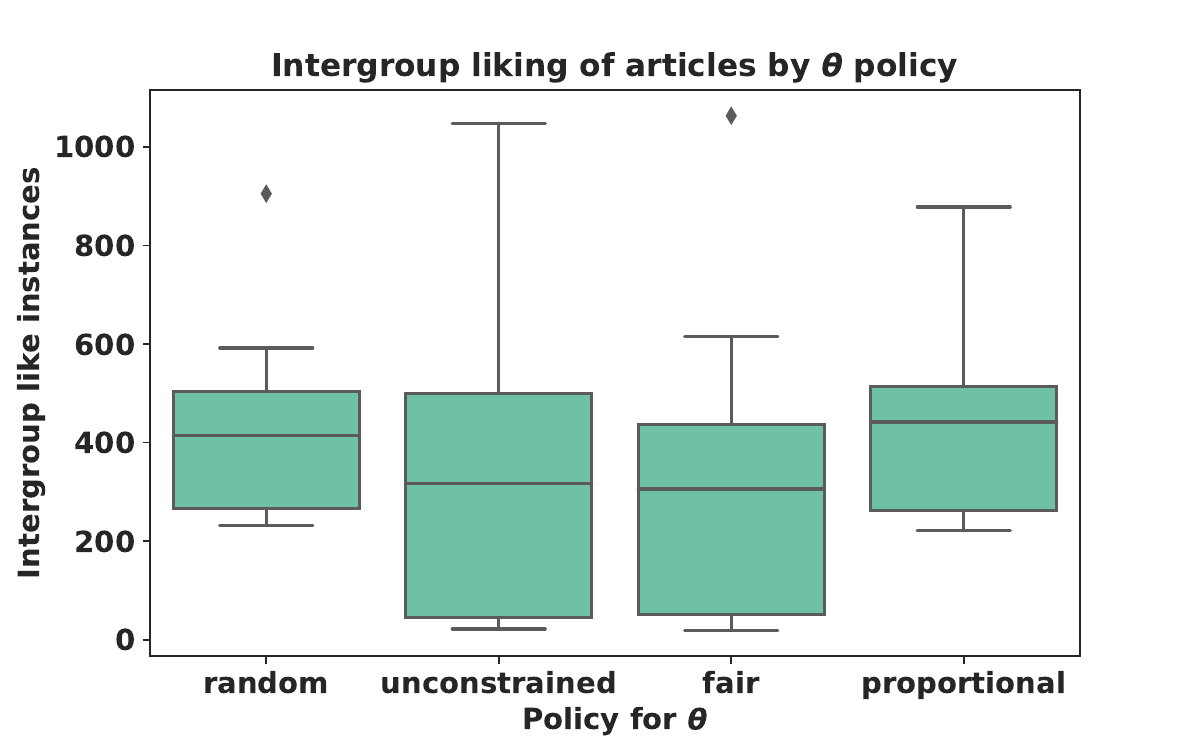}
         \caption{}
         \label{fig:boxplot_facebook_likes}
     \end{subfigure}
     \hfill
     \begin{subfigure}[b]{0.33\textwidth}
         \centering
         \includegraphics[width=\textwidth]{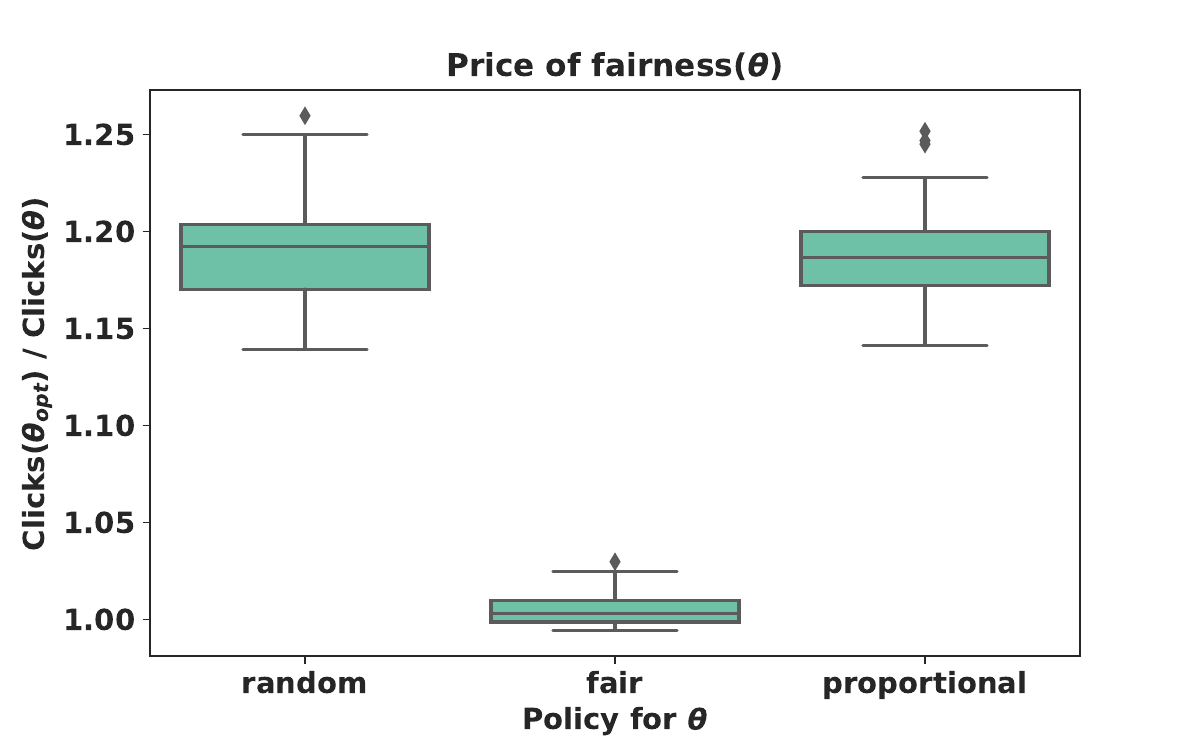}
         \caption{}
         \label{fig:facebook_pof_boxplot}
     \end{subfigure}
        \caption{Intergroup exposure (a) and liking (b), as well as the price of fairness (c), using model parameters from~\citet{ReplicationExposure}.}
        \label{fig:facebook}
\end{figure*}

\paragraph{Engagement Disparity}
Perhaps unsurprisingly, we can see in Figure~\ref{fig:boxplot_facebook_exposure} that intergroup exposure is significantly higher when randomizing exposure than when optimizing exposure as in the fairness-agnostic (\textit{unconstrained}) and fairness-aware (\textit{fair}) settings.
When evaluating the number of \emph{likes} across groups in Figure~\ref{fig:boxplot_facebook_likes}, this arises as an artifact of the model more generally, though the gap significantly decreases.
Optimizing in fairness-aware and -agnostic settings yield relatively similar distributions of intergroup likes on articles.


\paragraph{Price of Fairness}
We consider the price of fairness similar to that of~\citet{bertsimas2011price}, given in~\eqref{eq:price-of-fairness}.
Here, a lower price of fairness for a given policy is better, as it indicates being closer to the fairness-agnostic optimization problem.

\footnotesize
\begin{equation}\label{eq:price-of-fairness}
    POF(\theta) = \frac{\# clicks(\theta_{opt})}{\# clicks(\theta)}
\end{equation}
\normalsize

We can see in Figure~\ref{fig:facebook_pof_boxplot} that the price of fairness for the fairness-aware optimization problem is close to $1$ in most trials, which is observationally lower than the price of fairness for a uniformly randomized or proportional policy.
This suggests that our fairness-aware optimization problem yields approximately the same number of clicks as the fairness-agnostic solution. 
Figures~\ref{fig:twitter_uselections_pof_boxplot}, \ref{fig:twitter_brexit_pof_boxplot}, and~\ref{fig:twitter_abortion_pof_boxplot} in \S~\ref{subapp:additional-experiments} show the price of fairness for adding constraints compared to a uniformly random policy using the parameters estimated from~\citet{balancedExposure}. 
%


\section{Discussion and Conclusion}\label{sec:discussion_and_limitations}

Motivated by the concerning increase in polarization in social media platforms, this paper introduces the fair exposure problem and develops a theoretical dynamic model to study its implications. 
Albeit simple and intuitive, our model is highly stylized (as other models in the literature \citep{papanastasiou2020fake, allon2019information}).
One simplification is the propagation scheme which aims at approximating article sharing and user exposure in a computationally tractable way. 
Thus, our framework offers novel insights about the propagation in expectation across groups (instead of propagation from individual to individual).
Nevertheless, a theoretical analysis using an underlying graph structure would be a natural extension; we study this more realistic scenario through simulations in \S~\ref{sec:one-to-one}.
Another assumption of our model is that each user can see only one article.
We make this modeling choice merely for technical simplicity that offers tractability and clearer insights.
However, a partial interpretation of this assumption would be that the platform has limited slots for promoted content or that users most likely click on the first article they see~\citep{robertson1978ranking,wang2013incorporating,craswell2008experimental}.

As the Fairness Doctrine was introduced to ensure opinion diversity, a modern version of this policy could be similarly introduced in online platforms \citep{WP_FairnessDoctrine_2021}.
Although both the Fairness Doctrine and our model expose users of different groups to a diverse set of news articles at the same time, our model might expose each individual to only one article.
Such platform interventions have the great potential to ensure diversity of viewpoints; however, the design of such policies entails the careful examination of any ethical concerns. 
A question that naturally arises is whether it is ethical for the platform to algorithmically control and potentially randomize the content that a user sees and, ultimately, who---if anyone---has the responsibility to ensure fair exposure in online spaces. 
This question has been under close scrutiny in interpreting Section 230 of the Communications Decency Act in the United States~\citep{sec230}. 
For example, given that news sharing and discussions in social media can determine important political outcomes and thus the passive or more restrictive role that the platform chooses to undertake matters (see, e.g., \citet{BBC_Trump, FB_NYT2020}), it is unclear how a fair representation of content should be defined.
Thus, we acknowledge that balancing exposure to different ideologies of content might not actually be fair in a given context. 
For instance, \citet{Bail9216} suggest that showing people opposing viewpoints makes them more polarized, whereas \citet{Becker10717} show that echo chambers do not necessarily increase polarization.
Furthermore, considering the amount of disinformation and the technical challenges in identifying problematic content (e.g.,  fake news, hate speech) in platforms, the fair exposure constraints should not be applied to all content.
Implementing fair exposure can thus become particularly challenging, and more interdisciplinary research is needed to understand where to draw the boundary.
Our work is an initial step towards this broader goal.

Finally, our framework highlights how the introduction of fairness constraints can only partially mitigate group-homogeneous targeting and points to problematic outcomes, as sometimes only one group incurs the ``price of fairness'' while the other pays the ``cost of user engagement.''
It also gives rise to a series of emerging, challenging directions for future research related to platforms and algorithmic fairness.
These include the study of fair exposure notions, the design of dynamic interventions and more sophisticated targeting, ad pricing and revenue maximization under fair exposure constraints, and their implications on the competition among different platforms.

\section*{Acknowledgements}
This project has been part of the MD4SG working
group on Bias, Discrimination, and Fairness.
The material is based on work supported by the National Science Foundation under Graduate Research Fellowship No.\,DGE-1650115 and National Science Foundation Award No.\,2202898 (Jessie Finocchiaro).
Keziah Naggita was supported in part by the National Science Foundation under Grant No.\,CCF-1815011 and by the Simons Foundation under the Simons Collaboration on the Theory of Algorithmic Fairness.
Any opinions, findings, and conclusions or recommendations expressed in this material are those of the authors and do not necessarily reflect the views of the funding sources.

\bibliography{bibliography}

\onecolumn
\newpage
\appendix

\section{Visualization of Network Propagation}\label{sec:visualization_network}

\begin{figure}[h]
\centering
\resizebox{\textwidth}{!}{
\begin{tikzpicture}[scale=1,auto=center]
  
\node (a15) at (-1.5,0) {$t=1:$};
\node (a16) at (-2.5,-2) {$t=2:$};
\node (a17) at (-4.5,-4) {$t=3:$};
\node (a18) at (-8.5,-6) {$t=4:$};
   
\node (a0) at (0,0) [rectangle,draw] {$\pi_A \psi_{A,a}$};

\node (a1) at (-1,-2) [rectangle,draw] {$q_A \psi_{A,a}$};
\node (a2) at (1,-2) [rectangle,draw] {$(1-q_A) \psi_{B,a}$};

\node (a3) at (-3,-4) [rectangle,draw] {$q_A \psi_{A,a}$};
\node (a4) at (-1,-4) [rectangle,draw] {$(1-q_A) \psi_{B,a}$};
\node (a5) at (1,-4) [rectangle,draw] {$q_B \psi_{B,a}$};
\node (a6) at (3,-4) [rectangle,draw] {$(1-q_B) \psi_{A,a}$};

\node (a7) at (-7,-6) [rectangle,draw,fill=gray!20] {$q_A \psi_{A,a}$};
\node (a8) at (-5,-6) [rectangle,draw] {$(1-q_A) \psi_{B,a}$};
\node (a9) at (-3,-6) [rectangle,draw] {$q_B \psi_{B,a}$};
\node (a10) at (-1,-6) [rectangle,draw,fill=gray!20] {$(1-q_B) \psi_{A,a}$};
\node (a11) at (1,-6) [rectangle,draw] {$q_B \psi_{B,a}$};
\node (a12) at (3,-6) [rectangle,draw,fill=gray!20] {$(1-q_B) \psi_{A,a}$};
\node (a13) at (5,-6) [rectangle,draw,fill=gray!20] {$q_A \psi_{A,a}$};
\node (a14) at (7,-6) [rectangle,draw] {$(1-q_A) \psi_{B,a}$};

\draw[->, ultra thick] (a0) -- (a1);
\draw[->, ultra thick] (a0) -- (a2);

\draw[->, ultra thick] (a1) -- (a3);
\draw[->, ultra thick] (a1) -- (a4);

\draw[->, ultra thick] (a2) -- (a5);
\draw[->, ultra thick] (a2) -- (a6);

\draw[->, ultra thick] (a3) -- (a7);
\draw[->] (a3) -- (a8);

\draw[->] (a4) -- (a9);
\draw[->, ultra thick] (a4) -- (a10);

\draw[->] (a5) -- (a11);
\draw[->, ultra thick] (a5) -- (a12);

\draw[->, ultra thick] (a6) -- (a13);
\draw[->] (a6) -- (a14);

\end{tikzpicture}
}
\caption{\textit{Example visualization of the network propagation for $l_{A,a}$, with $T=4$, $\theta_{A,a}=1$, $\theta_{B,a}=0$}: The value of $l_{A,a}(t,\theta)$ is obtained by multiplying along the thick edges and adding up all the $2^{T-2}$ paths leading into leaves that include $q_A$ (intragroup propagation within $A$) or $1-q_B$ (intergroup propagation from $B$ to $A$) as a factor (depicted in gray).}
\label{fig.visualization_propagation}
\end{figure}
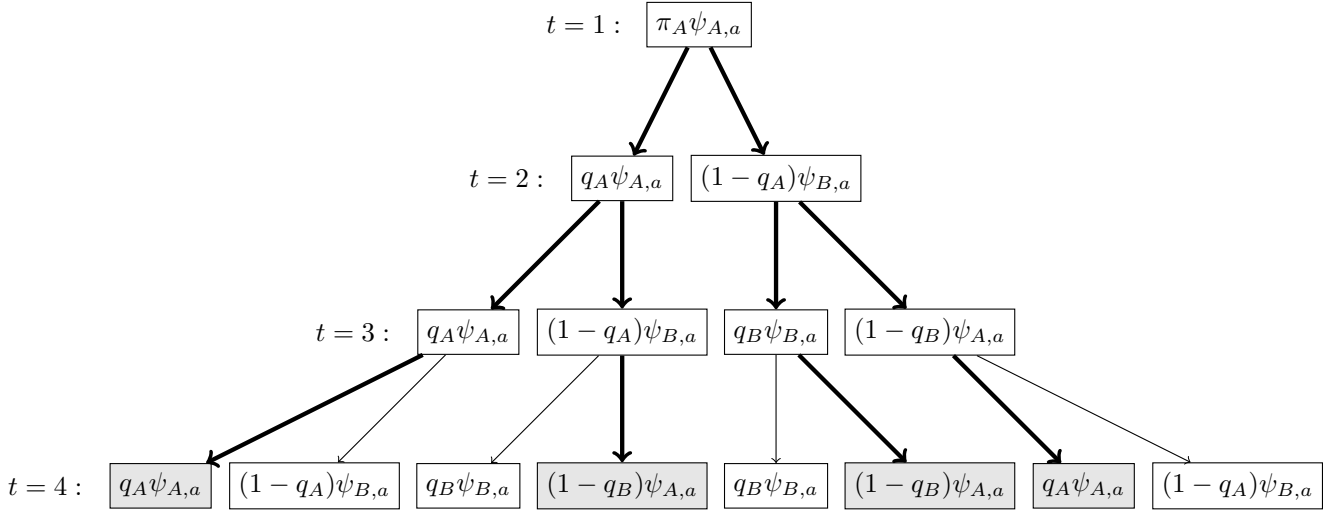

A visualization of a network propagation is given in Figure~\ref{fig.visualization_propagation}.

\section{Proofs for Theoretical Results in the Main Text}\label{sec:proofs}

\paragraph{Proof for Lemma \ref{lemma:incr_linear}}
\begin{proof}
The proof proceeds by induction. For the base case, let $t=2$ and note that
\begin{align*}
    l_{g,s}(2,\theta) &= \theta_{g,s}q_g \pi_g \psi_{g,s}^2 + \theta_{g',s}(1-q_{g'})\pi_{g'}\psi_{g',s}\psi_{g,s} \\
    l_{g',s}(2,\theta) &= \theta_{g,s}(1-q_{g})\pi_{g}\psi_{g',s}\psi_{g,s} + \theta_{g',s}q_{g'} \pi_{g'} \psi_{g',s}^2 
\end{align*}
are strictly increasing linear functions of $\theta_{g,s}$ and $\theta_{g',s}$.

Now suppose for some $t \geq 2$
\begin{align*}
    l_{g,s}(t,\theta) &= \theta_{g,s}w_1(t) + \theta_{g',s}w_2(t) \\
    l_{g',s}(t,\theta) &= \theta_{g,s}u_1(t) + \theta_{g',s}u_2(t), 
\end{align*}
where $w_1(t), w_2(t), u_1(t), u_2(t) > 0$. It follows from~\eqref{eq:mass_sh} that
\begin{align*}
    l_{g,s}(t+1,\theta) &= \theta_{g,s}\left( q_g \psi_{g,s} w_1(t) + (1-q_{g'})\psi_{g',s}u_1(t)\right) \\
    &\quad + \theta_{g',s}\left( q_g \psi_{g,s} w_2(t) + (1-q_{g'})\psi_{g',s}u_2(t)\right),
\end{align*}
which is a strictly increasing linear function of $\theta_{g,s}$ and $\theta_{g',s}$. We conclude that $l_{g,s}(t,\theta)$ is a strictly increasing linear function of $\theta_{g,s}$ and $\theta_{g',s}$ for all $t\geq 2$. Note that for $t=1$, $l_{g,s}(1,\theta) = \theta_{g,s}\pi_g \psi_{g,s}$, which is a strictly increasing linear function of $\theta_{g,s}$ and does not depend on $\theta_{g',s}$. The case for $l_{g',s}(t,\theta)$ is similar. This completes the proof.
\end{proof}

\paragraph{Proof for Theorem \ref{theorem:closed_form_Z_transform_of_coeffs}}
\begin{proof}
We will go through the proof for $w_{g,s}$, noting that the case for $u_{g,s}$ is similar. 

We first mention that $w_{g,s}(t)$ is a right-tailed sequence, so we will be using the one-sided $\mathcal{Z}$-transform. 
This is important because~\eqref{eq:mass_sh} only holds for $t \geq 1$. 
We first rewrite~\eqref{eq:mass_sh} with different initial conditions so that the relation holds for all $t \geq 0$,
\begin{align}
    w_{g,s}(t) = &\psi_{g,s} q_g w_{g,s}(t-1) + \psi_{g',s}(1-q_{g'}) w_{g',s}(t-1) \label{eq:lgs_mod} + \delta(t) \pi_g \psi_{g,s}.
\end{align}
Here, $\delta(t)$ is the Kronecker delta function and we take $w_{g,s}(-1) = w_{g',s}(-1) = 0$.
Here we begin the time index at $t=0$ to use the $\mathcal{Z}$-transform, and the index will later be shifted to start at $t=1$.

Denote the one-sided $\mathcal{Z}$-transform of~\eqref{eq:lgs_mod} as $\mathcal{W}_{g,s}(z)$, which is given by
\begin{align}
   \mathcal{W}_{g,s}(z) &= z^{-1} \psi_{g,s}q_g \mathcal{W}_{g,s}(z) + z^{-1} \psi_{g',s}(1-q_{g'}) \mathcal{W}_{g',s}(z) + \pi_g \psi_{g,s} \nonumber \\
   &= \frac{\psi_{g',s}(1-q_{g'})z^{-1}}{1 - \psi_{g,s}q_gz^{-1}}\mathcal{W}_{g',s}(z) + \frac{\pi_g \psi_{g,s}}{1 - \psi_{g,s}q_gz^{-1}}. \label{eq:zlgs}
\end{align}
Similarly, we write the $\mathcal{Z}$-transform of $w_{g',s}$ using the definition of $w_{g',s}$ analogous to~\eqref{eq:lgs_mod}:
\begin{align}
   \mathcal{W}_{g',s}(z) &= z^{-1} \psi_{g',s}q_{g'} \mathcal{W}_{g',s}(z) + z^{-1} \psi_{g,s}(1-q_{g}) \mathcal{W}_{g,s}(z) + \pi_{g'} \psi_{g',s} \nonumber\\
   &= \frac{\psi_{g,s}(1-q_{g})z^{-1}}{1 - \psi_{g',s}q_{g'}z^{-1}}\mathcal{W}_{g,s}(z) + \frac{\pi_{g'} \psi_{g',s}}{1 - \psi_{g',s}q_{g'}z^{-1}}. \label{eq:zlg's}
\end{align}

Substituting~\eqref{eq:zlg's} into~\eqref{eq:zlgs} yields
\begin{align*}
    \mathcal{W}_{g,s}(z) &= \left( 1 - \frac{\psi_{g',s}(1-q_{g'})z^{-1}}{1 - \psi_{g,s}q_gz^{-1}} \frac{\psi_{g,s}(1-q_{g})z^{-1}}{1 - \psi_{g',s}q_{g'}z^{-1}} \right)^{-1} \\
    &\quad \cdot \left(\frac{\psi_{g',s}(1-q_{g'})z^{-1}}{1 - \psi_{g,s}q_gz^{-1}}\frac{ \pi_{g'} \psi_{g',s}}{1 - \psi_{g',s}q_{g'}z^{-1}} +  \frac{ \pi_g \psi_{g,s}}{1 - \psi_{g,s}q_gz^{-1}} \right) \\
    &= \frac{ \pi_{g'} \psi_{g',s}^2(1-q_{g'})z^{-1} + \pi_g \psi_{g,s} (1 - \psi_{g',s}q_{g'}z^{-1})}{(1 - \psi_{g,s}q_gz^{-1})(1 - \psi_{g',s}q_{g'}z^{-1}) - \psi_{g,s}\psi_{g',s}(1-q_{g})(1-q_{g'})z^{-2}} \\
    &= \frac{\pi_g \psi_{g,s} + \left( \pi_{g'} \psi_{g',s}^2(1-q_{g'}) -  \pi_g \psi_{g,s}\psi_{g',s}q_{g'}\right)z^{-1} }{1 - (\psi_{g,s}q_g + \psi_{g',s}q_{g'})z^{-1} + \psi_{g,s}\psi_{g',s}(q_{g} + q_{g'} - 1)z^{-2}}  \\
    &= \frac{ \pi_g \psi_{g,s} + \psi_{g',s}\left( \pi_{g'} \psi_{g',s}(1-q_{g'}) - \pi_g\psi_{g,s} q_{g'}\right)z^{-1} }{(1 - a_{1,s} z^{-1})(1 - a_{2,s} z^{-1})}, 
\end{align*}
where $a_{1,s}$ and $a_{2,s}$ are as defined in the statement of the theorem. At this point, our goal is to take the inverse $\mathcal{Z}$-transform of the preceding expression. We will take the standard approach of inverting the partial fraction expansion of the above, which is given by
\begin{equation*}
    \mathcal{L}_{g,s} = \frac{A_{1,g,s}^w}{1-a_{1,s} z^{-1}} + \frac{A_{2,g,s}^w}{1 - a_{2,s} z^{-1}},
\end{equation*}
where $A_{1,g,s}^w$ and $A_{2,g,s}^w$ are as defined in the statement of the theorem.
The inverse transform is now easily computed by referencing a standard $\mathcal{Z}$-transform table, yielding
\begin{equation*}
    w_{g,s}(t) = 
    A_{1,g,s}^wa_{1,s}^t + A_{2,g,s}^wa_{2,s}^t, \quad  t\geq 0.
\end{equation*}
To transform back to the case where the sequence begins at $t=1$, rather than $t=0$ as we have used in this proof, we simply shift the sequence forward by a single time step, yielding
\begin{equation*}
    w_{g,s}(t) = 
    A_{1,g,s}^wa_{1,s}^{t-1} + A_{2,g,s}^wa_{2,s}^{t-1}, \quad  t\geq 1,
\end{equation*}
which completes the proof.
\end{proof}

\paragraph{Proof for Proposition \ref{prop:unconstrained_optimization}}

\begin{proof}
From Corollary~\ref{cor:total_mass_linear} in \S~\ref{appendix:additional_theoretical_results} we can immediately write the platform objective~\eqref{eq:platform-objective} as
\begin{equation*}
    \sum_{g \in \{A,B\}} \sum_{s \in \{ a,b\}} \sum_{t=1}^T l_{g,s}(t,\theta) = c_{A,a} \theta_{A,a} + c_{B,a} \theta_{B,a},
\end{equation*}
where
\begin{align*}
    c_{A,a} &:= \sum_{t=1}^T (w_{A,a}(t) - w_{A,b}(t) + u_{B,a}(t) - u_{B,b}(t)) \\
    c_{B,a} &:= \sum_{t=1}^T (u_{A,a}(t) - u_{A,b}(t) + w_{B,a}(t) - w_{B,b}(t)).
\end{align*}
Since the objective \eqref{eq:platform-objective} is linear in $\theta_{A,a}$ and $\theta_{B,a}$, the optimal $\theta_{A,a}$ is the largest allowable value if $c_{A,a} > 0$, and the smallest allowable value if $c_{A,a} < 0$. The case is similar for $\theta_{B,a}$. This completes the proof.
\end{proof}

\paragraph{Proof for Corollary \ref{cor:feasible_never_optimal}}

\begin{proof}

By Proposition \ref{prop:unconstrained_optimization_old} in \S~\ref{appendix:additional_theoretical_results}, the solution to the fairness-agnostic optimization problem~\eqref{eq:platform-objective} can be written as
\begin{equation*}
\theta^*_{g,a} = \mathbf{1} \left\{ \frac{\psi_{g,a} (z_{1,a} + \psi_{g',a}(1-q_{g'}-q_g) z_{2,a})}{\psi_{g,b} (z_{1,b} - \psi_{g',b}(1-q_{g'}-q_g) z_{2,b})} >1 \right\}.
\end{equation*}

Assume that $\theta_{A,a}^*=0$. Then, we will show that $\theta_{B,a}^*=0$ as well.

Recall that, by assumption, $\psi_{A,a} > \psi_{A,b}$ and $\psi_{B,b} > \psi_{B,a}$.
Therefore, given that $1-q_{A}-q_B <0$ and $z_{1,s}, z_{2,s} >0$, we get that
\begin{align*}
    1>&\frac{\psi_{A,a} (z_{1,a} + \psi_{B,a}(1-q_{A}-q_B) z_{2,a})}{\psi_{A,b} (z_{1,b} - \psi_{B,b}(1-q_{A}-q_B) z_{2,b})} = \frac{\psi_{A,a} (z_{1,a} + \psi_{B,a}(1-q_{A}-q_B) z_{2,a})}{\psi_{A,b} (z_{1,b} - \psi_{B,b}(1-q_{A}-q_B) z_{2,b})}\\
    >& \frac{\psi_{B,a} (z_{1,a} + \psi_{A,a}(1-q_{A}-q_B) z_{2,a})}{\psi_{B,b} (z_{1,b} - \psi_{A,b}(1-q_{A}-q_B) z_{2,b})}.
\end{align*}
Thus, $\theta_{B,a}^* =0$.
\end{proof}

\paragraph{Proof for Lemma \ref{lemma:condition_for_higher_intergroup_exposure}}
\begin{proof}
Consider the path representation of the mass computation problem as shown in Figure~\ref{fig:propagation_example}.
Since $\theta_{A,a} = 1$ and $\theta_{B,a}=0$, all $A$ users at time step $t=1$ are shown article $a$ and all $B$ users article $B$. 
As a result, all paths that constitute $l_{A,a}(t,\theta)$ start from node $A_1$ while all paths that constitute $l_{A,b}(t,\theta)$ start from node $B_1$; both end at node $A_t$.

There are two cases. Each such path $P$ of length $t>2$ can  include either  node $A_2$ or $B_2$. 

\paragraph{Case: $P$ includes $A_2$} Then,
\begin{equation*}
    \frac{l_{A,a}(t,\theta)}{l_{A,b}(t,\theta)} = \frac{\pi_A q_A}{\pi_B (1-q_B)} \cdot \frac{\psi_{A,a}}{\psi_{B,b}} \left(\frac{\psi_{A,a}}{\psi_{B,a}}\right)^i \left(\frac{\psi_{A,a}}{\psi_{B,a}}\right)^{t-1-i}, 
 \end{equation*}
where $t-1 \geq i\geq 2$ is the number of $A$ nodes in the path (\textit{after} the first node).
Observe that if the conditions in~\eqref{eq:conditions_A_more_exposed_to_b} hold, then the previous equation also holds. 
 
\paragraph{Case: $P$ includes $B_2$}
Then,
\begin{equation*}
    \frac{l_{A,a}(t,\theta)}{l_{A,b}(t,\theta)} = \frac{\pi_A (1-q_A)}{\pi_B q_B} \cdot \frac{\psi_{A,a}}{\psi_{B,b}} \left(\frac{\psi_{A,a}}{\psi_{B,a}}\right)^i \left(\frac{\psi_{A,a}}{\psi_{B,a}}\right)^{t-1-i}, 
 \end{equation*}
where $t-1>i\geq 1$ is the number of $A$ nodes in the path (\textit{after} the first node).
Observe that if $\frac{q_A \pi_A  }{(1-q_B) \pi_B} <1$ then $\frac{(1-q_A) \pi_A  }{q_B \pi_B} <1$ must also hold. 
Thus, if the conditions in~\eqref{eq:conditions_A_more_exposed_to_b} hold, then the previous equation also holds. 

From the previous two cases, it follows that $l_{A,a}(t,\theta) < l_{A,b}(t,\theta)$. This argument applies to any $t$. Taking the sum gives the desired result.
\end{proof} 

\paragraph{Proof for Lemma \ref{lemma:possibility_of_fair_exposure}}
\begin{proof}
If $\theta_{A,a} \neq 1-\theta_{B,a} $, then the lemma trivially holds.
Assume now that the platform manages to target equal fractions $\theta_{A,a} = 1-\theta_{B,a} = e$ at time $t=1$. Thus,
\begin{equation*}
    \frac{l_{A,s}(1,\theta)}{\pi_A} = \frac{l_{B,s'}(1,\theta)}{\pi_B}=e.
\end{equation*}
The rest of the proof is by induction.

Assume that the platform has achieved equal, constant exposure up to time $t-1$. To achieve the same level of exposure $e$ at time $t\geq2$, it must hold that 
\begin{equation}
\label{eq:equal_exposure_condition}
\frac{l_{A,s}(t,\theta)}{\pi_A} = \frac{l_{B,s'}(t,\theta)}{\pi_B}= \frac{l_{B,s'}(t,\theta)}{1-\pi_A}=e.
\end{equation}

Using~\eqref{eq:equal_exposure_condition} for $t-1$ (due to our induction hypothesis) in~\eqref{eq:mass_sh}, we get that 
\begin{align*}
    l_{A,s}(t,\theta) &= l_{A,s}(t-1,\theta) \left( q_A + \frac{1-\pi_A}{\pi_A} (1-q_B)  \right) \psi_{A,s} \\
    l_{B,s'}(t,\theta) &= l_{A,s}(t-1,\theta) \left( \frac{1-\pi_A}{\pi_A} q_B + (1-q_A)  \right) \psi_{B,s'}.
\end{align*}
Thus,~\eqref{eq:equal_exposure_condition} is equivalent to
\begin{align*}
    \psi_{A,s}\left(q_A  + \frac{1-\pi_A}{\pi_A} (1-q_B) \right) = \psi_{B,s'}\left( \frac{1-\pi_A}{\pi_A} q_B  + (1-q_A)\right).
\end{align*}
Therefore, we conclude that equal, constant exposure over time is possible if and only if~\eqref{eq:condition_equal_constant_exposure} holds and $\theta_{A,a}= \theta_{B,a}$.
\end{proof}

\paragraph{Proof for Lemma~\ref{lemma:avg_exp_1group}}
The proof makes use of Lemma~\ref{lemma:as_real} in \S~\ref{appendix:additional_theoretical_results}, which simply states that the relevant quantities in Theorem~\ref{theorem:closed_form_Z_transform_of_coeffs} are real numbers and that $a_{1,s} > a_{2,s} > 0$. 

\begin{proof}

Our goal is to show that there exist $\theta_{g,s}, \theta_{g',s}, \pi_g \in (0,1)$ such that, for any $e \in (0,1)$,
\begin{equation*}
    e = \frac{1}{T}\sum_{t=1}^T\frac{l_{g,s}(t,\theta)}{\pi_g} = \frac{1}{T \pi_g} \left(\theta_{g,s}\sum_{t=1}^T w_{g,s}(t) + \theta_{g',s}\sum_{t=1}^T u_{g,s}(t) \right),
\end{equation*}
where the rightmost expression comes from Corollary~\ref{cor:total_mass_linear} in \S~\ref{appendix:additional_theoretical_results}. 

From Lemma \ref{lemma:incr_linear} we know that $w_{g,s}(t) > 0$ for all $t\geq 1$, $u_{g,s}(t) > 0$ for all $t\geq 2$, and $u_{g,s}(1) = 0$. Therefore, $\sum_{t=1}^T w_{g,s}(t) > 0$, $\sum_{t=1}^T u_{g,s}(t) > 0$ and, from the constraint $\theta_{g,s} \in [0,1]$, it is clear that
\begin{align}
    0 \leq e &\leq \frac{1}{T \pi_g} \sum_{t=1}^T \left(w_{g,s}(t) + u_{g,s}(t) \right). \label{eqn:e_bounds}
\end{align}
This completes the proof.
\end{proof}

\paragraph{Proof for Theorem \ref{theorem:feasibility}}

\begin{proof}

Consider the hyperplanes $y_1$ to $y_4$, as introduced in \S~\ref{sec:fairness_constraints}, which induce the half-spaces containing the feasible region of the fairness-aware optimization problem~\eqref{eq:platform-objective}.
Note that $y_1$ and $y_4$ are upper bounds, and $y_2$ and $y_3$ are lower bounds.
Additionally, we have $\theta_{A,a},\theta_{B,a} \in [0,1]$.
Intuitively, the problem is infeasible if upper bounds are smaller than lower bounds, or if lower bounds are greater than upper bounds---each over the unit box.
Or, in other words, if the half-spaces induced by two bounding hyperplanes do not intersect over the unit box.
Note that $y_1$ to $y_4$ all have negative slope, as can be seen in \S~\ref{sec:fairness_constraints}.
Hence the fairness-aware problem will be infeasible in one of the following cases:

\paragraph{Case: $y_1$ below $y_2$}
This is equivalent to $y_1(\theta_{A,a}=0) < y_2(\theta_{A,a}=0)$ and $\theta_{A,a}(y_1=0) < \theta_{A,a}(y_2=0)$.
But this cannot happen, as shown in~\eqref{eq:y_1_greater_y_2} in \S~\ref{sec:fairness_constraints}.

\paragraph{Case: $y_4$ below $y_2$}
This is equivalent to $y_4(\theta_{A,a}=0) < y_2(\theta_{A,a}=0)$ and $\theta_{A,a}(y_4=0) < \theta_{A,a}(y_2=0)$. Using the formulas from \S~\ref{sec:fairness_constraints}, we know that this is the case if and only if $\frac{m_{A,b}}{\underline{m}_{A,b}} < \underline{\delta} \frac{m_{B,b}}{\underline{m}_{A,a}}$ and $\frac{m_{A,b}}{\underline{n}_{A,b}} < \underline{\delta} \frac{m_{B,b}}{\underline{n}_{A,a}}$.

\paragraph{Case: $y_1$ below $y_3$}
This is equivalent to $y_1(\theta_{A,a}=0) < y_3(\theta_{A,a}=0)$ and $\theta_{A,a}(y_1=0) < \theta_{A,a}(y_3=0)$. Using the formulas from \S~\ref{sec:fairness_constraints}, we know that this is the case if and only if $\overline{\delta} \frac{m_{B,b}}{\overline{m}_{A,a}} < \frac{m_{A,b}}{\overline{m}_{A,b}}$ and $\overline{\delta} \frac{m_{B,b}}{\overline{n}_{A,a}} < \frac{m_{A,b}}{\overline{n}_{A,b}}$.

\paragraph{Case: $y_4$ below $y_3$}
This is equivalent to $y_4(\theta_{A,a}=0) < y_3(\theta_{A,a}=0)$ and $\theta_{A,a}(y_4=0) < \theta_{A,a}(y_3=0)$.
But this cannot happen, as shown in~\eqref{eq:intersect_magenta} in \S~\ref{sec:fairness_constraints}.

\paragraph{Case: $y_1$ below $0$}
Since $y_1$ is a line with negative slope, this is the case if and only if $y_1(\theta_{A,a}=0) < 0$. But this cannot happen because $y_1(\theta_{A,a}=0) = \overline{\delta} \frac{m_{B,b}}{\overline{m}_{A,a}} \geq 0$ due to $\overline{\delta}, m_{B,b}, \overline{m}_{A,a} \geq 0$ (see Lemma~\ref{lemma:incr_linear} and the definitions of $m_{g,s}$ and $\overline{m}_{g,s}$).

\paragraph{Case: $y_4$ below $0$}
With the similar argument to the previous case, this cannot happen.

\paragraph{Case: $1$ below $y_2$}
Note that $y_2$ can only be ``above'' $1$, i.e., north-east of the unit box, if $\theta_{A,a}(y_2=0) > 1$. This is the case if and only if $\underline{\delta} m_{B,b} > \underline{n}_{A,a}$, which in turn is equivalent to $\underline{\delta} \sum_{t=1}^T w_{B,b}(t) > \sum_{t=1}^T w_{A,a}(t)$.
Additionally, for $y_2$ to be outside the unit box, it must be that $y_2(\theta_{A,a}=0)$ is large enough such that $y_2$ runs just above $(\theta_{A,a},\theta_{B,a})=(1,1)$. To determine this critical value, note that a line with $\theta_{A,a}$-axis intercept $x$ and running through $(1,1)$ has a $\theta_{B,a}$-axis intercept of $\frac{x}{x-1}$.
Hence, the second requirement for $y_2$ to be ``above'' $1$ is that
\begin{align*}
    y_2(\theta_{A,a}=0) = \underline{\delta} \frac{m_{B,b}}{\underline{m}_{A,a}} > \frac{\theta_{A,a}(y_2=0)}{\theta_{A,a}(y_2=0) - 1} = \underline{\delta} \frac{m_{B,b}}{\underline{\delta} m_{B,b} - \underline{n}_{A,a}}.
\end{align*}

\paragraph{Case: $1$ below $y_3$}
With the similar argument to the previous case, we obtain the requirements $\sum_{t=1}^T u_{A,b}(t) > \overline{\delta} \sum_{t=1}^T u_{B,a}(t)$ and $\frac{m_{A,b}}{\overline{m}_{A,b}} > \frac{m_{A,b}}{m_{A,b} - \overline{n}_{A,b}}$.

This completes the proof as all possible cases for which the fairness-aware problem is infeasible are covered.
\end{proof}


\section{Additional Theoretical Results}\label{appendix:additional_theoretical_results}

We first state a corollary of Theorem \ref{theorem:closed_form_Z_transform_of_coeffs}:
\begin{corollary} \label{cor:total_mass_linear}
\begin{equation*}
    \sum_{t=1}^T l_{g,s}(t,\theta) = \theta_{g,s}\sum_{t=1}^T w_{g,s}(t) + \theta_{g',s}\sum_{t=1}^T u_{g,s}(t)
\end{equation*}
where $w_{g,s}(t)$ and $u_{g,s}(t)$ are as defined in Theorem \ref{theorem:closed_form_Z_transform_of_coeffs}  and do not depend on $\theta_{g,s}$ for any $(g,s)$.
\end{corollary}

\begin{proof}
    The proof follows immediately from Theorem~\ref{theorem:closed_form_Z_transform_of_coeffs} by adding over $t = 1,\dots,T$ and factoring out $\theta_{g,s}$ and $\theta_{g',s}$.
\end{proof}

\begin{lemma} \label{lemma:as_real}
The quantities
\begin{align*}
    a_{1,s} &\coloneqq \frac{1}{2} \big( \psi_{g,s}q_g + \psi_{g',s}q_{g'} + \left[ (\psi_{g,s}q_g + \psi_{g',s}q_{g'})^2 - 4 \psi_{g,s}\psi_{g',s}(q_g + q_{g'} -1) \right]^{\frac{1}{2}} \big), \\
    a_{2,s} &\coloneqq \psi_{g,s}q_g + \psi_{g',s}q_{g'} - a_{1,s},\\
    A^w_{1,g,s} &\coloneqq \frac{ \pi_g \psi_{g,s} + \psi_{g',s}\left( \pi_{g'} \psi_{g',s}(1-q_{g'}) - \pi_g \psi_{g,s} q_{g'}\right)a_{1,s}^{-1}}{1-a_{2,s}a_{1,s}^{-1}}, \\
    A^w_{2,g,s} &\coloneqq \frac{ \pi_g \psi_{g,s} + \psi_{g',s}\left( \pi_{g'} \psi_{g',s}(1-q_{g'}) - \pi_g \psi_{g,s} q_{g'}\right)a_{2,s}^{-1}}{1-a_{2,s}^{-1}a_{1,s}}, \\
    A^u_{g,s} &\coloneqq \frac{ \psi_{g',s}\left( \pi_{g'} \psi_{g',s}(1-q_{g'}) - \pi_g \psi_{g,s} q_{g'}\right)a_{1,s}^{-1}}{1-a_{2,s}a_{1,s}^{-1}}
\end{align*}
are real numbers for all $q_g,q_{g'}\in (\frac{1}{2}, 1)$, $\psi_{g,s}, \psi_{g',s} \in (0,1)$. Furthermore, $a_{1,s} > a_{2,s} > 0$.
\end{lemma}

\begin{proof}
Recall that
\begin{align*}
    a_{1,s} &\coloneqq \frac{1}{2} \big( \psi_{g,s}q_g + \psi_{g',s}q_{g'} + \left[ (\psi_{g,s}q_g + \psi_{g',s}q_{g'})^2 - 4 \psi_{g,s}\psi_{g',s}(q_g + q_{g'} -1) \right]^{\frac{1}{2}} \big), \\
    a_{2,s} &\coloneqq \psi_{g,s}q_g + \psi_{g',s}q_{g'} - a_{1,s}.
\end{align*}
These values are simply the roots of a quadratic that appears in the proof of Theorem \ref{theorem:closed_form_Z_transform_of_coeffs}. We can rewrite the discriminant
\begin{align*}
    (\psi_{g,s}q_g + \psi_{g',s}q_{g'})^2 - 4 \psi_{g,s}\psi_{g',s}(q_g + q_{g'} -1) &= q_{g'}(\underbrace{\psi_{g',s} + 2q_g - 4}_{>-3}) + q_g(\underbrace{\psi_{g,s} - 1}_{>-1}) + 4 > 0,
\end{align*}
where the lower bounds on intermediate quantities follow from allowable values of problem parameters. Therefore, $a_{1,s}$ and $a_{2,s}$ are real. Since the discriminant is strictly positive, we have $a_{1,s}>\frac{1}{2}( \psi_{g,s}q_g + \psi_{g',s}q_{g'})$. By substitution, we immediately have $a_{2,s}<\frac{1}{2}( \psi_{g,s}q_g + \psi_{g',s}q_{g'})$. Therefore, $a_{1,s} > a_{2,s}$. To show positivity, note that
\begin{align*}
    a_{1,s} &= \frac{1}{2} \big( \psi_{g,s}q_g + \psi_{g',s}q_{g'} + \left[ (\psi_{g,s}q_g + \psi_{g',s}q_{g'})^2 - 4 \psi_{g,s}\psi_{g',s}(q_g + q_{g'} -1) \right]^{\frac{1}{2}} \big) \\
    &< \frac{1}{2} \big( \psi_{g,s}q_g + \psi_{g',s}q_{g'} + \left[ (\psi_{g,s}q_g + \psi_{g',s}q_{g'})^2 \right]^{\frac{1}{2}} \big) \\
    &= \psi_{g,s}q_g + \psi_{g',s}q_{g'}.
\end{align*}
Again, by substitution into the definition of $a_{2,s}$ we have $a_{2,s}>0$. This completes the proof. 
\end{proof}

\begin{lemma}\label{lemma:closed_form_Z_transform_old}
For all $t\geq 1$, regardless of group $g$ and article $s$, 
\begin{equation}
    l_{g,s}(t,\theta) = 
    A_{1,g,s}a_{1,s}^{t-1} + A_{2,g,s}a_{2,s}^{t-1}, \quad  t\geq 1,
\end{equation}
where
\begin{align*}
    a_{1,s} &\coloneqq \frac{1}{2} \big( \psi_{g,s}q_g + \psi_{g',s}q_{g'} + \left[ (\psi_{g,s}q_g + \psi_{g',s}q_{g'})^2 - 4 \psi_{g,s}\psi_{g',s}(q_g + q_{g'} -1) \right]^{\frac{1}{2}} \big), \\
    a_{2,s} &\coloneqq \psi_{g,s}q_g + \psi_{g',s}q_{g'} - a_{1,s},\\
    A_{1,g,s} &\coloneqq \frac{\theta_{g,s} \pi_g \psi_{g,s} + \psi_{g,s}\psi_{g',s}\left(\theta_{g',s} \pi_{g'} (1-q_{g'}) - \theta_{g,s} \pi_g q_{g'}\right)a_{1,s}^{-1}}{1-a_{2,s}a_{1,s}^{-1}}, \\
    A_{2,g,s} &\coloneqq \frac{\theta_{g,s} \pi_g \psi_{g,s} + \psi_{g,s}\psi_{g',s}\left(\theta_{g',s} \pi_{g'} (1-q_{g'}) - \theta_{g,s} \pi_g q_{g'}\right)a_{2,s}^{-1}}{1-a_{2,s}^{-1}a_{1,s}}. 
\end{align*}
\end{lemma}

\begin{proof}
We first mention that $l_{g,s}(t,\theta)$ is a right-tailed sequence, so we will be using the one-sided $\mathcal{Z}$-transform. 
This is important because~\eqref{eq:mass_sh} only holds for $t \geq 1$.
We first rewrite~\eqref{eq:mass_sh} with different initial conditions so that the relation holds for all $t \geq 0$:
\begin{equation}
    l_{g,s}(t,\theta) = \psi_{g,s}\left( q_g l_{g,s}(t-1,\theta) + (1-q_{g'}) l_{g',s}(t-1,\theta) \right) \label{eq:lgs_mod_old} + \delta(t)\theta_{g,s} \pi_g \psi_{g,s}.
\end{equation}
Here, $\delta(t)$ is the Kronecker delta function and we take $l_{g,s}(-1,\theta) = l_{g',s}(-1,\theta) = 0$ for all $\theta$.

Denote the one-sided $\mathcal{Z}$-transform of~\eqref{eq:lgs_mod_old} as $\mathcal{L}_{g,s}(z)$, which is given by
\begin{align}
   \mathcal{L}_{g,s}(z) &= z^{-1} \psi_{g,s}q_g \mathcal{L}_{g,s}(z) + z^{-1} \psi_{g,s}(1-q_{g'}) \mathcal{L}_{g',s}(z) + \theta_{g,s} \pi_g \psi_{g,s} \nonumber \\
   &= \frac{\psi_{g,s}(1-q_{g'})z^{-1}}{1 - \psi_{g,s}q_gz^{-1}}\mathcal{L}_{g',s}(z) + \frac{\theta_{g,s} \pi_g \psi_{g,s}}{1 - \psi_{g,s}q_gz^{-1}}. \label{eq:zlgs_old}
\end{align}
Similarly, we write the $\mathcal{Z}$-transform of $l_{g',s}$ using the definition of $l_{g',s}$ analogous to~\eqref{eq:lgs_mod_old}:
\begin{align}
   \mathcal{L}_{g',s}(z) &= z^{-1} \psi_{g',s}q_{g'} \mathcal{L}_{g',s}(z) + z^{-1} \psi_{g',s}(1-q_{g}) \mathcal{L}_{g,s}(z) + \theta_{g',s} \pi_{g'} \psi_{g',s} \nonumber\\
   &= \frac{\psi_{g',s}(1-q_{g})z^{-1}}{1 - \psi_{g',s}q_{g'}z^{-1}}\mathcal{L}_{g,s}(z) + \frac{\theta_{g',s} \pi_{g'} \psi_{g',s}}{1 - \psi_{g',s}q_{g'}z^{-1}}. \label{eq:zlg's_old}
\end{align}

Substituting~\eqref{eq:zlg's_old} into~\eqref{eq:zlgs_old} yields
\begin{align*}
    \mathcal{L}_{g,s}(z) &= \left( 1 - \frac{\psi_{g,s}(1-q_{g'})z^{-1}}{1 - \psi_{g,s}q_gz^{-1}} \frac{\psi_{g',s}(1-q_{g})z^{-1}}{1 - \psi_{g',s}q_{g'}z^{-1}} \right)^{-1}\\
    &\quad \cdot \left(\frac{\psi_{g,s}(1-q_{g'})z^{-1}}{1 - \psi_{g,s}q_gz^{-1}}\frac{\theta_{g',s} \pi_{g'} \psi_{g',s}}{1 - \psi_{g',s}q_{g'}z^{-1}} +  \frac{\theta_{g,s} \pi_g \psi_{g,s}}{1 - \psi_{g,s}q_gz^{-1}} \right) \\
    &= \frac{\theta_{g',s} \pi_{g'} \psi_{g',s}\psi_{g,s}(1-q_{g'})z^{-1} + \theta_{g,s} \pi_g \psi_{g,s} (1 - \psi_{g',s}q_{g'}z^{-1})}{(1 - \psi_{g,s}q_gz^{-1})(1 - \psi_{g',s}q_{g'}z^{-1}) - \psi_{g,s}\psi_{g',s}(1-q_{g})(1-q_{g'})z^{-2}} \\
    &= \frac{\theta_{g,s} \pi_g \psi_{g,s} + \left(\theta_{g',s} \pi_{g'} \psi_{g',s}\psi_{g,s}(1-q_{g'}) - \theta_{g,s} \pi_g \psi_{g,s}\psi_{g',s}q_{g'}\right)z^{-1} }{1 - (\psi_{g,s}q_g + \psi_{g',s}q_{g'})z^{-1} + \psi_{g,s}\psi_{g',s}(q_{g} + q_{g'} - 1)z^{-2}}  \\
    &= \frac{\theta_{g,s} \pi_g \psi_{g,s} + \psi_{g,s}\psi_{g',s}\left(\theta_{g',s} \pi_{g'} (1-q_{g'}) - \theta_{g,s} \pi_g q_{g'}\right)z^{-1} }{(1 - a_{1,s} z^{-1})(1 - a_{2,s} z^{-1})}, 
\end{align*}
where $a_{1,s}$ and $a_{2,s}$ are as defined previously. At this point, our goal is to take the inverse $\mathcal{Z}$-transform of the preceding expression. We will take the standard approach of inverting the partial fraction expansion of the above, which is given by
\begin{equation*}
    \mathcal{L}_{g,s} = \frac{A_{1,g,s}}{1-a_{1,s} z^{-1}} + \frac{A_{2,g,s}}{1 - a_{2,s} z^{-1}},
\end{equation*}
where $A_{1,g,s}$ and $A_{2,g,s}$ are as previously defined. The inverse transform is now easily computed by referencing a standard $\mathcal{Z}$-transform table, yielding
\begin{equation*}
    l_{g,s}(t,\theta) = 
    A_{1,g,s}a_{1,s}^t + A_{2,g,s}a_{2,s}^t, \quad  t\geq 0.
\end{equation*}
To transform back to the case where the sequence begins at $t=1$, rather than $t=0$ as we have used in this proof, we simply shift the sequence forward by a single time step, yielding
\begin{equation*}
    l_{g,s}(t,\theta) = 
    A_{1,g,s}a_{1,s}^{t-1} + A_{2,g,s}a_{2,s}^{t-1}, \quad  t\geq 1,
\end{equation*}
which completes the proof.
\end{proof}


\begin{proposition}
\label{prop:unconstrained_optimization_old}
The solution to the fairness-agnostic optimization problem is
\begin{equation*}
\theta^*_{g,a} = \mathbf{1} \left\{ \frac{\psi_{g,a} (z_{1,a} + \psi_{g',a}(1-q_{g'}-q_g) z_{2,a})}{\psi_{g,b} (z_{1,b} - \psi_{g',b}(1-q_{g'}-q_g) z_{2,b})} >1 \right\}
\end{equation*}
for $g \in \{A,B\}$, where
\begin{align*}
    z_{1,s} &\coloneqq \sum_{t=1}^T \sum_{j=0}^{t-1} a_{1,s}^{t-j-1} a_{2,s}^j = \frac{(a_{2,s} -1) a_{1,s}^{T+1} - a_{1,s} a_{2,s}^{T+1} + a_{1,s} + a_{2,s}(a_{2,s}^T -1)}{(a_{1,s} -1) (a_{2,s} -1) (a_{1,s} - a_{2,s})},\\
    z_{2,s} &\coloneqq \sum_{t=1}^T \sum_{j=0}^{t-2} a_{1,s}^{t-j-2} a_{2,s}^j = \frac{(a_{2,s} -1) a_{1,s}^{T} - a_{1,s} a_{2,s}^{T} + a_{1,s} + a_{2,s}(a_{2,s}^{T-1} -1)}{(a_{1,s} -1) (a_{2,s} -1) (a_{1,s} - a_{2,s})}.
\end{align*}
\end{proposition}

\begin{proof}
Using Lemma \ref{lemma:closed_form_Z_transform_old}, we find the derivative of the objective $g(\theta_{A,a}, \theta_{B,a}) \coloneqq \sum_{t = 1}^T \sum_{g \in \{A, B\}} \sum_{s \in \{a,b\}}  l_{g,s}(t,\theta)$ with respect to $\theta_{g,a}$:
\begin{align*}
    \frac{\partial g}{\partial \theta_{g,a}} &= \sum_{t = 1}^T \frac{a_{1,a}^{t-1}}{a_{1,a}-a_{2,a}} \left(\pi_g \psi_{g,a} a_{1,a} - \psi_{g,a} \psi_{g',a} \pi_g q_{g'} \right) + \frac{a_{2,a}^{t-1}}{a_{2,a}-a_{1,a}} \left(\pi_g \psi_{g,a} a_{2,a} - \psi_{g,a} \psi_{g',a} \pi_g q_{g'} \right) \\
    &\quad - \frac{a_{1,b}^{t-1}}{a_{1,b}-a_{2,b}} \left(\pi_g \psi_{g,b} a_{1,b} - \psi_{g,b} \psi_{g',b} \pi_g q_{g'} \right) - \frac{a_{2,a}^{t-1}}{a_{2,a}-a_{1,a}} ( \pi_g \psi_{g,a} a_{2,a} - \psi_{g,a} \psi_{g',a} \pi_g q_{g'}) \\
    &\quad + \frac{a_{1,a}^{t-1}}{a_{1,a} -a_{2,a}} \psi_{g',a} \psi_{g,a} \pi_g (1-q_g) + \frac{a_{2,a}^{t-1}}{a_{2,a}- a_{1,a}}  \psi_{g',a} \psi_{g,a} \pi_g (1-q_g) \\
    &\quad - \frac{a_{1,b}^{t-1}}{ a_{1,b}-a_{2,b}}   \psi_{g',b} \psi_{g,b} \pi_g (1-q_g) - \frac{a_{2,b}^{t-1}}{a_{2,b}- a_{1,b}} \psi_{g',b} \psi_{g,b} \pi_g (1-q_g) \\
    &=\sum_{t = 1}^T \frac{\pi_g \psi_{g,a}}{a_{1,a}-a_{2,a}} \left( a_{1,a}^t - a_{2,a}^t -  \psi_{g',a} q_{g'} (a_{1,a}^{t-1} - a_{2,a}^{t-1}) \right) - \frac{\pi_g \psi_{g,b}}{a_{1,b}-a_{2,b}} \left( a_{1,b}^t - a_{2,b}^t -  \psi_{g',b} q_{g'} (a_{1,b}^{t-1} - a_{2,b}^{t-1}) \right) \\
    &\quad + \frac{a_{1,a}^{t-1} - a_{2,a}^{t-1}}{a_{1,a} -a_{2,a}}  \psi_{g',a} \psi_{g,a} \pi_g (1-q_g) - \frac{a_{1,b}^{t-1} - a_{2,b}^{t-1}}{a_{1,b} -a_{2,b}}  \psi_{g',b} \psi_{g,b} \pi_g (1-q_g) \\
    &= \sum_{t = 1}^T \frac{\pi_g \psi_{g,a}}{a_{1,a}-a_{2,a}} \left( a_{1,a}^t - a_{2,a}^t +  \psi_{g',a} (1-q_{g'} -q_g) (a_{1,a}^{t-1} - a_{2,a}^{t-1}) \right) \\
    &\quad - \frac{\pi_g \psi_{g,b}}{a_{1,b}-a_{2,b}} \left( a_{1,b}^t - a_{2,b}^t -  \psi_{g',b} (1-q_{g'}-q_g) (a_{1,b}^{t-1} - a_{2,b}^{t-1}) \right) \\
    &= \sum_{t = 1}^T \pi_g \psi_{g,a} \left [ \sum_{j=0}^{t-1} a_{1,a}^{t-j-1} a_{2,a}^j +\psi_{g',a} (1-q_{g'} -q_g)  \sum_{j=0}^{t-2} a_{1,a}^{t-j-2} a_{2,a}^j  \right] \\
    &\quad - \pi_g \psi_{g,b} \left [ \sum_{j=0}^{t-1} a_{1,b}^{t-j-1} a_{2,b}^j -\psi_{g',b} (1-q_{g'} -q_g)  \sum_{j=0}^{t-2} a_{1,b}^{t-j-2} a_{2,b}^j  \right].
\end{align*}
Given the fact that $\frac{\partial g}{\partial \theta_{g,a}}$ is constant (and thus independent of $\theta_{g,a}, \theta_{g',a}$), it holds that, if $\frac{\partial g}{\partial \theta_{g,a}}>0$, then $\theta_{g,a} =1$ is optimal; otherwise, $\theta_{g,a} =0$ is optimal.

Using the definitions of $z_{1,s}$, $z_{2,s}$, we get that $\frac{\partial g}{\partial \theta_{g,a}}>0$ if and only if
\begin{equation*}
    \psi_{g,a} (z_{1,a} + \psi_{g',a}(1-q_{g'}-q_g) z_{2,a}) >
    \psi_{g,b} (z_{1,b} - \psi_{g',b}(1-q_{g'}-q_g) z_{2,b}).
\end{equation*}
Given that $z_{1,b} - \psi_{g',b}(1-q_{g'}-q_g) z_{2,b}>0$, the previous condition is equivalent to
\begin{equation*}
    \frac{\psi_{g,a} (z_{1,a} + \psi_{g',a}(1-q_{g'}-q_g) z_{2,a})}{\psi_{g,b} (z_{1,b} - \psi_{g',b}(1-q_{g'}-q_g) z_{2,b})} >1.
\end{equation*}
Consequently, the optimal $\theta^*_{g,a}$ to the fairness-agnostic optimization problem equals
\begin{equation*}
    \theta^*_{g,a} = \mathbf{1} \left\{ \frac{\psi_{g,a} (z_{1,a} + \psi_{g',a}(1-q_{g'}-q_g) z_{2,a})}{\psi_{g,b} (z_{1,b} - \psi_{g',b}(1-q_{g'}-q_g) z_{2,b})} >1 \right\}.
\end{equation*}
This completes the proof.
\end{proof}

\begin{theorem} \label{thm:constrained_soln}
For any admissible problem parameters, the optimal solution to the fairness-aware optimization problem is one of the following:
\begin{align*}
    &\theta_{A,a}^1 = \phi\left( \frac{m_{A,b}\overline{m}_{A,a} - \overline{\delta}m_{B,b}\underline{m}_{A,b}}
    {\underline{n}_{A,b}\overline{m}_{A,a} - \overline{n}_{A,a}\underline{m}_{A,b}} \right), \quad
    \theta_{B,a}^1 \in \left\{0, 1, \frac{\overline{\delta} m_{B,b} - \theta_{A,a}^1\overline{n}_{A,a}}{\overline{m}_{A,a}}, \frac{m_{A,b} - \theta_{A,a}^1\underline{n}_{A,b}}{\underline{m}_{A,b}}\right\} \\
    &\theta_{A,a}^2 = \phi\left( \frac{m_{A,b}\overline{m}_{A,a} - \overline{\delta}m_{B,b}\overline{m}_{A,b}}
    {\underline{n}_{A,b}\overline{m}_{A,a} - \overline{n}_{A,a}\overline{m}_{A,b}}\right), \quad
    \theta_{B,a}^2 \in \left\{0, 1, \frac{\overline{\delta} m_{B,b} - \theta_{A,a}^2\overline{n}_{A,a}}{\overline{m}_{A,a}}, \frac{m_{A,b} - \theta_{A,a}^2 \overline{n}_{A,b}}{\overline{m}_{A,b}}\right\} \\
    &\theta_{A,a}^3 = \phi\left( \frac{m_{A,b}\underline{m}_{A,a} - \underline{\delta}m_{B,b}\underline{m}_{A,b}}
    {\underline{n}_{A,b}\underline{m}_{A,a} - \underline{n}_{A,a}\underline{m}_{A,b}}\right), \quad
    \theta_{B,a}^3 \in \left\{0, 1, \frac{\underline{\delta}m_{B,b} - \theta_{A,a}^3 \underline{n}_{A,a}}{\underline{m}_{A,a}} , \frac{m_{A,b} - \theta_{A,a}^3\underline{n}_{A,b}}{\underline{m}_{A,b}}\right\}\\
    &\theta_{A,a}^4 = \phi\left( \frac{m_{A,b}\underline{m}_{A,a} - \underline{\delta}m_{B,b}\overline{m}_{A,b}}
    {\underline{n}_{A,b}\underline{m}_{A,a} - \underline{n}_{A,a}\overline{m}_{A,b}} \right), \quad
    \theta_{B,a}^4 \in \left\{0, 1, \frac{\underline{\delta}m_{B,b} - \theta_{A,a}^4 \underline{n}_{A,a}}{\underline{m}_{A,a}} , \frac{m_{A,b} - \theta_{A,a}^4 \overline{n}_{A,b}}{\overline{m}_{A,b}}\right\} \\
    &\theta_{B,a}^5 \in \{0,1\}, \quad
    \theta_{A,a}^5 \in \left\{0, 1, \frac{\overline{\delta}m_{B,b} - \theta_{B,a}^5 \overline{m}_{A,a}}{\overline{n}_{A,a}}, \frac{\underline{\delta}m_{B,b} - \theta_{B,a}^5 \underline{m}_{A,a}}{\underline{n}_{A,a}}, \frac{m_{A,b} - \theta_{B,a}^5 \overline{m}_{A,b}}{\overline{n}_{A,b}}, \frac{m_{A,b} - \theta_{B,a}^5 \underline{m}_{A,b}}{\underline{n}_{A,b}}\right\},
\end{align*}
where $\phi(\theta) \coloneqq \max(0, \min( 1, \theta))$, $\theta \in [0,1]$.
\end{theorem}

\begin{proof}
The solution to a linear program  with linear inequality constraints will lie on the vertices of the feasible region, which is a polytope. As referenced in the proof of Theorem \ref{theorem:feasibility}, and discussed in \S~\ref{sec:fairness_constraints}, the constraints \eqref{eq:platform-constraints-1} and \eqref{eq:platform-constraints-2}, along with the unit box constraint, can be rewritten as the system of linear inequalities
\begin{align*}
    y_3(\theta_{A,a}) = \frac{m_{A,b}}{\overline{m}_{A,b}} - \theta_{A,a} \frac{\overline{n}_{A,b}}{\overline{m}_{A,b}} \leq &\theta_{B,a}  \leq \frac{\overline{\delta}m_{B,b}}{\overline{m}_{A,a}} - \theta_{A,a} \frac{\overline{n}_{A,a}}{\overline{m}_{A,a}}  = y_1(\theta_{A,a}), \\
    y_2(\theta_{A,a}) = \frac{\underline{\delta}m_{B,b}}{\underline{m}_{A,a}} - \theta_{A,a} \frac{\underline{n}_{A,a}}{\underline{m}_{A,a}} \leq &\theta_{B,a}  \leq \frac{m_{A,b}}{\underline{m}_{A,b}} - \theta_{A,a} \frac{\underline{n}_{A,b}}{\underline{m}_{A,b}} = y_4(\theta_{A,a}) , \\
    0 \leq &\theta_{A,a}, \theta_{B,a} \leq 1,
\end{align*}
where we have used notation introduced in the main paper. The first two inequalities define four half-spaces. The intersection of these half-spaces and the unit box is a polytope defining the feasible region, which we call the constraint polytope. Assume the problem is feasible according to the conditions of Theorem \ref{theorem:feasibility}. There are two possibilities: (1) the intersection of the half-spaces defined by $y_1, y_2, y_3, y_4$ is contained in the unit box, or (2) it is only partially contained in the unit box. Since $y_1 > y_2$ and $y_4 > y_3$ on $0 \leq \theta_{A,a} \leq 1$ (as noted in \S~\ref{sec:fairness_constraints}), in the first case there will be at most four vertices of the constraint polytope. These vertices are solutions to the equations $y_1(\theta_{A,a}) = y_4(\theta_{A,a})$, $y_1(\theta_{A,a}) = y_3(\theta_{A,a})$, $y_2(\theta_{A,a}) = y_4(\theta_{A,a})$, and $y_2(\theta_{A,a}) = y_3(\theta_{A,a})$, respectively given by
\begin{align*}
    &\theta_{A,a}^1 = \frac{m_{A,b}\overline{m}_{A,a} - \overline{\delta}m_{B,b}\underline{m}_{A,b}}
    {\underline{n}_{A,b}\overline{m}_{A,a} - \overline{n}_{A,a}\underline{m}_{A,b}} , \quad
    \theta_{B,a}^1 = \frac{\overline{\delta} m_{B,b}}{\overline{m}_{A,a}} - \theta_{A,a}^1 \frac{\overline{n}_{A,a}}{\overline{m}_{A,a}} = y_1(\theta_{A,a}^1) = y_4(\theta_{A,a}^1)\\
    &\theta_{A,a}^2 = \frac{m_{A,b}\overline{m}_{A,a} - \overline{\delta}m_{B,b}\overline{m}_{A,b}}
    {\underline{n}_{A,b}\overline{m}_{A,a} - \overline{n}_{A,a}\overline{m}_{A,b}}, \quad
    \theta_{B,a}^2 = \frac{\overline{\delta} m_{B,b}}{\overline{m}_{A,a}} - \theta_{A,a}^2 \frac{\overline{n}_{A,a}}{\overline{m}_{A,a}} = y_1(\theta_{A,a}^2) = y_3(\theta_{A,a}^2) \\
    &\theta_{A,a}^3 = \frac{m_{A,b}\underline{m}_{A,a} - \underline{\delta}m_{B,b}\underline{m}_{A,b}}
    {\underline{n}_{A,b}\underline{m}_{A,a} - \underline{n}_{A,a}\underline{m}_{A,b}}, \quad
    \theta_{B,a}^3 = \frac{\underline{\delta} m_{B,b}}{\underline{m}_{A,a}} - \theta_{A,a}^3 \frac{\underline{n}_{A,a}}{\underline{m}_{A,a}} = y_2(\theta_{A,a}^3) = y_4(\theta_{A,a}^3) \\
    &\theta_{A,a}^4 = \frac{m_{A,b}\underline{m}_{A,a} - \underline{\delta}m_{B,b}\overline{m}_{A,b}}
    {\underline{n}_{A,b}\underline{m}_{A,a} - \underline{n}_{A,a}\overline{m}_{A,b}} , \quad
    \theta_{B,a}^4 = \frac{\underline{\delta} m_{B,b}}{\underline{m}_{A,a}} - \theta_{A,a}^4 \frac{\underline{n}_{A,a}}{\underline{m}_{A,a}} = y_2(\theta_{A,a}^4) = y_3(\theta_{A,a}^4),
\end{align*}
where $(\theta_{A,a}^i, \theta_{B,a}^i)$ represents the coordinate of a vertex. 

In the second case, if one or more of these vertices lie outside of the unit box, possible vertices exist where the $y_i$ intersect the boundary of the unit box, i.e., where $\theta_{A,a} \in \{0,1\}$ and/or $\theta_{B,a} \in \{0,1\}$. This set might also include vertices of the unit box as vertices of the feasible region. This set of possible solutions is given by
\begin{align*}
    &\widehat\theta_{A,a} \in \{0,1\}, \quad \widehat\theta_{B,a} \in \{0,1, y_1(\widehat\theta_{A,a}), y_2(\widehat\theta_{A,a}), y_3(\widehat\theta_{A,a}), y_4(\widehat\theta_{A,a})\}, \\
    &\widetilde\theta_{B,a} \in \{0,1\}, \quad \widetilde\theta_{A,a} \in \{0, 1, y_1^{-1}(\widetilde\theta_{B,a}), y_2^{-1}(\widetilde\theta_{B,a}), y_3^{-1}(\widetilde\theta_{B,a}), y_4^{-1}(\widetilde\theta_{B,a})\}.
\end{align*}
Collectively the six sets of possible vertices given to this point represent every possible vertex of the constraint set. We can simplify the solution classes by including the cases where one or both of $\theta_{A,a}$ and $\theta_{B,a}$ are $0$ or $1$. This yields
\begin{align*}
    &\theta_{A,a}^1 = \phi\left( \frac{m_{A,b}\overline{m}_{A,a} - \overline{\delta}m_{B,b}\underline{m}_{A,b}}
    {\underline{n}_{A,b}\overline{m}_{A,a} - \overline{n}_{A,a}\underline{m}_{A,b}} \right), \quad
    \theta_{B,a}^1 \in \{0, 1, y_1(\theta_{A,a}^1), y_4(\theta_{A,a}^1)\} \\
    &\theta_{A,a}^2 = \phi\left( \frac{m_{A,b}\overline{m}_{A,a} - \overline{\delta}m_{B,b}\overline{m}_{A,b}}
    {\underline{n}_{A,b}\overline{m}_{A,a} - \overline{n}_{A,a}\overline{m}_{A,b}}\right), \quad
    \theta_{B,a}^2 \in \{0, 1, y_1(\theta_{A,a}^2), y_3(\theta_{A,a}^2)\}\\
    &\theta_{A,a}^3 = \phi\left( \frac{m_{A,b}\underline{m}_{A,a} - \underline{\delta}m_{B,b}\underline{m}_{A,b}}
    {\underline{n}_{A,b}\underline{m}_{A,a} - \underline{n}_{A,a}\underline{m}_{A,b}}\right), \quad
    \theta_{B,a}^3 \in \{0, 1, y_2(\theta_{A,a}^3), y_4(\theta_{A,a}^3)\}\\
    &\theta_{A,a}^4 = \phi\left( \frac{m_{A,b}\underline{m}_{A,a} - \underline{\delta}m_{B,b}\overline{m}_{A,b}}
    {\underline{n}_{A,b}\underline{m}_{A,a} - \underline{n}_{A,a}\overline{m}_{A,b}} \right), \quad
    \theta_{B,a}^4 \in \{0, 1, y_2(\theta_{A,a}^4), y_3(\theta_{A,a}^4)\}, \\
    &\theta_{B,a}^5 \in \{0,1\}, \quad
    \theta_{A,a}^5 \in \{0, 1, y_1^{-1}(\theta_{B,a}^5), y_2^{-1}(\theta_{B,a}^5), y_3^{-1}(\theta_{B,a}^5), y_4^{-1}(\theta_{B,a}^5)\}.
\end{align*}
Plugging in the expressions for $y_i$ and $y_i^{-1}$ gives the result:
\begin{align*}
    &\theta_{A,a}^1 = \phi\left( \frac{m_{A,b}\overline{m}_{A,a} - \overline{\delta}m_{B,b}\underline{m}_{A,b}}
    {\underline{n}_{A,b}\overline{m}_{A,a} - \overline{n}_{A,a}\underline{m}_{A,b}} \right), \quad
    \theta_{B,a}^1 \in \left\{0, 1, \frac{\overline{\delta} m_{B,b} - \theta_{A,a}^1\overline{n}_{A,a}}{\overline{m}_{A,a}}, \frac{m_{A,b} - \theta_{A,a}^1\underline{n}_{A,b}}{\underline{m}_{A,b}}\right\} \\
    &\theta_{A,a}^2 = \phi\left( \frac{m_{A,b}\overline{m}_{A,a} - \overline{\delta}m_{B,b}\overline{m}_{A,b}}
    {\underline{n}_{A,b}\overline{m}_{A,a} - \overline{n}_{A,a}\overline{m}_{A,b}}\right), \quad
    \theta_{B,a}^2 \in \left\{0, 1, \frac{\overline{\delta} m_{B,b} - \theta_{A,a}^2\overline{n}_{A,a}}{\overline{m}_{A,a}}, \frac{m_{A,b} - \theta_{A,a}^2 \overline{n}_{A,b}}{\overline{m}_{A,b}}\right\} \\
    &\theta_{A,a}^3 = \phi\left( \frac{m_{A,b}\underline{m}_{A,a} - \underline{\delta}m_{B,b}\underline{m}_{A,b}}
    {\underline{n}_{A,b}\underline{m}_{A,a} - \underline{n}_{A,a}\underline{m}_{A,b}}\right), \quad
    \theta_{B,a}^3 \in \left\{0, 1, \frac{\underline{\delta}m_{B,b} - \theta_{A,a}^3 \underline{n}_{A,a}}{\underline{m}_{A,a}} , \frac{m_{A,b} - \theta_{A,a}^3\underline{n}_{A,b}}{\underline{m}_{A,b}}\right\}\\
    &\theta_{A,a}^4 = \phi\left( \frac{m_{A,b}\underline{m}_{A,a} - \underline{\delta}m_{B,b}\overline{m}_{A,b}}
    {\underline{n}_{A,b}\underline{m}_{A,a} - \underline{n}_{A,a}\overline{m}_{A,b}} \right), \quad
    \theta_{B,a}^4 \in \left\{0, 1, \frac{\underline{\delta}m_{B,b} - \theta_{A,a}^4 \underline{n}_{A,a}}{\underline{m}_{A,a}} , \frac{m_{A,b} - \theta_{A,a}^4 \overline{n}_{A,b}}{\overline{m}_{A,b}}\right\} \\
    &\theta_{B,a}^5 \in \{0,1\}, \quad
    \theta_{A,a}^5 \in \left\{0, 1, \frac{\overline{\delta}m_{B,b} - \theta_{B,a}^5 \overline{m}_{A,a}}{\overline{n}_{A,a}}, \frac{\underline{\delta}m_{B,b} - \theta_{B,a}^5 \underline{m}_{A,a}}{\underline{n}_{A,a}}, \frac{m_{A,b} - \theta_{B,a}^5 \overline{m}_{A,b}}{\overline{n}_{A,b}}, \frac{m_{A,b} - \theta_{B,a}^5 \underline{m}_{A,b}}{\underline{n}_{A,b}}\right\},
\end{align*}
\normalsize
\end{proof}

\section{Addendum to the Platform Optimization Problem}\label{sec:addendum_constrained_problem}

\subsection{Fairness-Aware Optimization Problem}\label{sec:addendum_fairness_aware_problem}
According to Theorem \ref{theorem:closed_form_Z_transform_of_coeffs}, we have
\begin{align*}
    l_{A,a}(t,\theta) &= \theta_{A,a} w_{A,a}(t) + \theta_{B,a} u_{A,a}(t) \\
    l_{A,b}(t,\theta) &= \theta_{A,b} w_{A,b}(t) + \theta_{B,b} u_{A,b}(t) \\
    l_{B,a}(t,\theta) &= \theta_{B,a} w_{B,a}(t) + \theta_{A,a} u_{B,a}(t) \\
    l_{B,b}(t,\theta) &= \theta_{B,b} w_{B,b}(t) + \theta_{A,b} u_{B,b}(t).
\end{align*}
Define
\begin{align*}
    c_{A,a} &:= \sum_{t=1}^T (w_{A,a}(t) - w_{A,b}(t) + u_{B,a}(t) - u_{B,b}(t)) \\
    c_{B,a} &:= \sum_{t=1}^T (u_{A,a}(t) - u_{A,b}(t) + w_{B,a}(t) - w_{B,b}(t)).
\end{align*}
Then the fairness-aware optimization problem can be written as
\begin{alignat*}{2}
    \max_{\theta_{A,a}, \theta_{B,a}} \ & c_{A,a} \theta_{A,a} + c_{B,a} \theta_{B,a} && \\
    \text{s.t.} \ & \frac{\theta_{A,a} \sum_{t=1}^T w_{A,a}(t) + \theta_{B,a} \sum_{t=1}^T u_{A,a}(t)}{\theta_{A,b} \sum_{t=1}^T u_{B,b}(t) + \theta_{B,b} \sum_{t=1}^T w_{B,b}(t)} &&\leq \overline{\delta}  \\
    & \frac{-\theta_{A,a} \sum_{t=1}^T w_{A,a}(t) - \theta_{B,a} \sum_{t=1}^T u_{A,a}(t)}{\theta_{A,b} \sum_{t=1}^T u_{B,b}(t) + \theta_{B,b} \sum_{t=1}^T w_{B,b}(t)} &&\leq -\underline{\delta} \\
    & \frac{\theta_{A,b} \sum_{t=1}^T w_{A,b}(t) + \theta_{B,b} \sum_{t=1}^T u_{A,b}(t)}{\theta_{A,a} \sum_{t=1}^T u_{B,a}(t) + \theta_{B,a} \sum_{t=1}^T w_{B,a}(t)} &&\leq \overline{\delta} \\
    & \frac{(\theta_{A,a}-1) \sum_{t=1}^T w_{A,b}(t) + (\theta_{B,a}-1) \sum_{t=1}^T u_{A,b}(t)}{\theta_{A,a} \sum_{t=1}^T u_{B,a}(t) + \theta_{B,a} \sum_{t=1}^T w_{B,a}(t)} &&\leq -\underline{\delta} \\
    &\theta_{A,a}, \theta_{B,a} &&\leq 1 \\
    &\theta_{A,a}, \theta_{B,a} &&\geq 0.
\end{alignat*}

This is equivalent to:
\begin{alignat*}{2}
    \max_{\theta_{A,a}, \theta_{B,a}} \ & c_{A,a} \theta_{A,a} + c_{B,a} \theta_{B,a} && \\
    \text{s.t.} \ & \theta_{A,a} \left( \sum_{t=1}^T w_{A,a}(t) + \overline{\delta} \sum_{t=1}^T u_{B,b}(t)\right) + \theta_{B,a} \left( \sum_{t=1}^T u_{A,a}(t) + \overline{\delta} \sum_{t=1}^T w_{B,b}(t)\right) &&\leq \overline{\delta} \sum_{t=1}^T u_{B,b}(t) + \overline{\delta} \sum_{t=1}^T w_{B,b}(t) \\
    & \theta_{A,a} \left(- \underline{\delta} \sum_{t=1}^T u_{B,b}(t) - \sum_{t=1}^T w_{A,a}(t)\right) + \theta_{B,a} \left(- \underline{\delta} \sum_{t=1}^T w_{B,b}(t) - \sum_{t=1}^T u_{A,a}(t)\right) &&\leq -\underline{\delta} \sum_{t=1}^T u_{B,b}(t) -\underline{\delta} \sum_{t=1}^T w_{B,b}(t) \\
    & \theta_{A,a} \left(- \sum_{t=1}^T w_{A,b}(t) - \overline{\delta} \sum_{t=1}^T u_{B,a}(t)\right) + \theta_{B,a} \left(- \sum_{t=1}^T u_{A,b}(t) - \overline{\delta} \sum_{t=1}^T w_{B,a}(t)\right) &&\leq - \sum_{t=1}^T w_{A,b}(t) - \sum_{t=1}^T u_{A,b}(t) \\
    & \theta_{A,a} \left( \sum_{t=1}^T w_{A,b}(t) + \underline{\delta} \sum_{t=1}^T u_{B,a}(t)\right) + \theta_{B,a} \left( \sum_{t=1}^T u_{A,b}(t) + \underline{\delta} \sum_{t=1}^T w_{B,a}(t)\right) &&\leq \sum_{t=1}^T w_{A,b}(t) + \sum_{t=1}^T u_{A,b}(t) \\
    &\theta_{A,a}, \theta_{B,a} &&\leq 1 \\
    &\theta_{A,a} \theta_{B,a} &&\geq 0.
\end{alignat*}
\normalsize

\subsection{Fairness Constraints}\label{sec:fairness_constraints}

From constraints \eqref{eq:platform-constraints-1} and \eqref{eq:platform-constraints-2} and using Theorem \ref{theorem:closed_form_Z_transform_of_coeffs}, we infer the following feasible bounds on $\theta_{B,a}$ (dependent on $\theta_{A,a}$), in addition to $\theta_{A,a}, \theta_{B,a} \in [0,1]$:
\begin{align*}
    \theta_{B,a} &\leq \frac{\overline{\delta} \sum_{t=1}^T u_{B,b}(t) + \overline{\delta} \sum_{t=1}^T w_{B,b}(t)}{\sum_{t=1}^T u_{A,a}(t) + \overline{\delta} \sum_{t=1}^T w_{B,b}(t)} - \theta_{A,a} \frac{\overline{\delta} \sum_{t=1}^T u_{B,b}(t) + \sum_{t=1}^T w_{A,a}(t)}{\sum_{t=1}^T u_{A,a}(t) + \overline{\delta} \sum_{t=1}^T w_{B,b}(t)} =: y_1 \\
    \theta_{B,a} &\geq \frac{\underline{\delta} \sum_{t=1}^T u_{B,b}(t) + \underline{\delta} \sum_{t=1}^T w_{B,b}(t)}{\sum_{t=1}^T u_{A,a}(t) + \underline{\delta} \sum_{t=1}^T w_{B,b}(t)} - \theta_{A,a} \frac{\underline{\delta} \sum_{t=1}^T u_{B,b}(t) + \sum_{t=1}^T w_{A,a}(t)}{\sum_{t=1}^T u_{A,a}(t) + \underline{\delta} \sum_{t=1}^T w_{B,b}(t)} =: y_2 \\
    \theta_{B,a} &\geq \frac{\sum_{t=1}^T u_{A,b}(t) + \sum_{t=1}^T w_{A,b}(t)}{\sum_{t=1}^T u_{A,b}(t) + \overline{\delta} \sum_{t=1}^T w_{B,a}(t)} - \theta_{A,a} \frac{\overline{\delta} \sum_{t=1}^T u_{B,a}(t) + \sum_{t=1}^T w_{A,b}(t)}{\sum_{t=1}^T u_{A,b}(t) + \overline{\delta} \sum_{t=1}^T w_{B,a}(t)} =: y_3 \\
    \theta_{B,a} &\leq \frac{\sum_{t=1}^T u_{A,b}(t) + \sum_{t=1}^T w_{A,b}(t)}{\sum_{t=1}^T u_{A,b}(t) + \underline{\delta} \sum_{t=1}^T w_{B,a}(t)} - \theta_{A,a} \frac{\underline{\delta} \sum_{t=1}^T u_{B,a}(t) + \sum_{t=1}^T w_{A,b}(t)}{\sum_{t=1}^T u_{A,b}(t) + \underline{\delta} \sum_{t=1}^T w_{B,a}(t)} =: y_4.
\end{align*}
Using the substitutions introduced in the main paper, we then obtain
\begin{align*}
    y_1 &= \overline{\delta}\frac{m_{B,b}}{\overline{m}_{A,a}} - \theta_{A,a} \frac{\overline{n}_{A,a}}{\overline{m}_{A,a}} \\
    y_2 &= \underline{\delta}\frac{m_{B,b}}{\underline{m}_{A,a}} - \theta_{A,a} \frac{\underline{n}_{A,a}}{\underline{m}_{A,a}} \\
    y_3 &= \frac{m_{A,b}}{\overline{m}_{A,b}} - \theta_{A,a} \frac{\overline{n}_{A,b}}{\overline{m}_{A,b}} \\
    y_4 &= \frac{m_{A,b}}{\underline{m}_{A,b}} - \theta_{A,a} \frac{\underline{n}_{A,b}}{\underline{m}_{A,b}}.
\end{align*}

Note that the bounding hyperplanes $y_1$ to $y_4$ all have negative slope.
In a 2-dimensional plot, the axes intersects of those hyperplanes are given as follows:
\begin{align*}
    y_1(\theta_{A,a}=0) &= \frac{\sum_{t=1}^T u_{B,b}(t) + \sum_{t=1}^T w_{B,b}(t)}{\frac{1}{\overline{\delta}}\sum_{t=1}^T u_{A,a}(t) + \sum_{t=1}^T w_{B,b}(t)}, \quad \theta_{A,a}(y_1=0) = \frac{\sum_{t=1}^T u_{B,b}(t) + \sum_{t=1}^T w_{B,b}(t)}{\sum_{t=1}^T u_{B,b}(t) + \frac{1}{\overline{\delta}} \sum_{t=1}^T w_{A,a}(t)} \\
    y_2(\theta_{A,a}=0) &= \frac{\sum_{t=1}^T u_{B,b}(t) + \sum_{t=1}^T w_{B,b}(t)}{\frac{1}{\underline{\delta}}\sum_{t=1}^T u_{A,a}(t) + \sum_{t=1}^T w_{B,b}(t)}, \quad \theta_{A,a}(y_2=0) = \frac{\sum_{t=1}^T u_{B,b}(t) + \sum_{t=1}^T w_{B,b}(t)}{\sum_{t=1}^T u_{B,b}(t) + \frac{1}{\underline{\delta}} \sum_{t=1}^T w_{A,a}(t)} \\
    y_3(\theta_{A,a}=0) &= \frac{\sum_{t=1}^T u_{A,b}(t) + \sum_{t=1}^T w_{A,b}(t)}{\sum_{t=1}^T u_{A,b}(t) + \overline{\delta}\sum_{t=1}^T w_{B,a}(t)}, \quad \theta_{A,a}(y_3=0) = \frac{\sum_{t=1}^T u_{A,b}(t) + \sum_{t=1}^T w_{A,b}(t)}{\overline{\delta}\sum_{t=1}^T u_{B,a}(t) + \sum_{t=1}^T w_{A,b}(t)} \\
    y_4(\theta_{A,a}=0) &= \frac{\sum_{t=1}^T u_{A,b}(t) + \sum_{t=1}^T w_{A,b}(t)}{\sum_{t=1}^T u_{A,b}(t) + \underline{\delta}\sum_{t=1}^T w_{B,a}(t)}, \quad \theta_{A,a}(y_4=0) = \frac{\sum_{t=1}^T u_{A,b}(t) + \sum_{t=1}^T w_{A,b}(t)}{\underline{\delta}\sum_{t=1}^T u_{B,a}(t) + \sum_{t=1}^T w_{A,b}(t)}.
\end{align*}
\normalsize
As a sanity check, note that as we let $\overline{\delta} \to \infty$ and $\underline{\delta} \to 0$ (and using Lemma \ref{lemma:incr_linear}), the axes intersects of $y_1$ and $y_4$ will become greater or equal to $1$, whereas the axes intersects of $y_2$ and $y_3$ will go to $0$, rendering constraints \eqref{eq:platform-constraints-1} and \eqref{eq:platform-constraints-2} redundant, as desired.
We further see that
\begin{align}\label{eq:intersect_magenta}
\begin{split}
    y_4(\theta_{A,a}=0) &> y_3(\theta_{A,a}=0) \\
    \theta_{A,a}(y_4=0) &> \theta_{A,a}(y_3=0)
\end{split}
\end{align}
as well as
\begin{align}\label{eq:y_1_greater_y_2}
\begin{split}
    y_1(\theta_{A,a}=0) &> y_2(\theta_{A,a}=0) \\
    \theta_{A,a}(y_1=0) &> \theta_{A,a}(y_2=0).
\end{split}
\end{align}

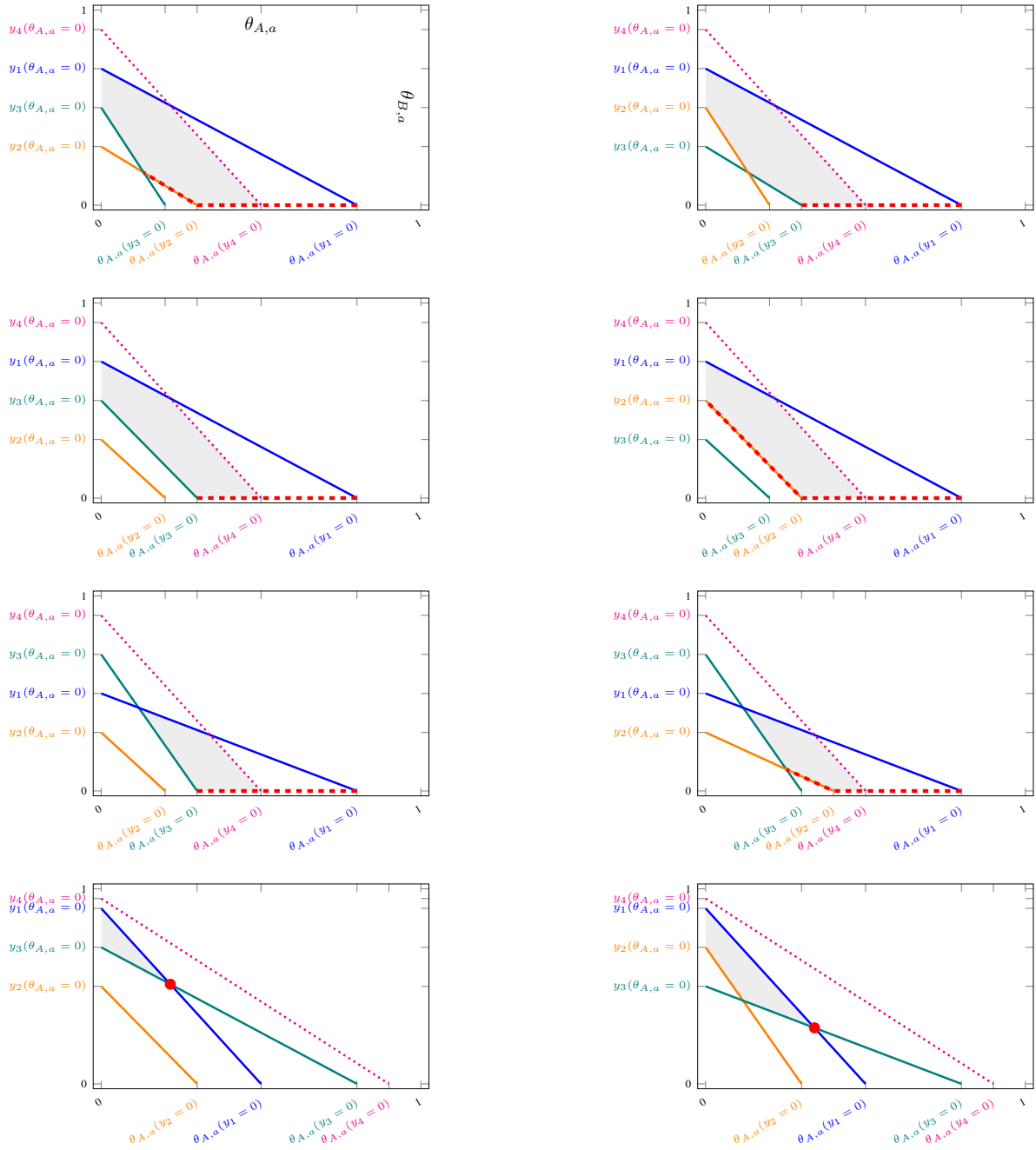
\begin{figure}
\centering
\begin{minipage}{0.48\textwidth}
    \centering
    \begin{tikzpicture}[scale=0.8]
    \begin{axis}[
        x label style={at={(axis description cs:0.5,0.975)},anchor=north},
        y label style={at={(axis description cs:0.975,0.5)},rotate=180,anchor=north},
        xlabel={$\theta_{A,a}$},
        ylabel={$\theta_{B,a}$},
        height=5.5cm,
        width=8cm,
        xmin=-0.025, xmax=1.025,
        ymin=-0.025, ymax=1.025,
        xtick={0,0.2,0.3,0.8,0.5,1},
        ytick={0,0.5,0.3,0.7,0.9,1},
        xticklabels={0,\textcolor{teal}{$\theta_{A,a}(y_3=0)$},\textcolor{orange}{$\theta_{A,a}(y_2=0)$},\textcolor{blue}{$\theta_{A,a}(y_1=0)$},\textcolor{magenta}{$\theta_{A,a}(y_4=0)$},1},
        yticklabels={0,\textcolor{teal}{$y_3(\theta_{A,a}=0)$},\textcolor{orange}{$y_2(\theta_{A,a}=0)$},\textcolor{blue}{$y_1(\theta_{A,a}=0)$},\textcolor{magenta}{$y_4(\theta_{A,a}=0)$},1},
        x tick label style = {font = \tiny, align = center, rotate = 30, anchor = north east},
        y tick label style = {font = \tiny, align = center},
    ]
    
    \fill[fill=gray!20,opacity=.7]
        (0.3,0) -- (0.5,0) -- (0.2/0.925,189/370) -- (0,0.7) -- (0,0.5) -- (2/15,1/6) -- cycle;
    
    \addplot[
        color=orange,
        style=very thick,
        name path = orange
        ]
        plot coordinates {
        (0,0.3)(0.3,0)
        };
        
    \addplot[
        color=blue,
        style=very thick,
        name path = blue
        ]
        plot coordinates {
        (0,0.7)(0.8,0)
        };
        
    \addplot[
        color=teal,
        style=very thick,
        name path = teal
        ]
        plot coordinates {
        (0,0.5)(0.2,0)
        };
        
    \addplot[
        color=magenta,
        style=very thick, dotted,
        name path = magenta
        ]
        plot coordinates {
        (0,0.9)(0.5,0)
        };
        
    \addplot[
        color=red,
        mark options=solid,
        mark indices = {1,2},
        line width = 2pt,
        style = dashed,
        name path = red
        ]
        plot coordinates {
        (0.8,0)(0.3,0)(2/15,1/6)
        };
    
    \end{axis}
    \end{tikzpicture}
\end{minipage}
\hspace*{\fill}
\begin{minipage}{0.48\textwidth}
    \centering
    \begin{tikzpicture}[scale=0.8]
    \begin{axis}[
        x label style={at={(axis description cs:0.5,0.975)},anchor=north},
        y label style={at={(axis description cs:0.975,0.5)},rotate=180,anchor=north},
        height=5.5cm,
        width=8cm,
        xmin=-0.025, xmax=1.025,
        ymin=-0.025, ymax=1.025,
        xtick={0,0.2,0.3,0.8,0.5,1},
        ytick={0,0.5,0.3,0.7,0.9,1},
        xticklabels={0,\textcolor{orange}{$\theta_{A,a}(y_2=0)$},\textcolor{teal}{$\theta_{A,a}(y_3=0)$},\textcolor{blue}{$\theta_{A,a}(y_1=0)$},\textcolor{magenta}{$\theta_{A,a}(y_4=0)$},1},
        yticklabels={0,\textcolor{orange}{$y_2(\theta_{A,a}=0)$},\textcolor{teal}{$y_3(\theta_{A,a}=0)$},\textcolor{blue}{$y_1(\theta_{A,a}=0)$},\textcolor{magenta}{$y_4(\theta_{A,a}=0)$},1},
        x tick label style = {font = \tiny, align = center, rotate = 30, anchor = north east},
        y tick label style = {font = \tiny, align = center},
    ]
    
    \fill[fill=gray!20,opacity=.7]
        (0.3,0) -- (0.5,0) -- (0.2/0.925,189/370) -- (0,0.7) -- (0,0.5) -- (2/15,1/6) -- cycle;
    
    \addplot[
        color=teal,
        style=very thick,
        name path = orange
        ]
        plot coordinates {
        (0,0.3)(0.3,0)
        };
        
    \addplot[
        color=blue,
        style=very thick,
        name path = blue
        ]
        plot coordinates {
        (0,0.7)(0.8,0)
        };
        
    \addplot[
        color=orange,
        style=very thick,
        name path = teal
        ]
        plot coordinates {
        (0,0.5)(0.2,0)
        };
        
    \addplot[
        color=magenta,
        style=very thick, dotted,
        name path = magenta
        ]
        plot coordinates {
        (0,0.9)(0.5,0)
        };
        
    \addplot[
        color=red,
        mark options=solid,
        line width = 2pt,
        mark indices = {1},
        style = dashed,
        name path = red
        ]
        plot coordinates {
        (0.8,0)(0.3,0)
        };
    
    \end{axis}
    \end{tikzpicture}
\end{minipage}

\medskip
\begin{minipage}{0.48\textwidth}
    \centering
    \begin{tikzpicture}[scale=0.8]
    \begin{axis}[
        x label style={at={(axis description cs:0.5,0.975)},anchor=north},
        y label style={at={(axis description cs:0.975,0.5)},rotate=180,anchor=north},
        height=5.5cm,
        width=8cm,
        xmin=-0.025, xmax=1.025,
        ymin=-0.025, ymax=1.025,
        xtick={0,0.2,0.3,0.8,0.5,1},
        ytick={0,0.5,0.3,0.7,0.9,1},
        xticklabels={0,\textcolor{orange}{$\theta_{A,a}(y_2=0)$},\textcolor{teal}{$\theta_{A,a}(y_3=0)$},\textcolor{blue}{$\theta_{A,a}(y_1=0)$},\textcolor{magenta}{$\theta_{A,a}(y_4=0)$},1},
        yticklabels={0,\textcolor{teal}{$y_3(\theta_{A,a}=0)$},\textcolor{orange}{$y_2(\theta_{A,a}=0)$},\textcolor{blue}{$y_1(\theta_{A,a}=0)$},\textcolor{magenta}{$y_4(\theta_{A,a}=0)$},1},
        x tick label style = {font = \tiny, align = center, rotate = 30, anchor = north east},
        y tick label style = {font = \tiny, align = center},
    ]
    
    \fill[fill=gray!20,opacity=.7]
        (0.3,0) -- (0.5,0) -- (0.2/0.925,189/370) -- (0,0.7) -- (0,0.5) -- cycle;
    
    \addplot[
        color=teal,
        style=very thick,
        name path = orange
        ]
        plot coordinates {
        (0,0.5)(0.3,0)
        };
        
    \addplot[
        color=blue,
        style=very thick,
        name path = blue
        ]
        plot coordinates {
        (0,0.7)(0.8,0)
        };
        
    \addplot[
        color=orange,
        style=very thick,
        name path = teal
        ]
        plot coordinates {
        (0,0.3)(0.2,0)
        };
        
    \addplot[
        color=magenta,
        style=very thick, dotted,
        name path = magenta
        ]
        plot coordinates {
        (0,0.9)(0.5,0)
        };
        
    \addplot[
        color=red,
        mark options=solid,
        mark indices = {1},
        line width = 2pt,
        style = dashed,
        name path = red
        ]
        plot coordinates {
        (0.8,0)(0.3,0)
        };
    
    \end{axis}
    \end{tikzpicture}
\end{minipage}
\hspace*{\fill}
\begin{minipage}{0.48\textwidth}
    \centering
    \begin{tikzpicture}[scale=0.8]
    \begin{axis}[
        x label style={at={(axis description cs:0.5,0.975)},anchor=north},
        y label style={at={(axis description cs:0.975,0.5)},rotate=180,anchor=north},
        height=5.5cm,
        width=8cm,
        xmin=-0.025, xmax=1.025,
        ymin=-0.025, ymax=1.025,
        xtick={0,0.2,0.3,0.8,0.5,1},
        ytick={0,0.5,0.3,0.7,0.9,1},
        xticklabels={0,\textcolor{teal}{$\theta_{A,a}(y_3=0)$},\textcolor{orange}{$\theta_{A,a}(y_2=0)$},\textcolor{blue}{$\theta_{A,a}(y_1=0)$},\textcolor{magenta}{$\theta_{A,a}(y_4=0)$},1},
        yticklabels={0,\textcolor{orange}{$y_2(\theta_{A,a}=0)$},\textcolor{teal}{$y_3(\theta_{A,a}=0)$},\textcolor{blue}{$y_1(\theta_{A,a}=0)$},\textcolor{magenta}{$y_4(\theta_{A,a}=0)$},1},
        x tick label style = {font = \tiny, align = center, rotate = 30, anchor = north east},
        y tick label style = {font = \tiny, align = center},
    ]
    
    \fill[fill=gray!20,opacity=.7]
        (0.3,0) -- (0.5,0) -- (0.2/0.925,189/370) -- (0,0.7) -- (0,0.5) -- cycle;
    
    \addplot[
        color=orange,
        style=very thick,
        name path = orange
        ]
        plot coordinates {
        (0,0.5)(0.3,0)
        };
        
    \addplot[
        color=blue,
        style=very thick,
        name path = blue
        ]
        plot coordinates {
        (0,0.7)(0.8,0)
        };
        
    \addplot[
        color=teal,
        style=very thick,
        name path = teal
        ]
        plot coordinates {
        (0,0.3)(0.2,0)
        };
        
    \addplot[
        color=magenta,
        style=very thick, dotted,
        name path = magenta
        ]
        plot coordinates {
        (0,0.9)(0.5,0)
        };
        
    \addplot[
        color=red,
        mark options=solid,
        mark indices = {1,2,3},
        line width = 2pt,
        style = dashed,
        name path = red
        ]
        plot coordinates {
        (0.8,0)(0.3,0)(0,0.5)
        };
    
    \end{axis}
    \end{tikzpicture}
\end{minipage}

\medskip
\begin{minipage}{0.48\textwidth}
    \centering
    \begin{tikzpicture}[scale=0.8]
    \begin{axis}[
        x label style={at={(axis description cs:0.5,0.975)},anchor=north},
        y label style={at={(axis description cs:0.975,0.5)},rotate=180,anchor=north},
        height=5.5cm,
        width=8cm,
        xmin=-0.025, xmax=1.025,
        ymin=-0.025, ymax=1.025,
        xtick={0,0.2,0.3,0.8,0.5,1},
        ytick={0,0.5,0.3,0.7,0.9,1},
        xticklabels={0,\textcolor{orange}{$\theta_{A,a}(y_2=0)$},\textcolor{teal}{$\theta_{A,a}(y_3=0)$},\textcolor{blue}{$\theta_{A,a}(y_1=0)$},\textcolor{magenta}{$\theta_{A,a}(y_4=0)$},1},
        yticklabels={0,\textcolor{blue}{$y_1(\theta_{A,a}=0)$},\textcolor{orange}{$y_2(\theta_{A,a}=0)$},\textcolor{teal}{$y_3(\theta_{A,a}=0)$},\textcolor{magenta}{$y_4(\theta_{A,a}=0)$},1},
        x tick label style = {font = \tiny, align = center, rotate = 30, anchor = north east},
        y tick label style = {font = \tiny, align = center},
    ]
    
    \fill[fill=gray!20,opacity=.7]
        (0.3,0) -- (0.5,0) -- (0.4/1.175,0.5-0.25/1.175) -- (24/205,0.5-15/205) -- cycle;
    
    \addplot[
        color=teal,
        style=very thick,
        name path = orange
        ]
        plot coordinates {
        (0,0.7)(0.3,0)
        };
        
    \addplot[
        color=blue,
        style=very thick,
        name path = blue
        ]
        plot coordinates {
        (0,0.5)(0.8,0)
        };
        
    \addplot[
        color=orange,
        style=very thick,
        name path = teal
        ]
        plot coordinates {
        (0,0.3)(0.2,0)
        };
        
    \addplot[
        color=magenta,
        style=very thick, dotted,
        name path = magenta
        ]
        plot coordinates {
        (0,0.9)(0.5,0)
        };
        
    \addplot[
        color=red,
        mark options=solid,
        mark indices = {1},
        line width = 2pt,
        style = dashed,
        name path = red
        ]
        plot coordinates {
        (0.8,0)(0.3,0)
        };
    
    \end{axis}
    \end{tikzpicture}
\end{minipage}
\hspace*{\fill}
\begin{minipage}{0.48\textwidth}
    \centering
    \begin{tikzpicture}[scale=0.8]
    \begin{axis}[
        x label style={at={(axis description cs:0.5,0.975)},anchor=north},
        y label style={at={(axis description cs:0.975,0.5)},rotate=180,anchor=north},
        height=5.5cm,
        width=8cm,
        xmin=-0.025, xmax=1.025,
        ymin=-0.025, ymax=1.025,
        xtick={0,0.4,0.3,0.8,0.5,1},
        ytick={0,0.5,0.3,0.7,0.9,1},
        xticklabels={0,\textcolor{orange}{$\theta_{A,a}(y_2=0)$},\textcolor{teal}{$\theta_{A,a}(y_3=0)$},\textcolor{blue}{$\theta_{A,a}(y_1=0)$},\textcolor{magenta}{$\theta_{A,a}(y_4=0)$},1},
        yticklabels={0,\textcolor{blue}{$y_1(\theta_{A,a}=0)$},\textcolor{orange}{$y_2(\theta_{A,a}=0)$},\textcolor{teal}{$y_3(\theta_{A,a}=0)$},\textcolor{magenta}{$y_4(\theta_{A,a}=0)$},1},
        x tick label style = {font = \tiny, align = center, rotate = 30, anchor = north east},
        y tick label style = {font = \tiny, align = center},
    ]
    
    \fill[fill=gray!20,opacity=.7]
        (0.4,0) -- (0.5,0) -- (0.4/1.175,0.5-0.25/1.175) -- (24/205,0.5-15/205) -- (24/95,0.7-168/285) -- cycle;
    
    \addplot[
        color=teal,
        style=very thick,
        name path = orange
        ]
        plot coordinates {
        (0,0.7)(0.3,0)
        };
        
    \addplot[
        color=blue,
        style=very thick,
        name path = blue
        ]
        plot coordinates {
        (0,0.5)(0.8,0)
        };
        
    \addplot[
        color=orange,
        style=very thick,
        name path = teal
        ]
        plot coordinates {
        (0,0.3)(0.4,0)
        };
        
    \addplot[
        color=magenta,
        style=very thick, dotted,
        name path = magenta
        ]
        plot coordinates {
        (0,0.9)(0.5,0)
        };
        
    \addplot[
        color=red,
        mark options=solid,
        mark indices = {1,2},
        line width = 2pt,
        style = dashed,
        name path = red
        ]
        plot coordinates {
        (0.8,0)(0.4,0)(24/95,0.7-168/285)
        };
    
    \end{axis}
    \end{tikzpicture}
\end{minipage}

\medskip
\begin{minipage}{0.48\textwidth}
    \centering
    \begin{tikzpicture}[scale=0.8]
    \begin{axis}[
        x label style={at={(axis description cs:0.5,0.975)},anchor=north},
        y label style={at={(axis description cs:0.975,0.5)},rotate=180,anchor=north},
        height=5.5cm,
        width=8cm,
        xmin=-0.025, xmax=1.025,
        ymin=-0.025, ymax=1.025,
        xtick={0,0.3,0.8,0.5,1,0.9},
        ytick={0,0.5,0.7,0.9,1,0.95},
        xticklabels={0,\textcolor{orange}{$\theta_{A,a}(y_2=0)$},\textcolor{teal}{$\theta_{A,a}(y_3=0)$},\textcolor{blue}{$\theta_{A,a}(y_1=0)$},1,\textcolor{magenta}{$\theta_{A,a}(y_4=0)$}},
        yticklabels={0,\textcolor{orange}{$y_2(\theta_{A,a}=0)$},\textcolor{teal}{$y_3(\theta_{A,a}=0)$},\textcolor{blue}{$y_1(\theta_{A,a}=0)$},1,\textcolor{magenta}{$y_4(\theta_{A,a}=0)$}},
        x tick label style = {font = \tiny, align = center, rotate = 30, anchor = north east},
        y tick label style = {font = \tiny, align = center},
    ]
    
    \fill[fill=gray!20,opacity=.7]
        (0.2/0.925,189/370) -- (0,0.9) -- (0,0.7) -- cycle;
    
    \addplot[
        color=orange,
        style=very thick,
        name path = orange
        ]
        plot coordinates {
        (0,0.5)(0.3,0)
        };
        
    \addplot[
        color=teal,
        style=very thick,
        name path = blue
        ]
        plot coordinates {
        (0,0.7)(0.8,0)
        };
        
    \addplot[
        color=magenta,
        mark options=solid,
        style=very thick, dotted,
        name path = teal
        ]
        plot coordinates {
        (0,0.95)(0.9,0)
        };
        
    \addplot[
        color=blue,
        style=very thick,
        name path = magenta
        ]
        plot coordinates {
        (0,0.9)(0.5,0)
        };
        
    \addplot[
        color=red,
        mark=*,
        mark options = solid,
        line width = 2pt,
        name path = red
        ]
        plot coordinates {
        (0.2/0.925,189/370)
        };
    
    \end{axis}
    \end{tikzpicture}
\end{minipage}
\hspace*{\fill}
\begin{minipage}{0.48\textwidth}
    \centering
    \begin{tikzpicture}[scale=0.8]
    \begin{axis}[
        x label style={at={(axis description cs:0.5,0.975)},anchor=north},
        y label style={at={(axis description cs:0.975,0.5)},rotate=180,anchor=north},
        height=5.5cm,
        width=8cm,
        xmin=-0.025, xmax=1.025,
        ymin=-0.025, ymax=1.025,
        xtick={0,0.3,0.8,0.5,1,0.9},
        ytick={0,0.7,0.5,0.9,1,0.95},
        xticklabels={0,\textcolor{orange}{$\theta_{A,a}(y_2=0)$},\textcolor{teal}{$\theta_{A,a}(y_3=0)$},\textcolor{blue}{$\theta_{A,a}(y_1=0)$},1,\textcolor{magenta}{$\theta_{A,a}(y_4=0)$}},
        yticklabels={0,\textcolor{orange}{$y_2(\theta_{A,a}=0)$},\textcolor{teal}{$y_3(\theta_{A,a}=0)$},\textcolor{blue}{$y_1(\theta_{A,a}=0)$},1,\textcolor{magenta}{$y_4(\theta_{A,a}=0)$}},
        x tick label style = {font = \tiny, align = center, rotate = 30, anchor = north east},
        y tick label style = {font = \tiny, align = center},
    ]
    
    \fill[fill=gray!20,opacity=.7]
        (24/205,0.5-15/205) -- (0,0.7) -- (0,0.9) -- (16/47,0.5-80/376) -- cycle;
    
    \addplot[
        color=orange,
        style=very thick,
        name path = orange
        ]
        plot coordinates {
        (0,0.7)(0.3,0)
        };
        
    \addplot[
        color=teal,
        style=very thick,
        name path = blue
        ]
        plot coordinates {
        (0,0.5)(0.8,0)
        };
        
    \addplot[
        color=magenta,
        mark options=solid,
        style=very thick, dotted,
        name path = teal
        ]
        plot coordinates {
        (0,0.95)(0.9,0)
        };
        
    \addplot[
        color=blue,
        style=very thick,
        name path = magenta
        ]
        plot coordinates {
        (0,0.9)(0.5,0)
        };
        
    \addplot[
        color=red,
        mark=*,
        mark options = solid,
        line width = 2pt,
        name path = red
        ]
        plot coordinates {
        (16/47,0.5-80/376)
        };
    
    \end{axis}
    \end{tikzpicture}
\end{minipage}
\caption{Exemplary feasible regions (shaded gray) of the fairness-aware optimization problem for all (except ones with empty feasible regions) possible qualitative positions of lower-bound constraints (in orange and teal) in combination with the blue upper-bound constraint. Indicated by the red dashed line (or dot) is the trajectory of the optimal solution if $c_{A,a} > c_{B,a} > 0$, depending on the position of the second upper-bound constraint in dotted magenta.}
\label{fig:graphical_solution}
\end{figure}

Figure~\ref{fig:graphical_solution} depicts an example of non-empty feasible regions of the fairness-aware optimization problem for all eight possible qualitative positions of lower-bounding hyperplanes $y_2$ and $y_3$ in combination with the upper-bound constraint induced by $y_1$.
Given $y_1$, $y_2$, $y_3$, the second upper-bounding hyperplane $y_4$ can take on any position such that~\eqref{eq:intersect_magenta} is satisfied---otherwise, the feasible set will be empty.
Figure~\ref{fig:graphical_solution} also shows in dotted red the trajectory of the optimal solution to the fairness-aware problem for the given example and the case of $c_{A,a} > c_{B,a} > 0$, as a function of the position of $y_4$.
Note that the optimal solution in this case will involve the largest feasible value of $\theta_{A,a}$.

\section{Deferred Experimental Background and Results}
\label{app:experimental}

\paragraph{Parameter Values from Datasets}

\begin{table}
    \centering
    \resizebox{\textwidth}{!}{
    \begin{tabular}{l c c c c}
        \toprule
        Variable & \textsc{Twitter: US Elections} & \textsc{Twitter: Brexit} & \textsc{Twitter: Abortion} & \textsc{Facebook}\\ 
        \midrule
        $(\pi_A, \pi_B)$ & (0.432, 0.567) & (0.480, 0.520) & (0.623, 0.370) & (0.500, 0.500) \\
        $(q_A, q_B)$ & (0.9877, 1.0000) & (0.6800, 0.3840) & (0.5500, 0.8200) & (0.7200, 0.6800) \\
        $p_{A,a}$ params: $(\alpha, \beta)$ & (41.46, 556.87) & (1.64, 62.92) & (2.30, 27.59) & (0.95, 1.35) \\
        $p_{A,b}$ params: $(\alpha, \beta)$ & (0.75, 413.47) & (1.72, 380.14) & (0.16, 50.83) & (0.18, 2.76) \\
        $p_{B,a}$ params: $(\alpha, \beta)$ & (6.10, 1519.85) & (1.48, 27.40) & (0.25, 7.40) & (0.10, 3.09)  \\
        $p_{B,b}$ params: $(\alpha, \beta)$ & (2153.00, 23467.67) & (39.60, 505.90) & (2.20, 53.70) & (0.88, 1.62) \\
        \bottomrule
    \end{tabular}
    }
    \caption{Estimated model parameters from empirical datasets.}
    \label{tab:exp-param-vals}
\end{table}

Estimated parameters from empirical datasets are shown in Table~\ref{tab:exp-param-vals}.
Moreover, some of our model parameters are network-independent.
In that case, some of those default parameters are in Table~\ref{tab:default-params-data-indep}. 
\begin{table}
    \centering
    \begin{tabular}{c c}
    \toprule
    Variable & Value \\
    \midrule
    Number of trials & $25$ \\
    $T$ & $10$ \\
    $n$ & $10^5$ \\
    $c_{g,s}$ & $1$ \\
    $v_{g,g}$ & $2000$ \\
    $v_{g,g'}$ & $200$\\
    $\underline{\delta}$ & $\nicefrac{1}{4}$ \\
    $\overline{\delta}$ & 2 \\
    \bottomrule
    \end{tabular}
    \caption{Default network-independent parameters for experiments, unless otherwise changed.}
    \label{tab:default-params-data-indep}
\end{table}

\begin{figure}[ht!]
    \begin{minipage}{0.49\linewidth}
        \centering
        \includegraphics[width=\linewidth]{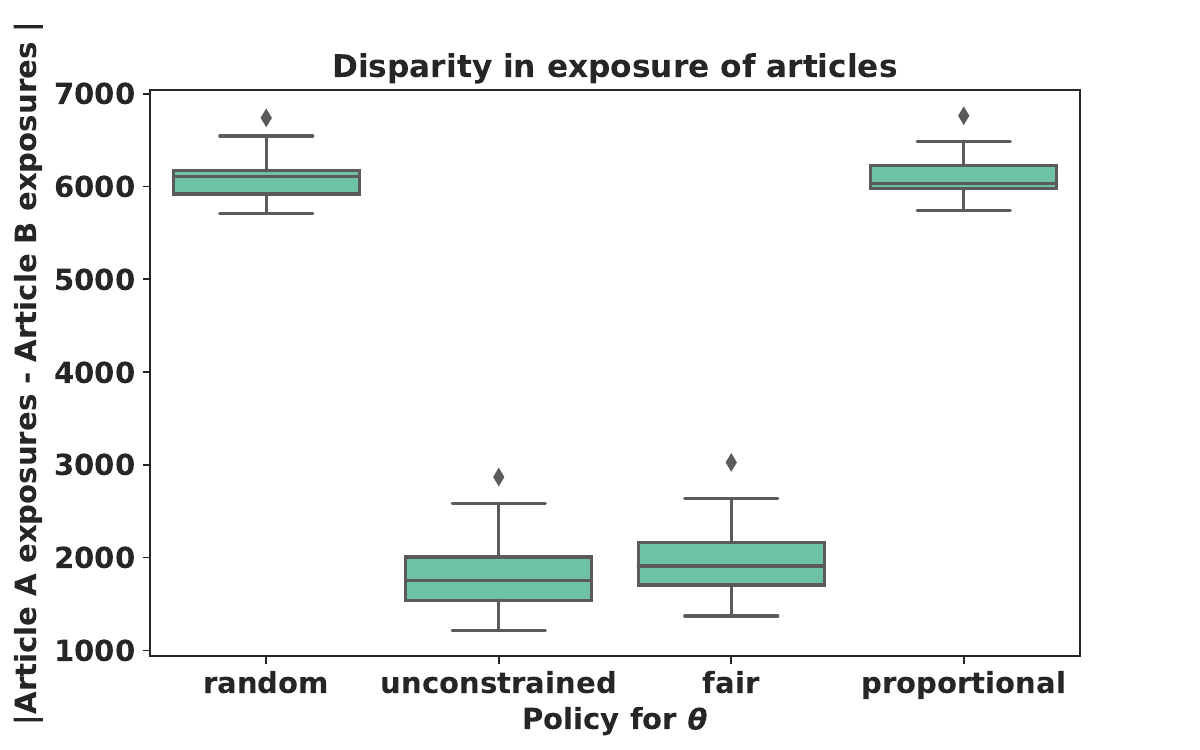}
    \end{minipage}  
    \hfill
    \begin{minipage}{0.49\linewidth}
        \centering
        \includegraphics[width=\linewidth]{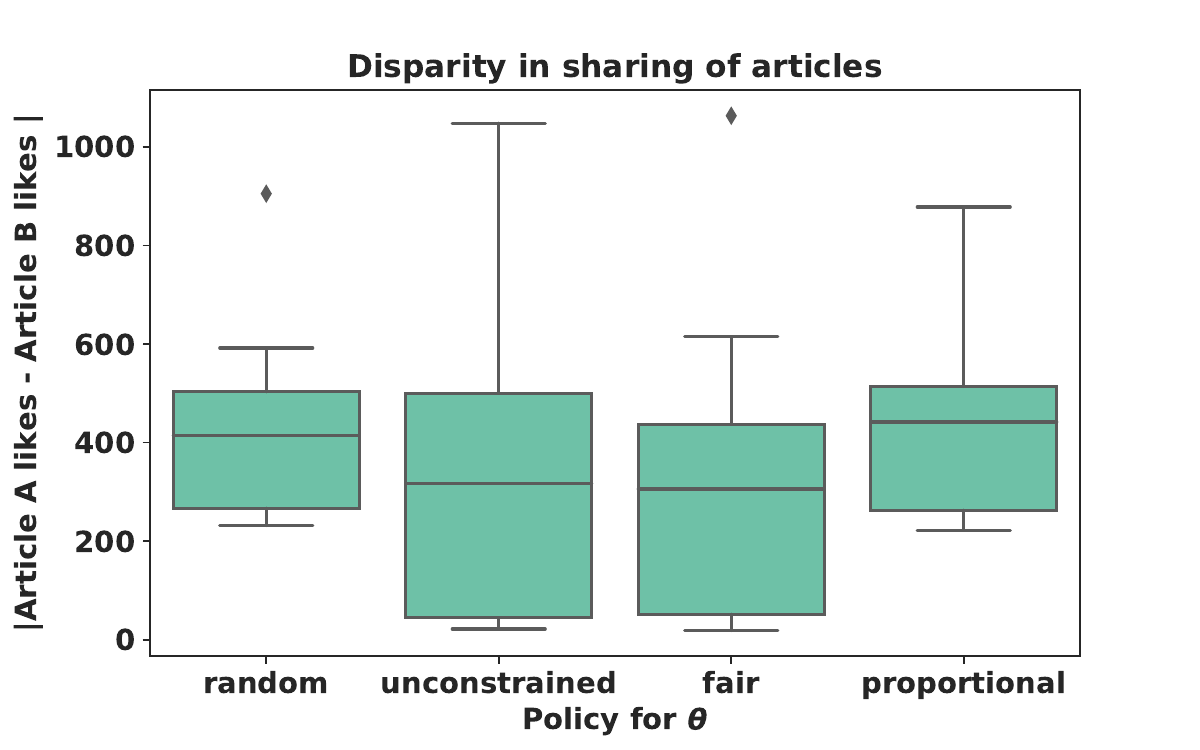}
    \end{minipage}
    \caption{Exposure disparity en masse and across groups.}
    \label{fig:facebook-boxplots-dispartity}
\end{figure}

\paragraph{Exposure Disparity}
We are interested in understanding how imposing balanced exposure constraints  might affect disparity in expected exposure and clicks.
Figure~\ref{fig:facebook-boxplots-dispartity} highlights the disparity in exposure for different optimization policies $\theta$.
%
We observe that a uniformly randomized policy (\textit{random}) and proportional policy (\textit{proportional}; $\theta_{g,s} = \pi_{g}$) yield a large disparity in article exposure between article $a$ and article $b$, while this disparity is lower in the fairness-agnostic (\textit{unconstrained}) and fairness-aware (\textit{fair}) settings, though there is no significant difference between the two.
When evaluating differences in how often the articles get liked (Figure~\ref{fig:facebook-boxplots-dispartity}, right) this gap closes across all four policies.
%

\begin{figure}[ht]
    \begin{minipage}{0.49\linewidth}
    \centering
    \includegraphics[width=\linewidth]{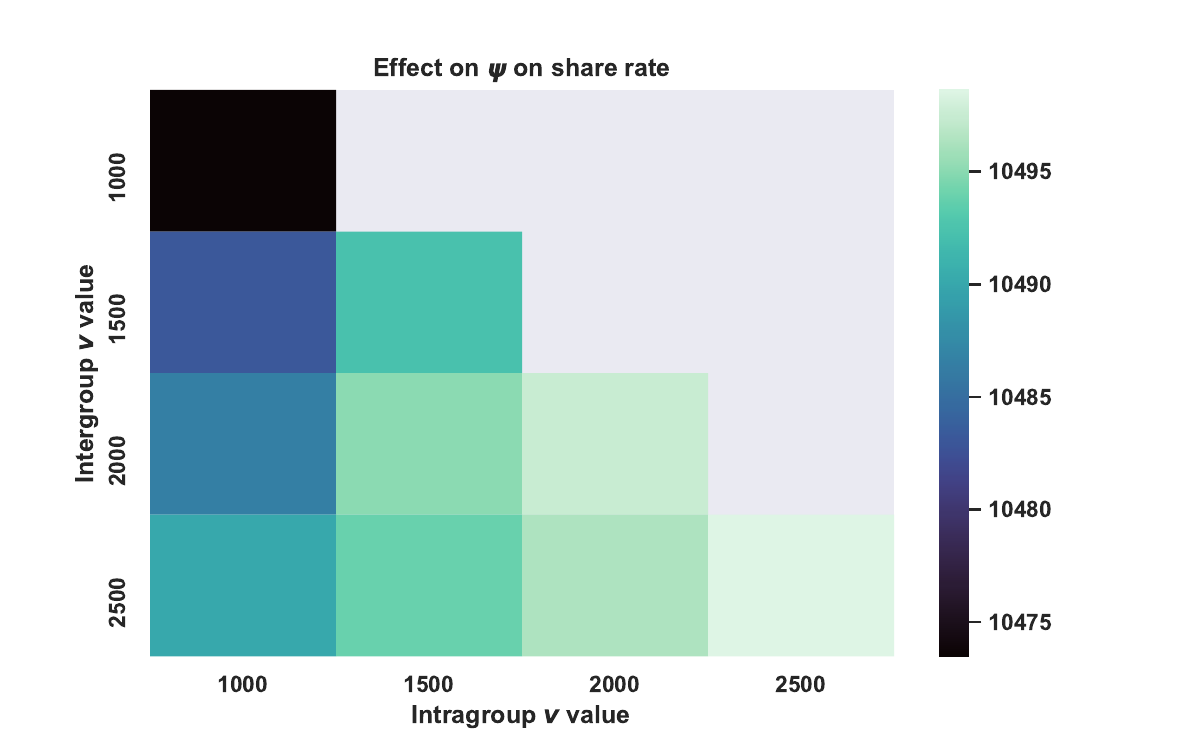}
    \label{fig:psi-heatmaps}
    \end{minipage}
    \hfill
    \begin{minipage}{0.49\linewidth}
    \centering
    \includegraphics[width=\linewidth]{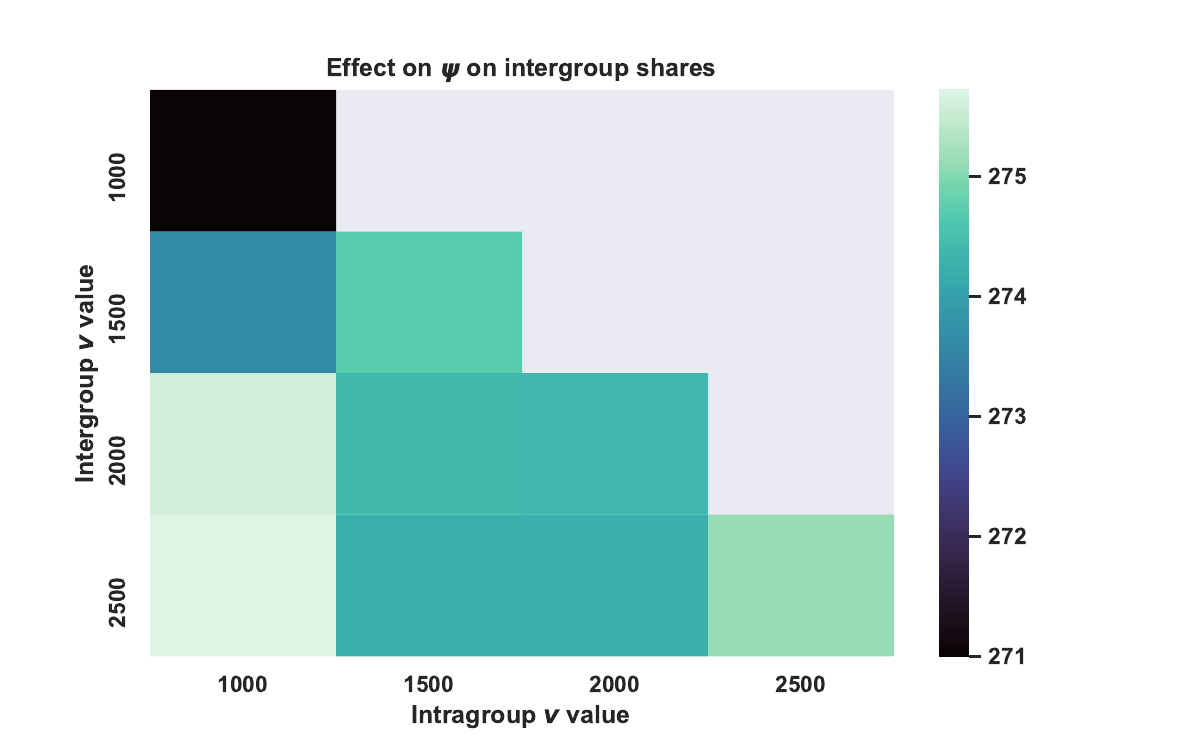}
    \label{fig:psi-intergroup-heatmaps}
    \end{minipage}
    \caption{The effect of click valuation $\psi_{g,s}$ impacts the total number of clicks in the model.}
    \label{fig:symmetric-psi-heatmaps}
\end{figure}

\paragraph{Population Interaction with Model: Varying Values for Clicking ($\psi$ via Proxy $\nicefrac{c}{v}$)}
While many of the model parameters can be approximated via maximum likelihood estimation on a given dataset, it is difficult to estimate values and costs for clicking on an article.
In Figure~\ref{fig:symmetric-psi-heatmaps} (left), we show a heatmap for the click rates in the model as a function of $v_{g,s}$ and $v_{g',s}$, where $c_{g,s} = 1$ for all $g$ and $s$, and preferences are symmetric.
Figure~\ref{fig:symmetric-psi-heatmaps} (right) shows intergroup click rates on the same axes. 
Results are averaged over $25$ trials.
In Figure~\ref{fig:symmetric-psi-heatmaps} (right), we can see how increasing the intergroup value $v_{g,g'}$ unsurprisingly increases the number of shares, and intragroup $v_{g,g}$ seems to have little effect. 

\begin{figure}
\begin{minipage}{0.49\linewidth}
    \centering
    \includegraphics[width=\linewidth]{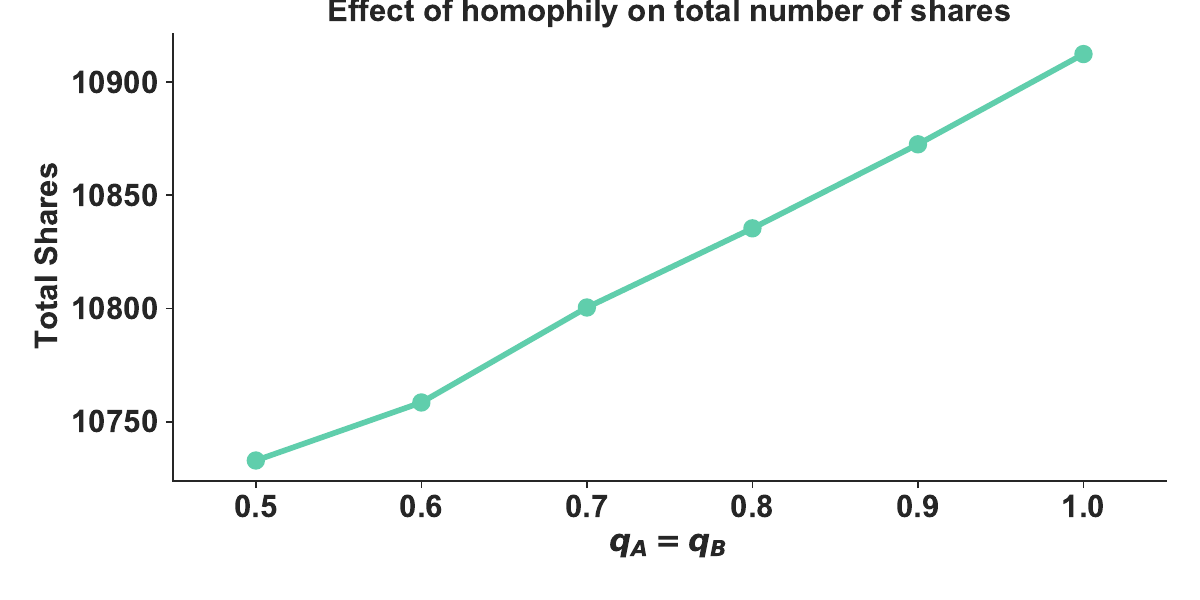}
\end{minipage}
\hfill
\begin{minipage}{0.49\linewidth}
    \centering
    \includegraphics[width=\linewidth]{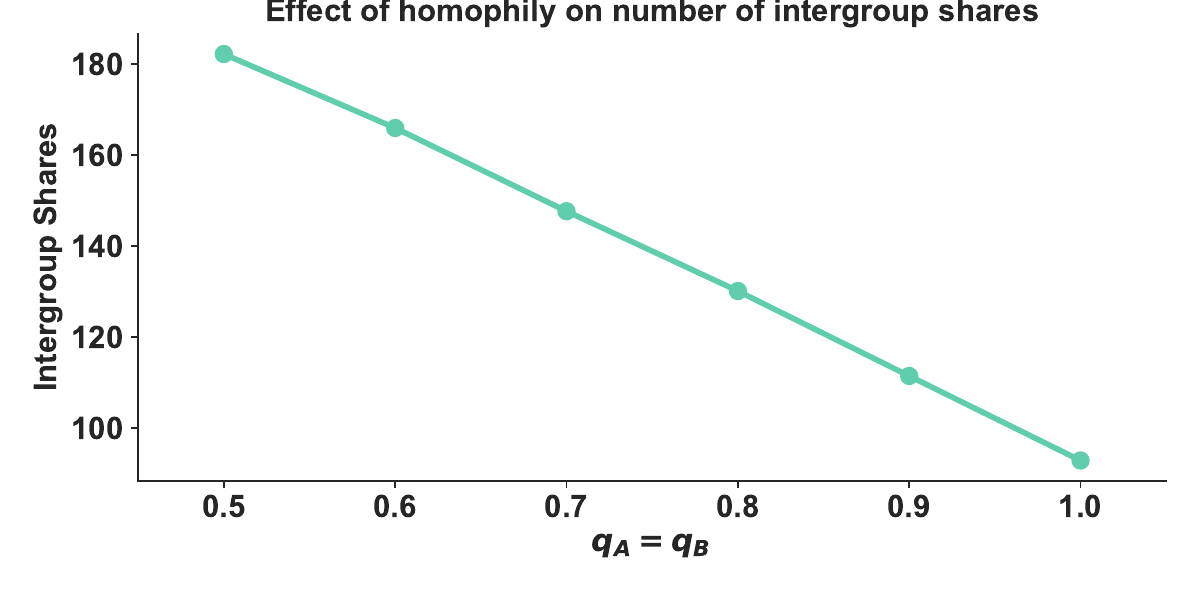}
\end{minipage}
\caption{Changing the homophily parameter effects on the number of shares is perhaps very intuitive. Increasing homophily increases total shares, but decreases intergroup shares.}
\label{fig:symmetric-homophily}
\end{figure}

\paragraph{Population Interaction with Model: Varying Homophily Constraints}  
Since homophily is such an important and well-studied aspect of information flow in a social network, we also consider the effect of estimating $q_A$ and $q_B$ from empirical datasets, it is worth observing that these variables might vary by topic, even on the same social network.\footnote{For example, the Brexit dataset of~\citet{ReplicationExposure} is not very homophilous for one of the positions, and we suspect this might be because opinion on Brexit tends to be correlated with age, but social networks often have many connections across generations.} 
These results are unsurprising: increasing homophily increases total shares, yet decreases intergroup shares, even when increased along both groups simultaneously (Figure~\ref{fig:symmetric-homophily}).

\section{Deferred Results for~\citet{balancedExposure}}\label{subapp:additional-experiments}

\citet{balancedExposure} studies user engagement with articles both inside and outside their preferred group across 14 different topics.
They found a stronger preference for like-minded articles in political topics than non-political (e.g., rooting for a football team to win the Super Bowl), and we estimate our model parameters by maximum likelihood estimation from their dataset.

\citet{balancedExposure} collected data on publicly available Tweets via the Twitter streaming API.
Consent is culturally assumed by Tweets being publicly available, though it is worth noting that this dataset being publicly available requires effort on the public's behalf to $(i)$ ascertain knowledge of their presence in the dataset, and $(ii)$ a request to be removed from the dataset does not prevent previous users of the dataset from having their Tweets.
Part of the dataset is available at \url{https://github.com/gvrkiran/BalancedExposure#readme} without license.
We only use publicly available data.

\paragraph{Twitter: US Elections~\citep{balancedExposure}}

\begin{figure}[ht]
    \centering
    \includegraphics[width=0.5\linewidth]{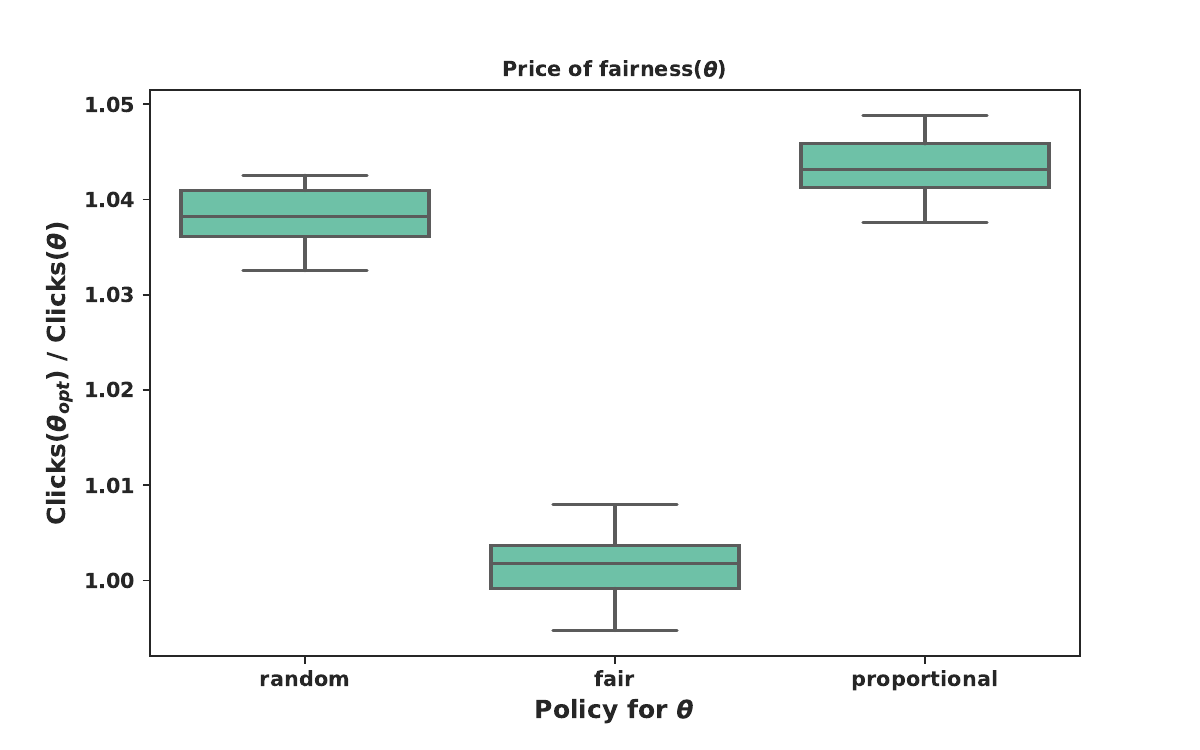}
    \caption{Price of fairness with parameters estimated from~\citet{balancedExposure}: US Elections.}
    \label{fig:twitter_uselections_pof_boxplot}
\end{figure}

\begin{figure}[ht]
    \begin{minipage}{0.49\linewidth}
        \centering
        \includegraphics[width=0.9\linewidth]{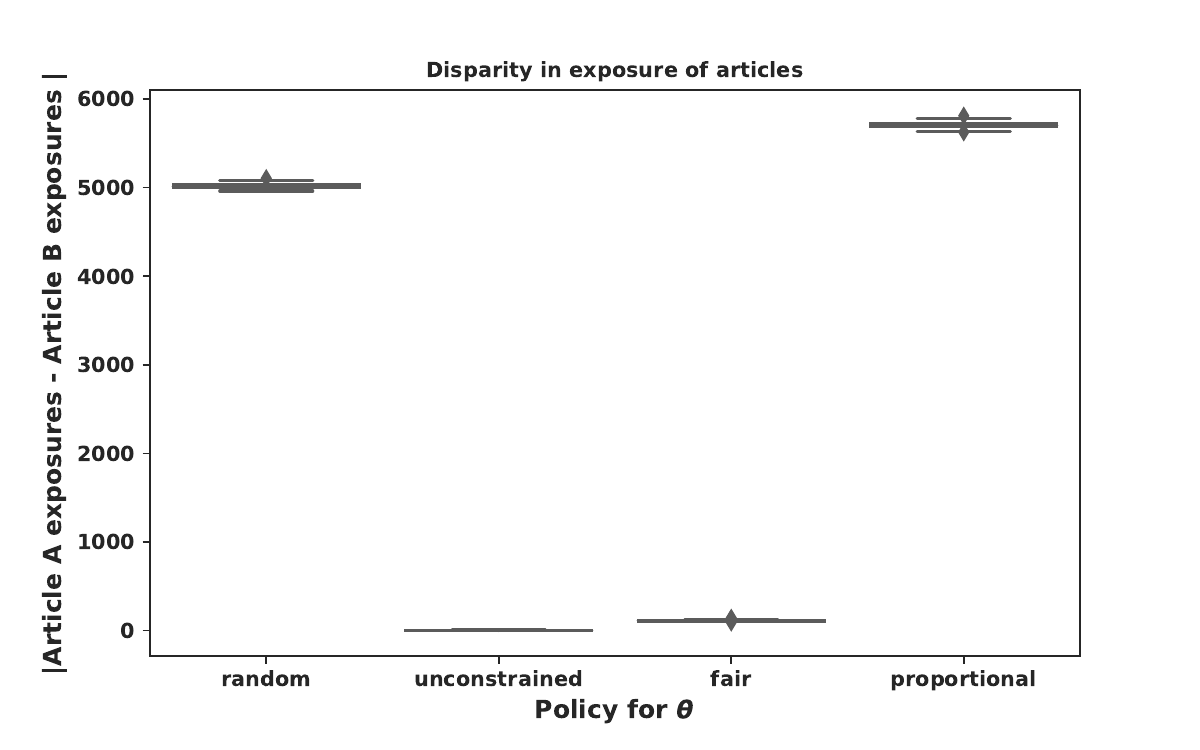}
    \end{minipage}  
    \hfill
    \begin{minipage}{0.49\linewidth}
        \centering
        \includegraphics[width=0.9\linewidth]{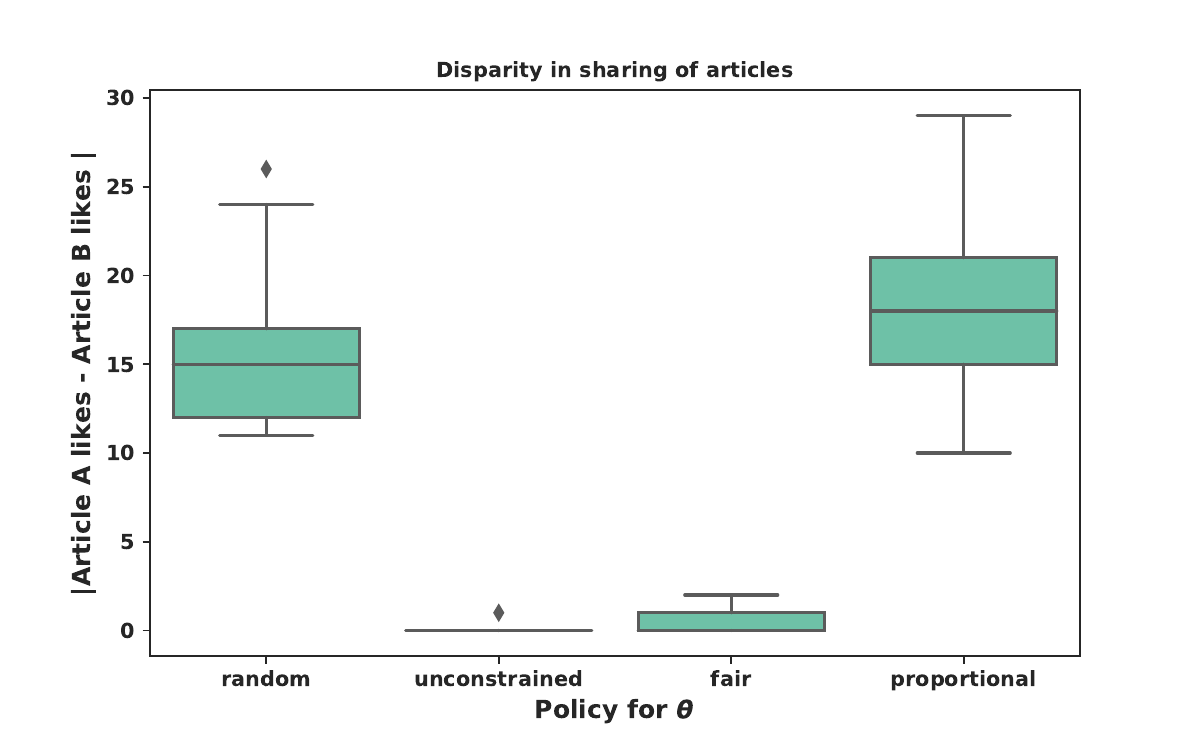}
    \end{minipage}
    \hfill
    \begin{minipage}{0.49\linewidth}
        \centering
        \includegraphics[width=0.9\linewidth]{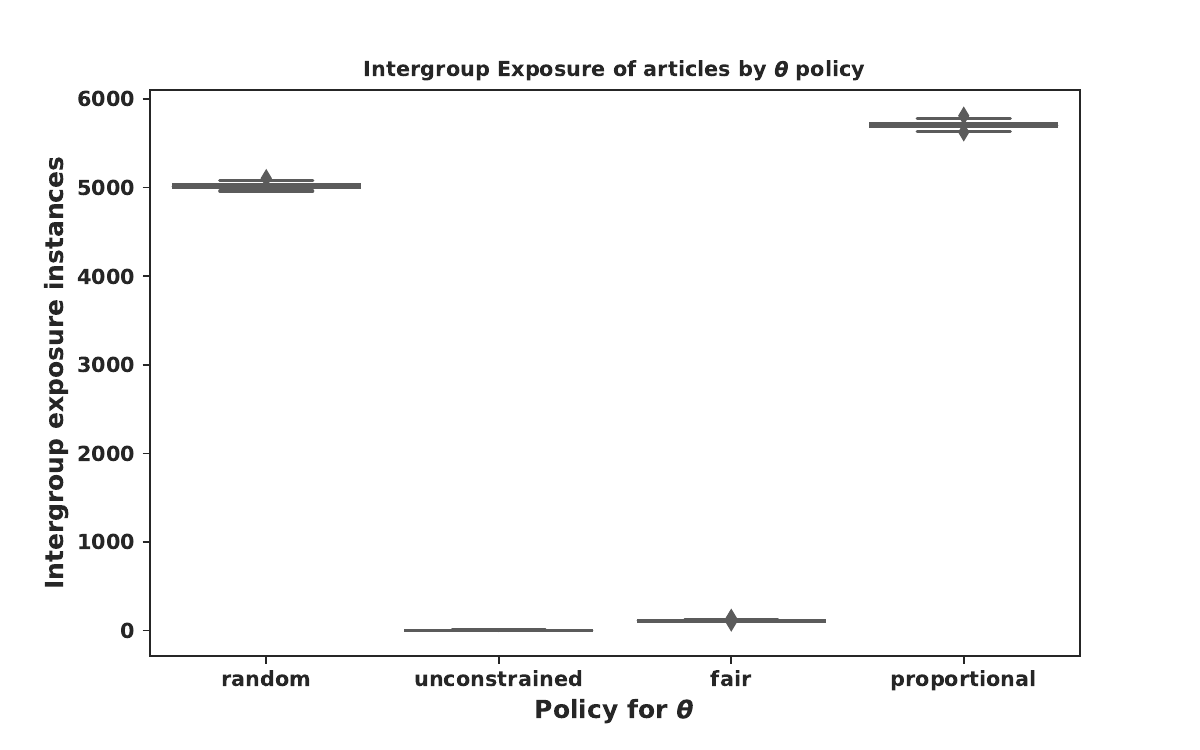}
    \end{minipage}
    \hfill
    \begin{minipage}{0.49\linewidth}
        \centering
        \includegraphics[width=0.9\linewidth]{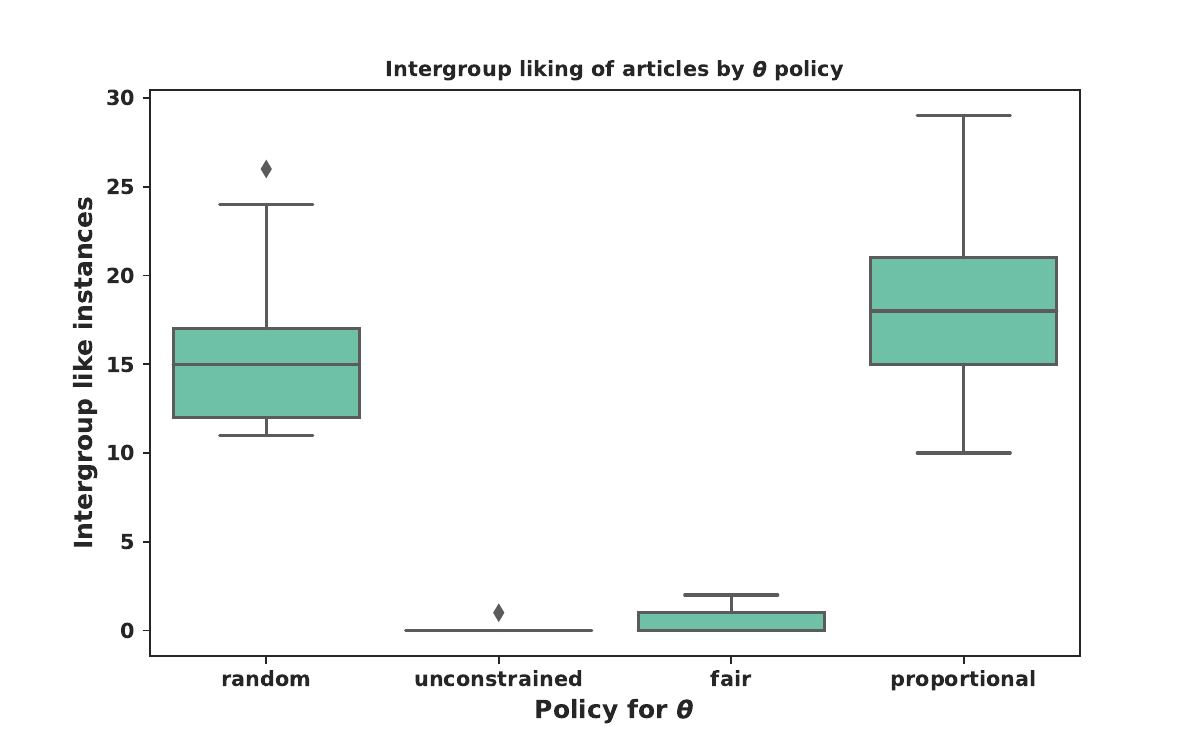}
    \end{minipage}
    \caption{Disparity in clicks and exposure, both en masse and across groups using model parameters from~\citet{balancedExposure}: US Elections.}
    \label{fig:twitter_uselections-boxplots-dispartity}
\end{figure}

The first topic of engagement studied is on articles about presidential candidates in the 2016 US presidential election. 
Figure~\ref{fig:twitter_uselections_pof_boxplot} shows the price of fairness for a uniformly randomized policy (\textit{random}), the fairness-aware optimized policy (\textit{fair}), and the proportional policy, respectively; the price of fairness for the fairness-aware optimization is close to $1$, meaning that performance (measured by number of clicks) of this policy is nearly optimal.
Figure~\ref{fig:twitter_uselections-boxplots-dispartity} shows intergroup and intragroup disparity in exposure and clicks on articles.
In line with the price of fairness, we can observe that the fairness-aware and fairness-agnostic optimization problems yield similar exposure and clicks.
This might be an artifact of the fairness parameters $\overline \delta$ and $\underline \delta$ being too non-restrictive.

\paragraph{Twitter: Brexit~\citep{balancedExposure}}

\begin{figure}[ht]
    \centering
    \includegraphics[width=0.5\linewidth]{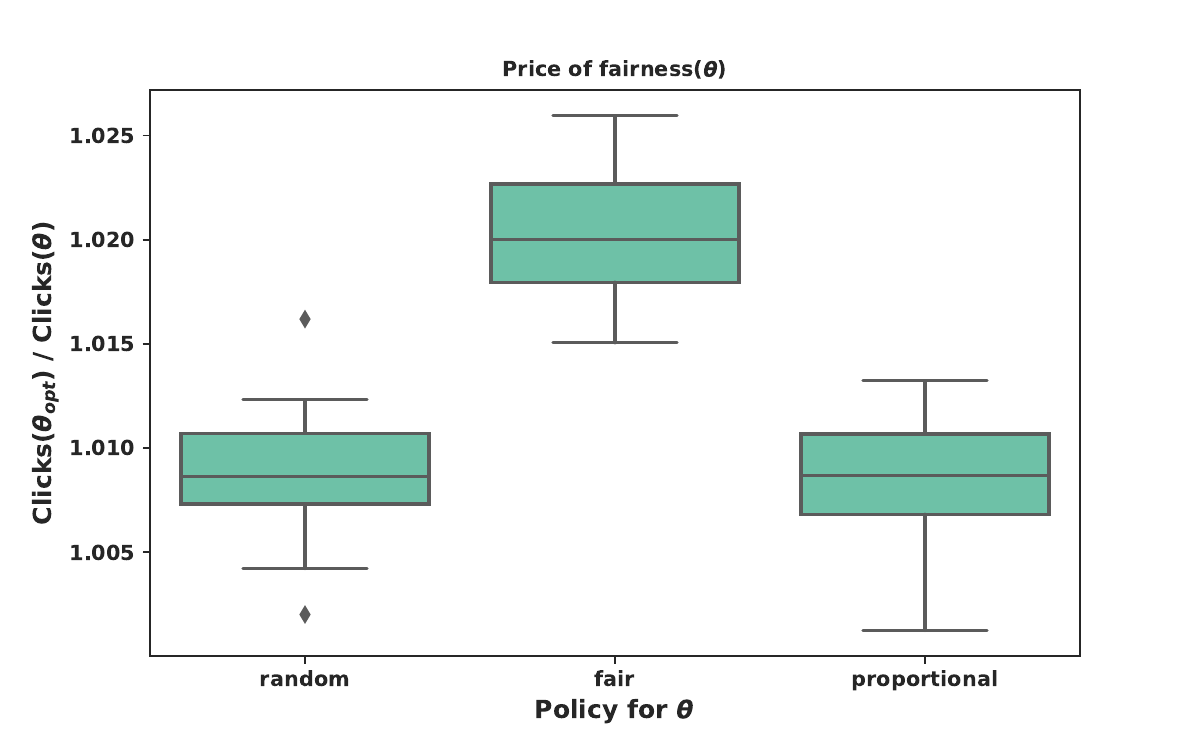}
    \caption{Price of fairness with parameters estimated from~\citet{balancedExposure}: Brexit.}
    \label{fig:twitter_brexit_pof_boxplot}
\end{figure}

\begin{figure}[ht]
    \begin{minipage}{0.49\linewidth}
        \centering
        \includegraphics[width=0.9\linewidth]{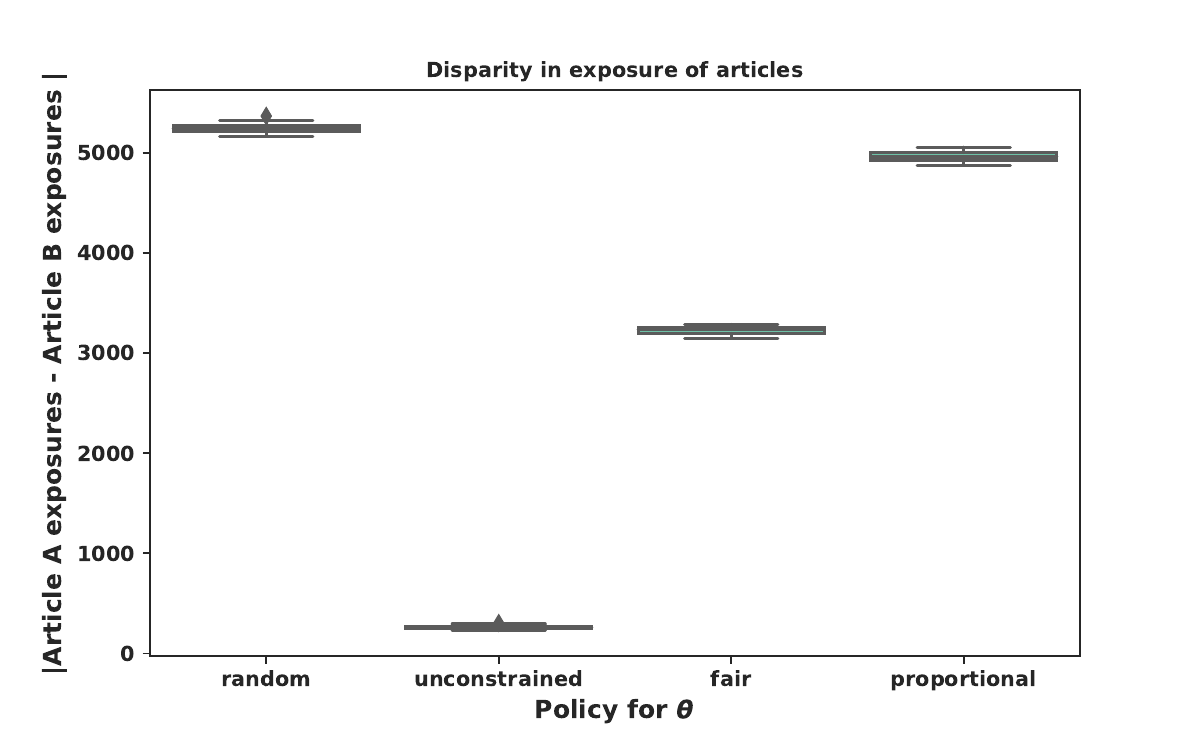}
    \end{minipage}  
    \hfill
    \begin{minipage}{0.49\linewidth}
        \centering
        \includegraphics[width=0.9\linewidth]{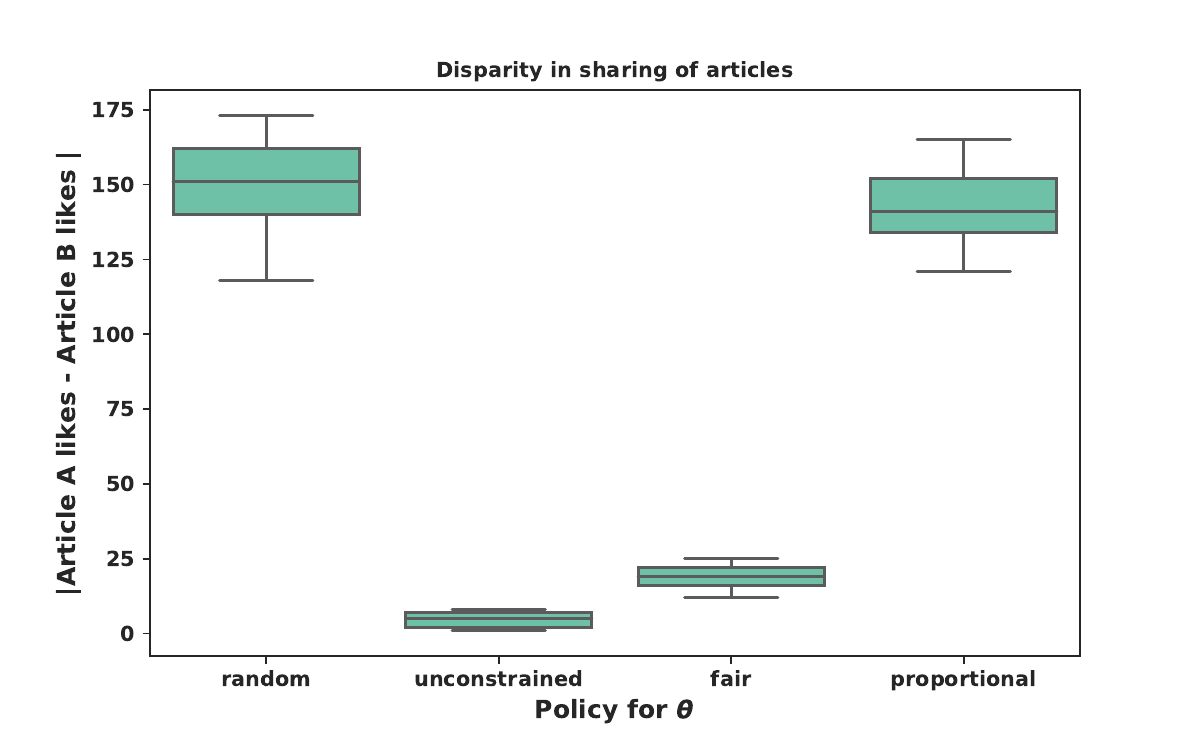}
    \end{minipage}
    \hfill
    \begin{minipage}{0.49\linewidth}
        \centering
        \includegraphics[width=0.9\linewidth]{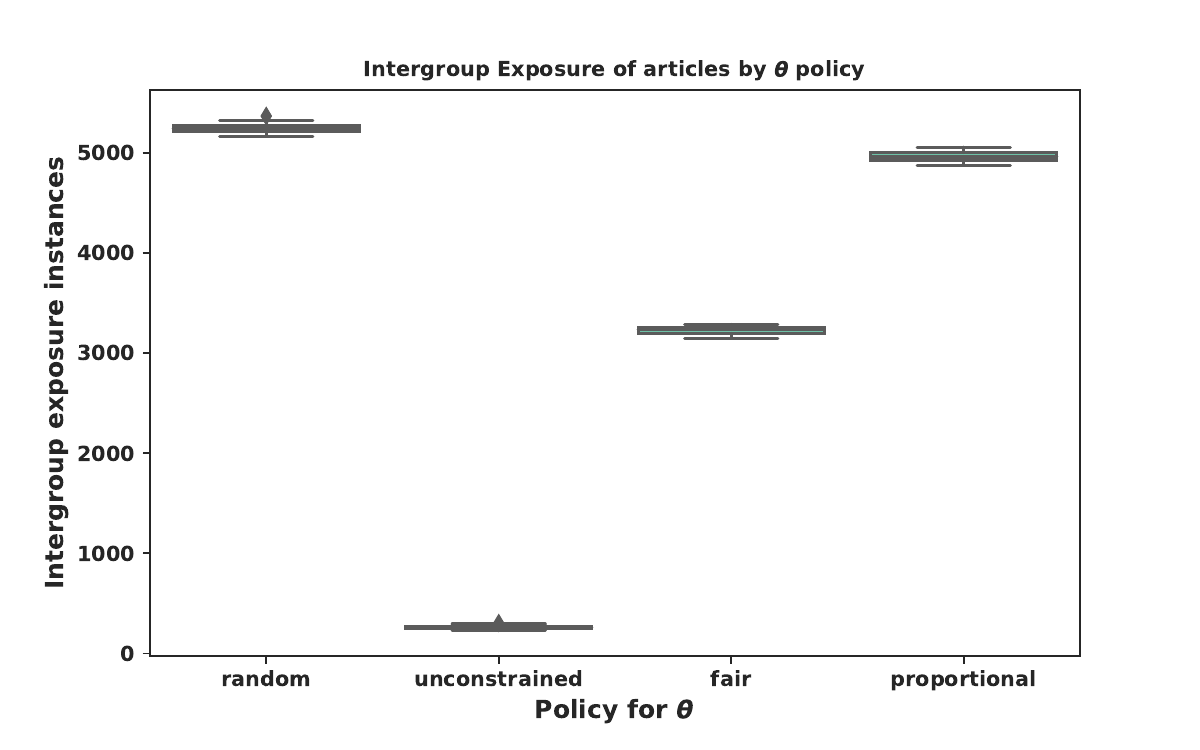}
    \end{minipage}
    \hfill
    \begin{minipage}{0.49\linewidth}
        \centering
        \includegraphics[width=0.9\linewidth]{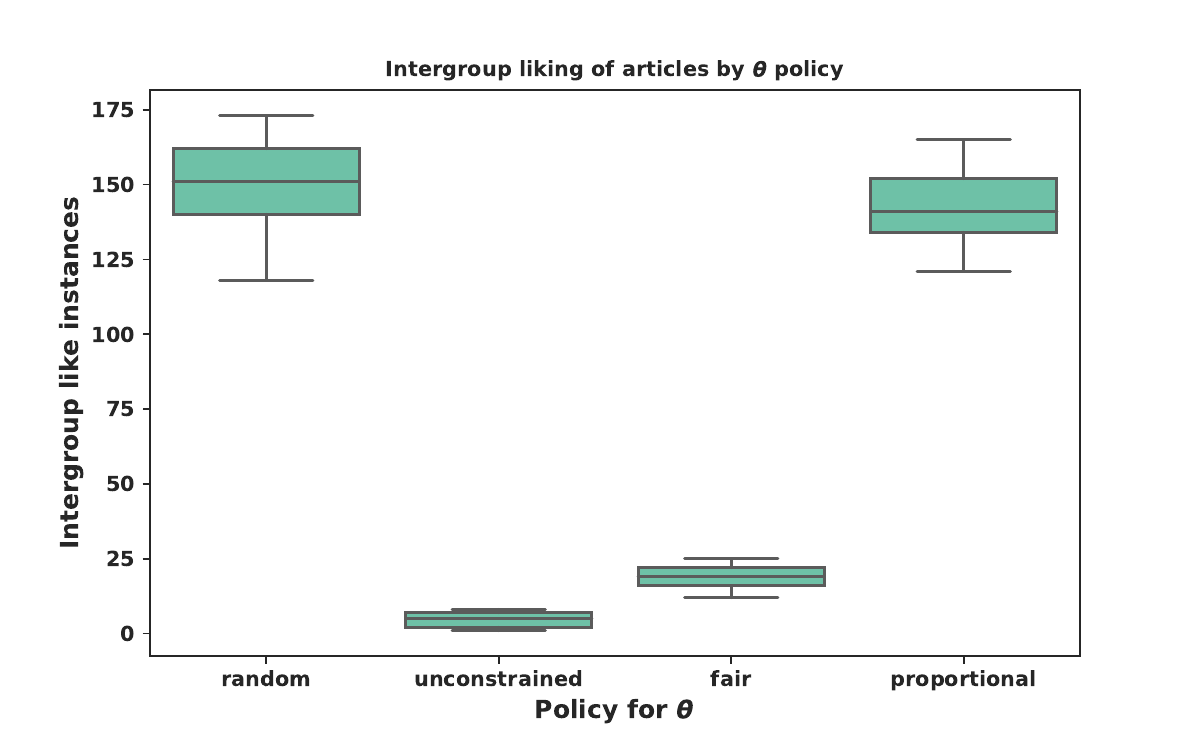}
    \end{minipage}
    \caption{Disparity in clicks and exposure, both en masse and across groups, for~\citet{balancedExposure}: Brexit.}
    \label{fig:twitter_brexit-boxplots-dispartity}
\end{figure}

The next topic studied is on articles about Brexit\footnote{\url{https://www.government.nl/topics/brexit/question-and-answer/what-is-brexit}}. 
Figure~\ref{fig:twitter_brexit_pof_boxplot} shows the price of fairness for three different policies; here, a randomized policy performs closer to optimal than a fairness-aware policy.
Figure~\ref{fig:twitter_brexit-boxplots-dispartity} shows intergroup and intragroup disparity in exposure and clicks on articles.
We conjecture these results might be this way because of the generational divide guiding many opinions on Brexit instead of a solely political one, so the homophily variable $q_B$ (see Table~\ref{tab:exp-param-vals}) is too low for one of the groups, and therefore groups change more often than not. 

\paragraph{Twitter: Abortion~\citep{balancedExposure}}

\begin{figure}[ht]
    \centering
    \includegraphics[width=0.5\linewidth]{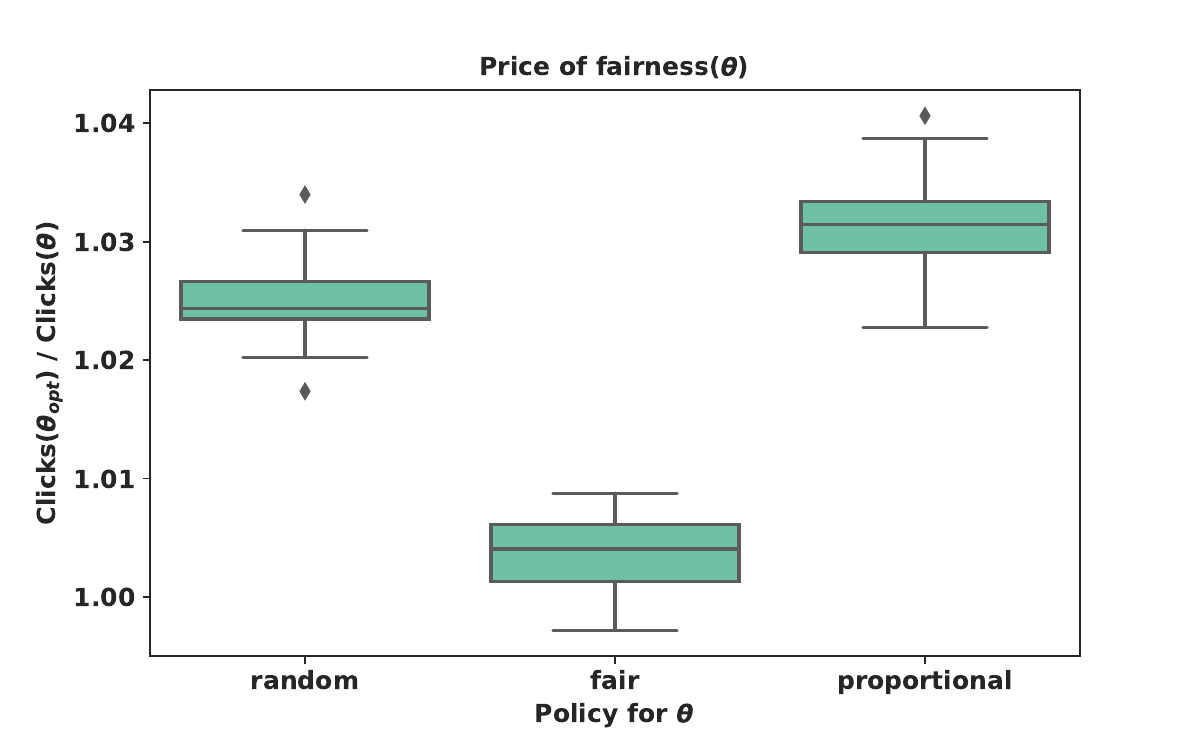}
    \caption{Price of fairness with parameters estimated from~\citet{balancedExposure}: Abortion.}
    \label{fig:twitter_abortion_pof_boxplot}
\end{figure}

\begin{figure}[ht]
    \begin{minipage}{0.49\linewidth}
        \centering
        \includegraphics[width=0.9\linewidth]{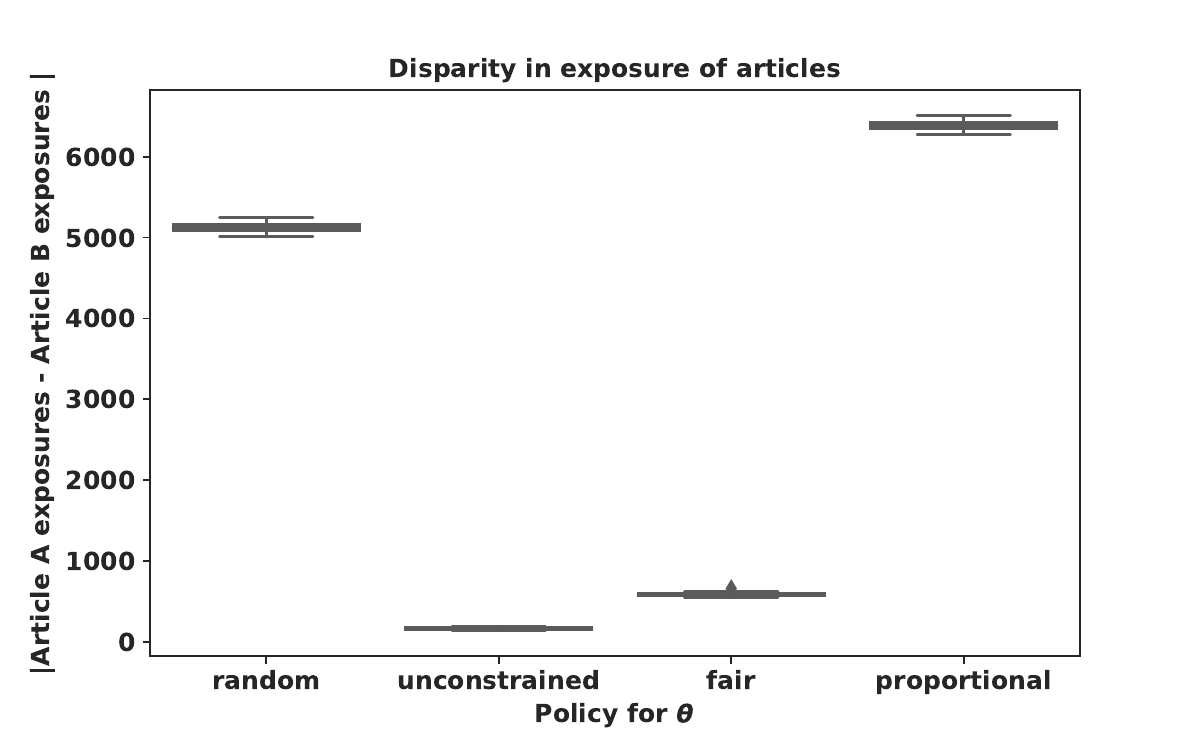}
    \end{minipage}  
    \hfill
    \begin{minipage}{0.49\linewidth}
        \centering
        \includegraphics[width=0.9\linewidth]{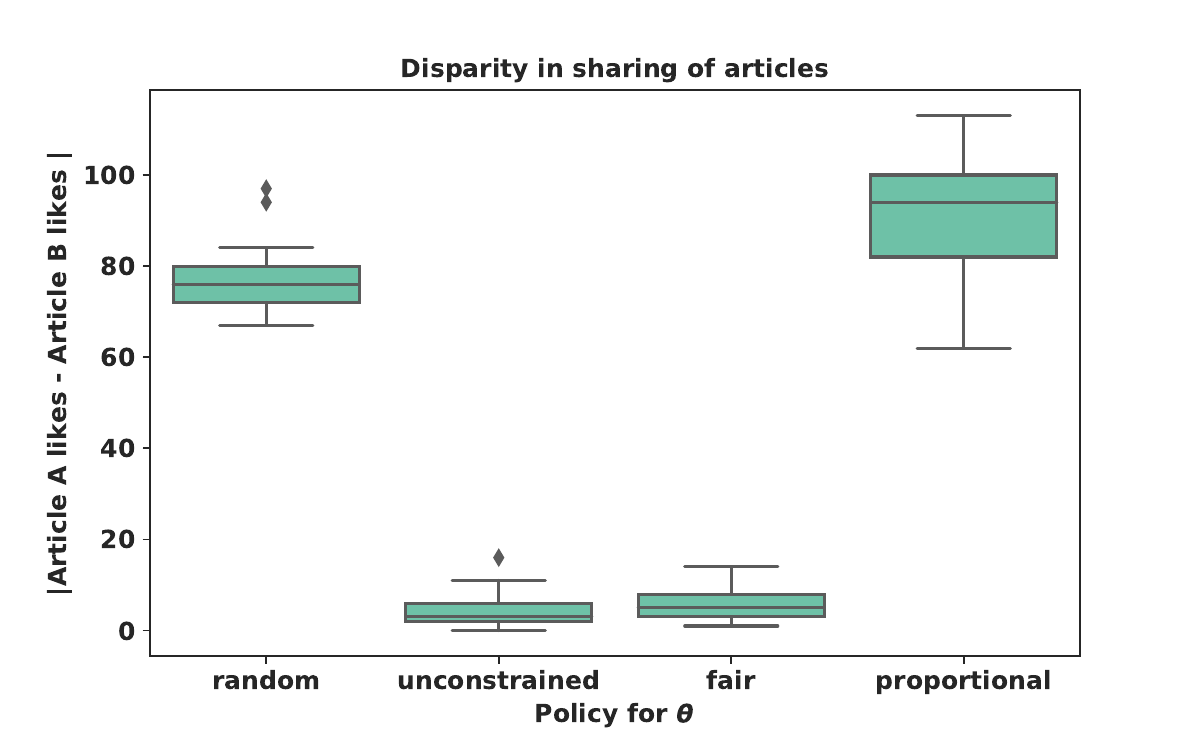}
    \end{minipage}
    \hfill
    \begin{minipage}{0.49\linewidth}
        \centering
        \includegraphics[width=0.9\linewidth]{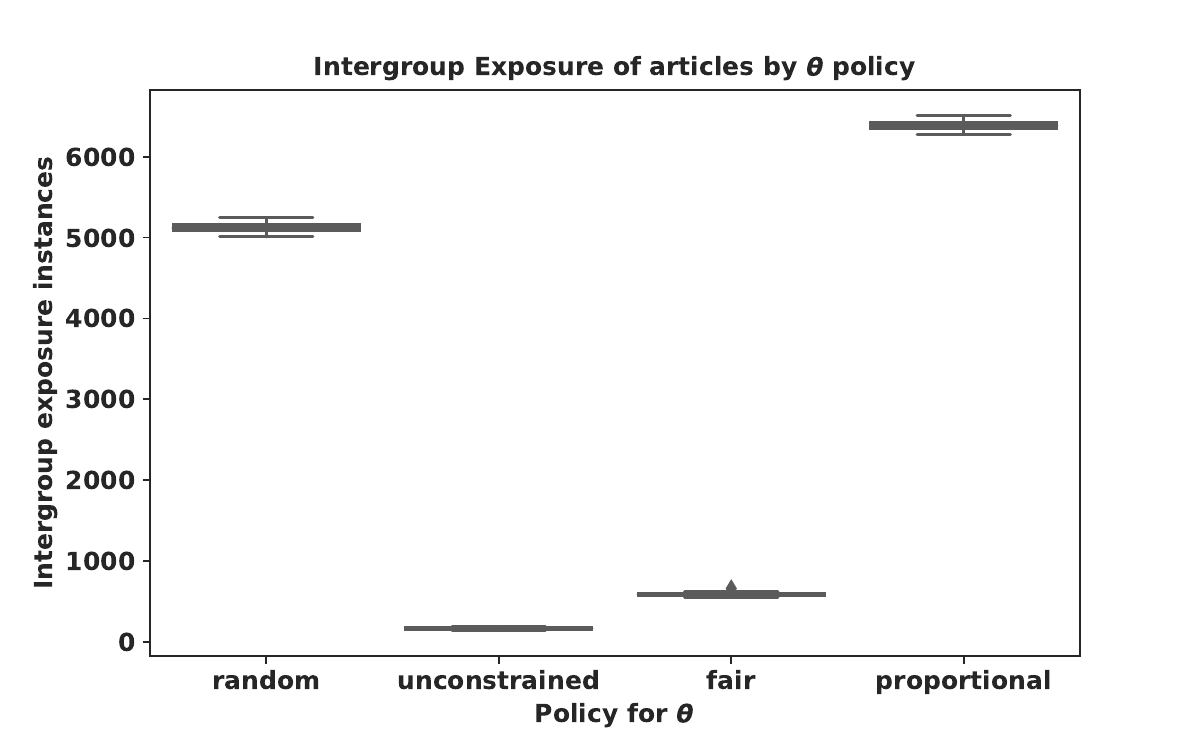}
    \end{minipage}
    \hfill
    \begin{minipage}{0.49\linewidth}
        \centering
        \includegraphics[width=0.9\linewidth]{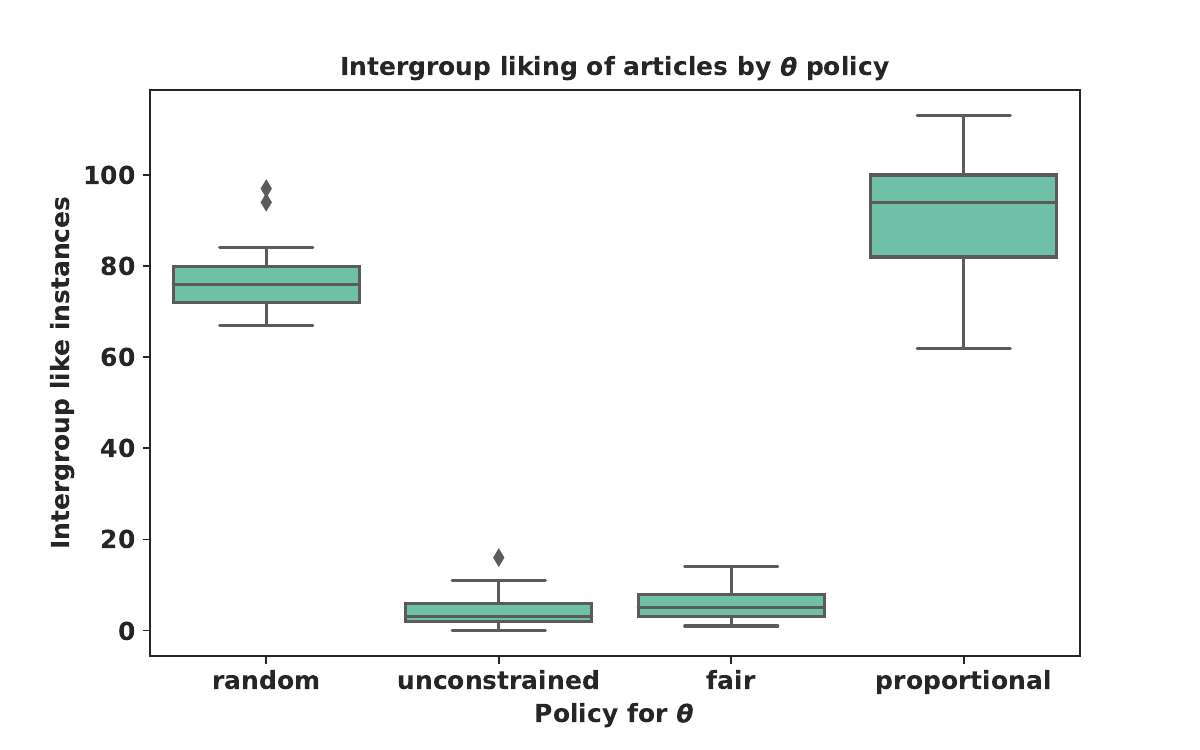}
    \end{minipage}
    \caption{Disparity in clicks and exposure, both en masse and across groups, for  ~\citet{balancedExposure}: Abortion.}
    \label{fig:twitter_abortion-boxplots-dispartity}
\end{figure}

The final topic of engagement studied is on articles about access to abortion services. 
Figure~\ref{fig:twitter_abortion_pof_boxplot} shows the price of fairness for three different policies; the price of fairness for the fairness-aware optimization is close to $1$, though noticably above it, meaning that performance (measured by number of clicks) of this policy is nearly optimal but still constrained.
Figure~\ref{fig:twitter_abortion-boxplots-dispartity} shows intergroup and intragroup disparity in exposure and clicks on articles.
In line with the price of fairness, we can observe that the fairness-aware and fairness-agnostic optimization problems yield similar exposure and clicks.

\section{Comparing Our Model with a Graph Model}\label{sec:one-to-one}

Even though most social network platforms allow users to broadcast content to a large subset of their \emph{friends}, our model imposes that a user can only share an article with one friend at a time. To understand the implications of this assumption, we compare the propagation of articles in our model with that of a network where the assumption does not hold.
We used the Abortion dataset \citep{balancedExposure} for this experiment---the median number of friends of a user in this dataset is 27. In the network setting, we allow every user in the dataset to share the article that they read with all of their friends (adjacent nodes). In the model setting, we uniformly select one friend with whom to share the article, while controlling for all the other parameters of the model (e.g., $\pi_g$, $p_{g,s}$, $\theta_{g,s}$, $M$). We set these parameters to the values in Table~\ref{tab:default-params-data-indep} and propagate two contrasting articles for both groups of the dataset. We repeat this experiment 100 times.

\begin{figure}[ht]
    \centering
    \includegraphics[width=\linewidth]{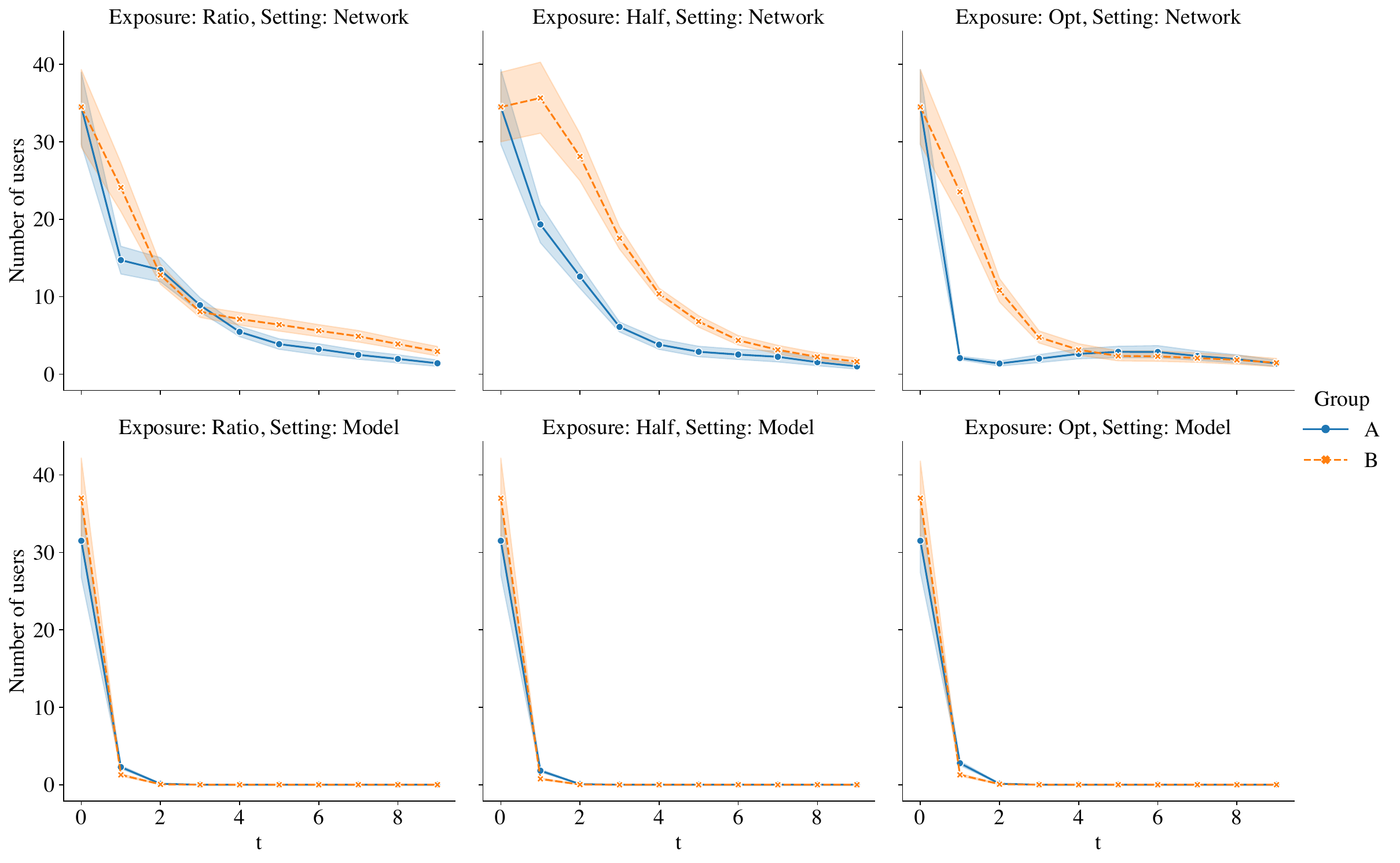}
    \caption{Number of users sharing an article at each time step $t$ in the network model (top) and our model (bottom). Each column is a different exposure level: the first, second and third columns are the solutions for the \textit{ratio} (fairness-aware), \textit{half} (randomized), and \textit{opt} (unconstrained) optimization problems, respectively. The error bands are the 95\% confidence intervals of the number of users.}
    \label{fig:network_vs_model}
\end{figure}

Figure~\ref{fig:network_vs_model} shows the number of users in the system over time in the network model (first row) compared to our model (second row) for different initial exposures (columns).
Note that in this analysis, \textit{half} refers to the random policy, \textit{opt} to the unconstrained policy, and \textit{ratio} to the fairness-aware policy.

It is not surprising that the number of users in the network model does not decrease monotonically as, for each user, we may have several friends sharing the article. By contrast, in our model, the article can only be shared---or not shared---by only one friend. As a result, the article campaigns die out much faster in our model than in the network.

Even though for specific time steps, we see large differences between the network and our model, some of the relative trends that we observe for our model in \S~\ref{sec:simulations} hold for the network model. 
In the top row, we see that the initial exposure levels found by solving the three optimization problems also apply to the network: the groups have similar numbers of users for all time steps in the \textit{ratio} exposure (first column), while optimizing for engagement (third column) leads to the lowest number of users for the least homophilous group ($A$).

Furthermore, in aggregate, we find fewer users that share articles in group $A$ than $B$ for any given exposure level, which is a consequence of the higher homophily of group $B$ in this dataset.
However, homophily has a greater effect on the network model, as shown by the greater number of users in group $B$ for the \textit{half} and \textit{opt} exposures.
Finally, in \S~\ref{subapp:additional-experiments} for the case of the Abortion dataset, we see that the total expected number of shares for \textit{half} exposure is significantly higher than the other exposure levels, and so is in the network model.

To conclude, even though we do not accurately model the propagation of the articles for each propagation time step, our model may be helpful to study the relative effects of homophily in expectation. In addition, the initial exposure obtained by solving the fair exposure optimization problem (\textit{ratio}) in our model also achieves balanced exposure in the network model.

\end{document}